\documentclass[a4paper,12pt]{article}
\usepackage{jcappub} % for details on the use of the package, please
\usepackage{cite}
\usepackage{graphicx}
\usepackage{enumerate}
\usepackage{hyperref}
\usepackage{slashed}

\usepackage{cancel}
\usepackage{color}
\usepackage{graphicx}
\usepackage{amsmath}
\usepackage{amsfonts}
\usepackage{amssymb}
\usepackage{rotating}
\usepackage{booktabs}
\usepackage{xcolor}
\usepackage{soul}

\def\red#1{\textcolor{red}{#1}}

\def\comment#1{}

\title{\boldmath 
{Spacetime-Foam Correlations as the Origin of Cosmological-Constant Dark Energy: Their Effects on Cosmological Evolution and Astrophysical Systems}
}

\author{She-Sheng Xue}

\affiliation{ICRANet Piazzale della Repubblica, 10 -65122, Pescara, Italy
\\ Physics Department, Sapienza University of Rome, 
Rome, Italy\\INFN, Sezione di Perugia, %Via A. Pascoli, I-06123, 
Perugia, Italy
\\ICTP-AP, University of Chinese Academy of Sciences, Beijing, China
}

\emailAdd{xue@icra.it, she-sheng.xue@cern.ch} 

\abstract{Wheeler’s spacetime foam at the Planck scale is characterized by
quantum-gravitational fluctuations that create, oscillate, and
annihilate microscopic spacetime structures. We model these structures
as a Planck-scale lattice using Euclidean quantum gravity on a
simplicial complex. Their long-wavelength collective excitations are
described by interacting scalar fields, which we call ``foamons,'' in
analogy with phonons in a crystal lattice.

Through Coleman--Hawking interactions, foamons couple to quantum or
classical field operators at scales much larger than the Planck length.
These operators provide physical infrared scales that determine the
characteristic correlation lengths of the foamon fields. We use a
Wilsonian renormalisation-group framework to study their
scaling-invariant behaviour near infrared and ultraviolet fixed points.
Integrating out the foamon degrees of freedom generates effective
actions for physical field operators, with the foamon correlation
lengths setting the corresponding macroscopic scales and effective
couplings.

We apply this framework to cosmology and astrophysics, where the
Ricci-scalar sector gives rise to an effective cosmological term
$\Lambda_R\sim\xi_R^{-2}$ with equation of state
$p_\Lambda\approx-\rho_\Lambda$. For a cosmological correlation length
$\xi_R\sim H^{-1}$, this yields
$\rho_{\Lambda_R}\sim M_{\rm pl}^2H^2$, providing a possible mechanism
for the small observed dark-energy scale. The effective cosmological
term $\Lambda_R$ varies in spacetime and interacts with matter through
its dependence on the Ricci scalar. We investigate its effects on
cosmic evolution, galactic rotation curves, and the deflection of
light grazing the solar surface.}

\begin{document}
\maketitle
\flushbottom

%%%%%%%%%%%%%%%%%%%%%%%%%%%%
%

%\vskip1cm
\section{Introduction}

\subsection{Background wormhole dynamics and physical constants}\label{nature}

\begin{figure*}[t]
\centering
\begin{center}
%\vspace{-1.5em}
\includegraphics[height=3.0cm,width=6.8cm]{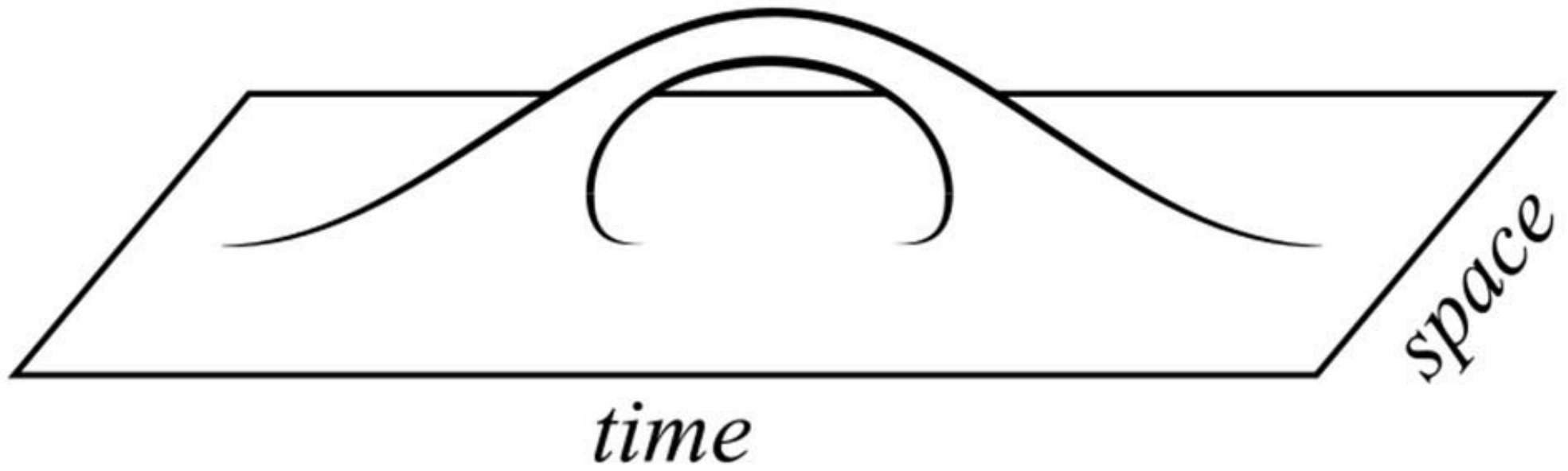}\hspace{0.5em}
\includegraphics[height=3.0cm,width=6.8cm]{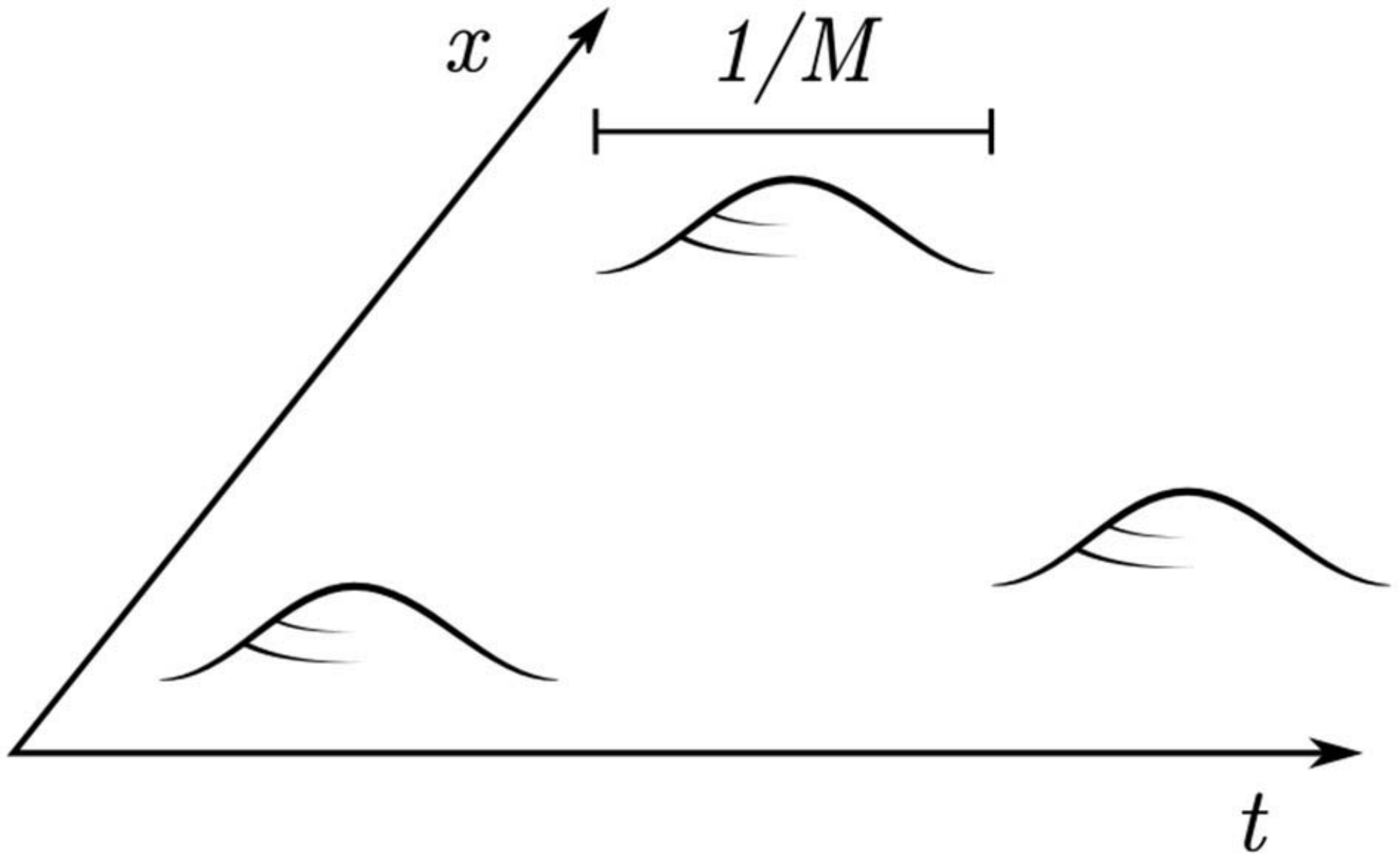}
\caption{The left panel illustrates a Euclidean spacetime wormhole solution, in which an
$R^3$ configuration evolves into an $R^3$ configuration connected to an $S^3$ wormhole,
which is subsequently absorbed, returning the system to an $R^3$ configuration.
The wormhole solution connecting two spacetime points with nontrivial topology is
quantitatively described by the corresponding Euclidean classical solution
\cite{Giddings1988}. The right panel illustrates a background distribution of wormholes
in spacetime. The ``{\it time}'' coordinate on the left and ``$t$'' on the right denote
Euclidean time, corresponding to the tunnelling scale $\beta\sim 1/M$, where $M$ is the
characteristic wormhole mass scale. The figures are adapted from Figs.~1 and 2 of
Ref.~\cite{Hebecker2018}.}  
\label{wormholefig}
\end{center}
\vspace{-2em}
\end{figure*}

\red{Spacetime provides the physical arena for spacetime foams, which may be viewed as a collection of microscopic wormhole-like structures generated by gravitational (spacetime) fluctuations. These structures connect different spacetime points or regions and give rise to nontrivial spacetime topology \cite{Wheeler1955,Wheeler1957,Fuller1962,Regge1961,Hawking1978}, as heuristically illustrated in Fig.~\ref{wormholefig}.}

The laws and constants of nature, as perceived by an ordinary observer in a macroscopic universe, may be associated with a background of wormhole-induced operators acting on the wave-function space of possible universes \cite{Hawking1980,Hawking1982,Strominger1984,Gross1984,Lavrelashvili1988,Banks1988,Coleman1988,COLEMAN1988867,Giddings1988a}. Coleman’s formulation of this idea in the context of the cosmological-constant problem \cite{Coleman1988} is based on the Euclidean path-integral formulation of quantum gravity and the corresponding effective action. In this framework, integrating over wormhole configurations generates nonlocal interactions between field operators,
\begin{eqnarray}
I^{\rm nonlocal}_{E} &=& \int\sqrt{g} d^4x_1 \sqrt{g} d^4x_2 O^\dagger_i(x_1)C_{ij}(x_1,x_2)O_j(x_2)+(\cdot\cdot\cdot),\label{nonlocal}\\
&\sim& \alpha_i^\dagger C^{-1}_{ij}\alpha_j
+\alpha_i\int \sqrt{g} d^4xO_i(x)+(\cdot\cdot\cdot).
\label{aaction}
\end{eqnarray}
The functional two-point kernel \(C_{ij}(x_1,x_2)\) and the functional \(n\)-point kernels \((\cdot\cdot\cdot)\) in the nonlocal action (\ref{nonlocal}) characterize the background dynamics of the wormholes. Their eigenvalues and eigenvectors are related to the auxiliary parameters \(\{\alpha_i\}\) introduced in (\ref{aaction}).\footnote{The path integration over the parameters \(\{\alpha_i\}\) in (\ref{aaction}) reproduces the nonlocal action (\ref{nonlocal}).} The corresponding quantization operators \(\hat{\alpha}_i\) and \(\hat{\alpha}_i^\dagger\) satisfy
\begin{equation}
[\hat\alpha_i,\hat\alpha^\dagger_j]=\delta_{ij},
\label{alphaq}
\end{equation}
and can be interpreted as describing the creation and annihilation of a wormhole of topological type \(i\), with amplitude \(e^{-\beta M/2}\) \cite{Coleman1989}; see Fig.~\ref{wormholefig}.

The ground state \(|\alpha_i\rangle\), or \(\alpha_i\)-vacuum, is defined by
\begin{equation}
(\hat\alpha_i+\hat\alpha_i^\dagger)|\alpha_i\rangle
=\alpha_i|\alpha_i\rangle,
\label{alphav}
\end{equation}
and is characterised by a set of eigenvalues \(\{\alpha_i\}\), which label different degenerate quantum states of the universe that can be connected through tunnelling. This construction is analogous to the topological \(\theta\)-vacuum in gauge theories \cite{Callan:1976je}.

The second term in Eq.~(\ref{aaction}) describes the Coleman--Hawking interaction of background wormholes with the field operators \(O_i\) at length scales much larger than the characteristic wormhole size \(M^{-1}\) \cite{COLEMAN1988867,Coleman1988,HAWKING198839,Hawking1987}:
\begin{equation}
\alpha_i\int \sqrt{g},d^4x,O_i(x).
\label{scalarc2}
\end{equation}
The subscript \(i\) of the field operator \(O_i\) specifies the type of operator associated with a wormhole of topological type \(i\). Consequently, the interaction in Eq.~(\ref{scalarc2}) allows spacetime wormholes to modify the effective couplings and physical constants associated with the corresponding effective field theories. The different operators \(O_i(x)\) represent local operators in effective theories, including the cosmological-constant term \(\rho_\Lambda\), the Einstein--Hilbert term proportional to \(R/G\), and higher-curvature operators such as \(R^2\). At scales much larger than the characteristic wormhole size, the nonlocal action (\ref{nonlocal}) can be represented by a local derivative expansion, yielding an effective gravitational action of the form
\begin{eqnarray}    
I^{\rm nonlocal}_{E}&\Rightarrow& \int\sqrt{g} d^4x \Big[\rho_{_\Lambda} - \frac{1}{16\pi G}R + \gamma R^2 +(\cdot\cdot\cdot)\Big].
\label{eaction}
\end{eqnarray}

\begin{figure*}[t]
\centering
\begin{center}
%\vspace{-1.5em}
\includegraphics[height=4.5cm,width=8.8cm]{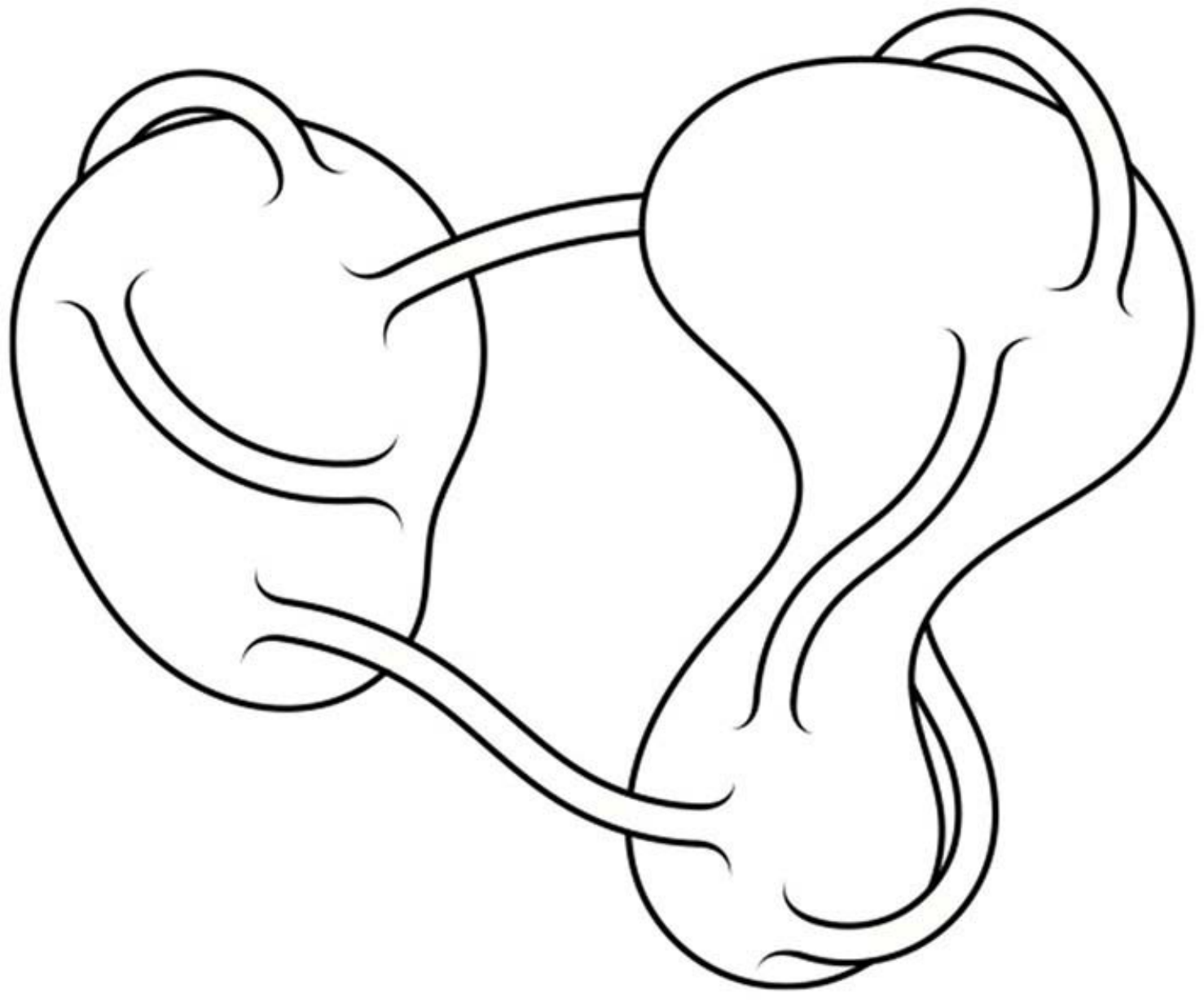}
\caption{The Coleman scenario for background wormhole dynamics. Two (or many) large smooth geometries (spherical universes) connected by wormholes. Large universes connected by wormholes. We adopt this figure from Figure 1 of the Reference 
\cite{Fischler1989}.}  
\label{wormb}
\end{center}
\vspace{-2em}
\end{figure*}

The Coleman formulation \cite{Coleman1988,Preskill:1988na}
assumes the following dynamics for the background wormholes:
\begin{enumerate}[{(i)}]
    \item The Euclidean path-integral amplitudes for different
    $\{\alpha_i\}$-parametrized universes, described by Eq.~(\ref{aaction}),
    are dominated by large-scale universes connected by background wormholes
    characterized by $C_{ij}(x_1,x_2)$, as schematically illustrated in
    Fig.~\ref{wormb}.

    \item For a large-scale spherical universe with a small cosmological
    constant $\Lambda$, the corresponding Euclidean amplitude is of order
    \[
    \exp\left\{\frac{1}{G^2\rho_{_\Lambda}}\right\},
    \]
    where $\rho_{_\Lambda}=\Lambda/(8\pi G)$ is the cosmological-constant
    energy density, which has been introduced {\it a priori} into the
    effective action (\ref{eaction}).

    \item After summing over wormholes connecting pairs of spacetime points
    $x_1$ and $x_2$, the probability distribution for the different
    $\alpha_i$-parametrized universes takes the form \cite{FISCHLER198948}
    \begin{equation}
    \sim e^{-C^{-1}_{ij}\alpha_i\alpha_j}
    \exp\Bigg\{
    \exp\left[
    \frac{3}{8(\Lambda+\alpha_1)(G+\alpha_2)}
    +(\gamma+\alpha_3)+\cdots
    \right]
    \Bigg\},
    \label{prob}
    \end{equation}
    where the effective constants $\Lambda$, $G$, $\gamma$, and $\cdots$
    depend on the parameters $\{\alpha_i\}$.
\end{enumerate}
The observed universe (\ref{eaction}) is therefore identified with the region of \(\{\alpha_i\}\) parameter space in which the probability distribution (\ref{prob}) is strongly peaked, favouring a small positive effective cosmological constant.

In such a formulation, however, the Euclidean path integral suffers from an infrared catastrophe, owing to the absence of physically sensible infrared regulators for the background-wormhole dynamics described by \(C_{ij}\), together with the unbounded behavior of the Euclidean action (\ref{eaction}) for large-scale universes. Consequently, arbitrarily large numbers of macroscopic wormholes \cite{FISCHLER198948} and large-scale spherical universes \cite{Fischler1989} can be nucleated as the infrared scale \(G^2\rho_{\Lambda}\) approaches zero, leading to an infrared catastrophe in Euclidean quantum gravity. The amplitude (\ref{prob}) for the \(\{\alpha_i\}\)-parametrized universes therefore becomes runaway, rendering the parameters \(\alpha_i\) unbounded and the associated probability functional ill-defined. Thus, the Coleman formulation in its original form does not provide a physically well-defined infrared description. Various alternative approaches have subsequently been proposed to address this problem; see, for example, the review in Ref.~\cite{Hebecker2018}. In the article, we address this infrared problem by introducing a physical long-distance regulator through the collective dynamics of spacetime-foam excitations. 

\subsection{\red{Spacetime foam dynamics at Planck distances}}\label{instanton}

\red{In our scenario, we focus exclusively on the background wormhole dynamics at the Planck length \(\ell_{\rm pl}\), which we refer to as spacetime-foam dynamics. As illustrated in Fig.~\ref{wormholefig}, the corresponding wormhole mass scale is taken to be \(M\approx M_{\rm pl}=\ell_{\rm pl}^{-1}\). These Planck-scale structures are attributed to quantum-gravitational fluctuations of spacetime and provide a natural ultraviolet (UV) cutoff, corresponding to a minimal characteristic length scale of order \(\ell_{\rm pl}\). Despite the absence of a well-established quantum field theory of gravity, we develop a heuristic framework below to describe the collective dynamics of these Planck-scale spacetime fluctuations.}

\red{Ref.~\cite{Gross1982} investigates unstable quantum fluctuations around a classical instanton solution in Euclidean quantum gravity, described by the action
\begin{equation}
{\mathcal A}_{E}
=\frac{1}{16\pi G}\int \sqrt{g}d^4xR,
\qquad
Z_E=\int {\mathcal D}[g_{\mu\nu}]e^{-{\mathcal A}_{E}} .
\label{insc}
\end{equation}
The authors find that the gravitational instanton corresponds to a classically unstable configuration that can nucleate through quantum fluctuations. Its nucleation (creation) and absorption (annihilation) \cite{GIDDINGS1989481,Lee1988} may be interpreted semiclassically as tunnelling processes, with an amplitude of the form
\begin{equation}
Z_E\propto e^{-I_E},
\qquad
I_E=\frac{\beta M}{2},
\label{insct}
\end{equation}
where \(I_E\) is the Euclidean gravitational instanton action, \(M\) is the characteristic mass scale of the instanton, and \(\beta\) denotes the characteristic Euclidean tunnelling scale.} 

\red{Viewing a gravitational instanton as a half-wormhole \cite{Preskill1989}, we may heuristically interpret a Planck-scale wormhole, with characteristic mass scale \(M\sim M_{\rm pl}\) (Fig.~\ref{wormholefig}), as a virtual quantum process involving the nucleation of an instanton followed by its absorption (annihilation) at the Planck scale, i.e., $\beta\sim  M^{-1}_{\rm pl}$. In this picture, spacetime-foam dynamics correspond to virtual quantum fluctuations involving the creation and annihilation of instanton--anti-instanton configurations, which contribute to the quantum ground state of Euclidean quantum gravity. These fluctuations are encoded in the spacetime correlation kernels \(C_{ij}(x_1,x_2)\) and higher-point functionals \((\cdot\cdot\cdot)\) appearing in the nonlocal Euclidean action (\ref{nonlocal}). By analogy, the QED vacuum contains virtual quantum fluctuations associated with charged particle--antiparticle pairs, whose contributions are represented by closed-loop Feynman diagrams at characteristic scales set by the Compton wavelength.}

\red{Therefore, we conjecture that the violent quantum-gravitational fluctuations at the Planck length,
\(\ell_{\rm pl}=M_{\rm pl}^{-1}\), can support a large population of Planck-scale wormholes, interpreted as instanton--anti-instanton configurations, with characteristic mass \(M\sim M_{\rm pl}\) and tunnelling length \(\beta\sim\ell_{\rm pl}\). For such configurations,
$ e^{-\beta M/2}\sim\mathcal{O}(1), $
so that the corresponding semiclassical tunnelling amplitude is not exponentially suppressed. If the nucleation of these Planck-scale configurations is energetically and entropically favoured, their number may become as large as permitted by the underlying spacetime geometry and topology. We therefore assume a characteristic wormhole number density of order
$ n_{\rm wh}\sim M_{\rm pl}^{3}, $
corresponding to a close-packed Planck-scale configuration.}

\red{This configuration is the ground state of the quantum spacetime that must exhibit the complex small-scale structure known as
``foam" \cite{Wheeler1955,Wheeler1957,Fuller1962}.
Thus, the physical spacetime made of fundamental ``constituents", foams, becomes a fluctuating foamy {\it Planck lattice} \cite{Preparata1991}, i.e., {\it spacetime foam dynamics,} endowed with a 
fundamental spacetime spacing length $\tilde\ell$, 
\begin{eqnarray}
\tilde\ell\gtrsim \ell_{\rm pl},
\label{flen}
\end{eqnarray}
rendering a natural ultraviolet cutoff $\tilde\Lambda_{\rm cutoff}\approx \pi/\tilde\ell$ for quantum field theories of field operators $O_i$. In this picture, the Planck lattice or spacetime foam dynamics constitutes the microscopic ground-state structure of quantum spacetime and provides the underlying medium whose long-wavelength collective excitations are described below as foamons.}

\subsection{\red{Spacetime foam's collective excitations  and physical constants}}

\red{In contrast to the Coleman formulation (\ref{prob}), our proposal proceeds as follows:
\begin{enumerate}[(i)]
    \item Within the framework of spacetime-foam dynamics, we study the long-wavelength collective excitations of Planck-scale spacetime-foam fluctuations and their Coleman--Hawking coupling (\ref{scalarc2}) to physical field operators $O_i$ at characteristic length scales $\xi_i \gg \ell_{\rm pl}$;
    \item Using the Wilsonian renormalisation-group framework to describe the scale dependence of the ratio $\xi_i/\tilde\ell$, we calculate the contributions of these long-wavelength excitations to the nonlocal correlation functional $C_{ij}(x_1,x_2)$ in the Euclidean path integral (\ref{nonlocal}), with the physical operators $O_i$ providing the relevant infrared (IR) regulators;
    \item Within scale-invariant domains of $C_{ij}(x_1,x_2)$, we construct effective local field theories for the physical operators $O_i$ at length scales $\xi_i \gg \ell_{\rm pl}$. In particular, when $O_i$ is associated with the Ricci scalar $R$ at a macroscopic scale $\xi_R$, we obtain the local Euclidean action (\ref{eaction}), in which the cosmological-constant term emerges \emph{a posteriori} from the long-distance scaling of the foam correlations, with \(\Lambda\propto\xi_R^{-2}\).
\end{enumerate}
}

\red{To further illustrate the spacetime-foam dynamics discussed in
Sect.~\ref{instanton}, we first briefly review the framework of
Euclidean quantum gravity and the correlation functional
$C_{ij}(x_1,x_2)$ defined on a simplicial complex representing a
discrete spacetime lattice with a cutoff
$\tilde\ell\gtrsim\ell_{\rm pl}$. 
For both physical motivation and mathematical tractability, we then
introduce a quasiparticle field, the ``foamon'' field, to provide an
effective description of the long-wavelength collective excitations
of spacetime-foam fluctuations, in analogy with phonons describing
collective vibrational modes in a crystal lattice. 
At this preliminary stage, we present a qualitative formulation of the
proposed framework, to gain further insight into the
long-standing problem of the origin and smallness of the cosmological
constant, which is a central issue in the Coleman formulation.
}

\red{We organise the article as follows. Section~\ref{complex} briefly
reviews Euclidean quantum gravity formulated on a simplicial complex
and discusses its fixed points and scale-invariant domains.
Section~\ref{eucli} examines the spectrum and correlation length of the
foamon field interacting with the physical operators $O_i$, whose characteristic scales provide the relevant infrared regulators.
Sections~\ref{coleman} and~\ref{infrared} employ the Wilsonian
renormalisation-group approach to identify the scale-invariant domains
associated with (i) the infrared fixed point in the symmetric phase and
(ii) the ultraviolet fixed point in the symmetry-breaking phase, where
numerous massive fermionic degrees of freedom are present.
Section~\ref{effeca} demonstrates the emergence of an effective action
for the physical field operators $O_i$.
In Sect.~\ref{effeic}, by identifying $O_i$ with the Ricci scalar $R$,
we derive the Einstein action with a scale-dependent
cosmological term $\Lambda_R$ determined by the macroscopic scale $\xi_R$, together with its equation of state and
its interaction with matter.
Section~\ref{effeia} investigates the effects of an inhomogeneous
$\Lambda_R$ on observables in astrophysical systems.
Finally, we summarise our results and discuss their implications for
vacuum energy.}

%In this article, $G=M^{-2}_{\rm pl}$ is denoted as the the Newton constant, where $M_{\rm pl}$ is the Planck scale. The reduced Planck scale $m_{\rm pl}\equiv (8\pi)^{-1/2} M_{\rm pl}=2.43\times 10^{18} $GeV.

\section{\red{Euclidean quantum gravity in a simplicial complex}}\label{complex}

\begin{figure*}[t]
\centering
\begin{center}
%\vspace{-1.5em}
\includegraphics[height=3.5cm,width=6.0cm]{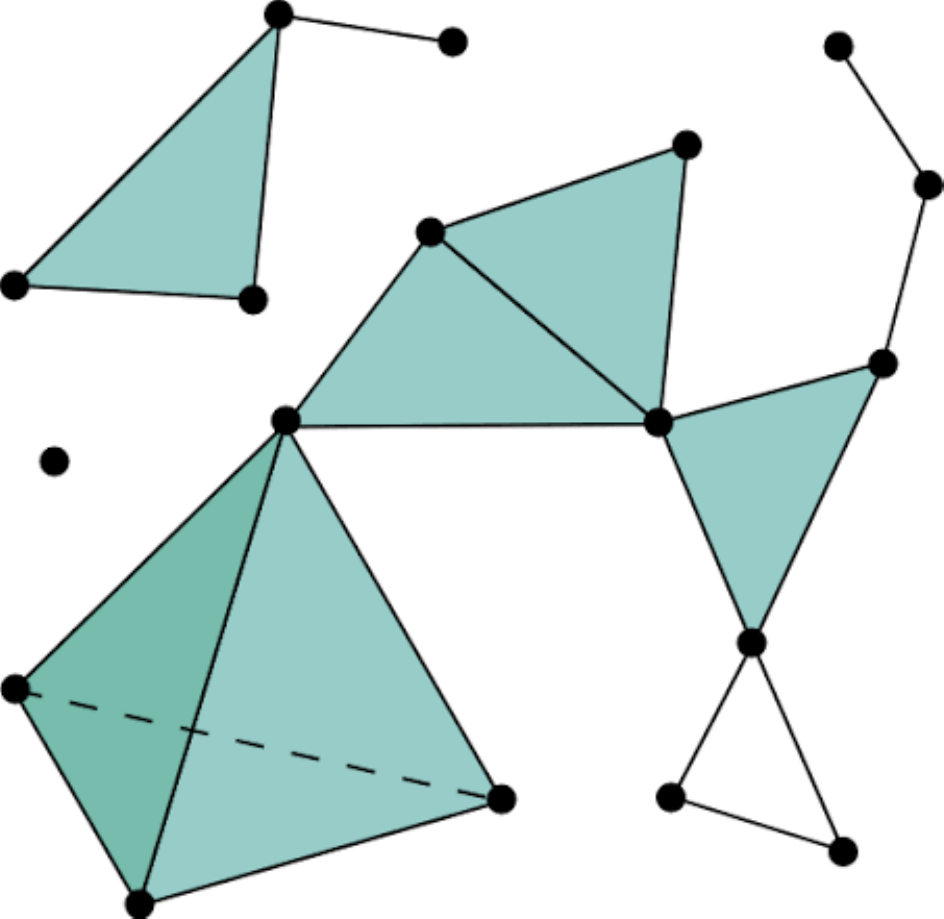}
~~~~~~~~~\includegraphics[height=3.5cm,width=6.8cm]{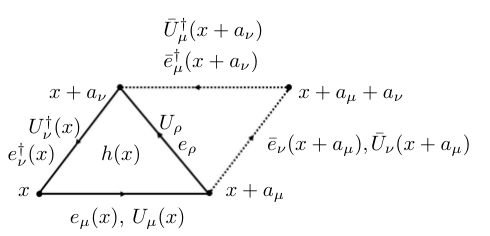}\hspace{0.5em}
\caption{The left figure (https://en.wikipedia.org/wiki/Simplicialcomplex) shows 0-,1-,2-, and 3-simplices of a discrete Euclidean spacetime, i.e., the simplicial complex ${\mathcal M}$ in four dimensions. 
The right figure (Fig.~1 of Ref.~\cite{Xue2009}) illustrates that fermion fields $\psi(x)$ is defined at vertex (0-simplex), tetrad $e_\mu$ and spin-connection fields $U_\mu$ associate to 1-simplex with length $(a_\mu\equiv \tilde\ell_\mu)$, and the smallest
holonomy field $X_h$ (\ref{pact}) associates to 2-simplex $h$.}
\label{simplex}
\end{center}
\vspace{-2em}
\end{figure*}

A complete theoretical description of spacetime-foam dynamics remains
beyond reach because the quantum gravitational dynamics and the
corresponding ground state at the Planck scale are not yet fully
understood. Nevertheless, in this section we develop a qualitative
framework for gaining insight into the dynamics of spacetime foam,
viewed as a foamy Planck lattice, based on nonperturbative Euclidean
quantum gravity \cite{Xue2009, Xue2010, Xue2012}. We formulate the
theory on a background-independent simplicial complex within the
framework of Regge calculus, as illustrated schematically in
Fig.~\ref{simplex}.

\subsection{\red{Invariant loop and line fields in a simplicial complex}}
\label{complexInv}

The basic gravitational variables in Einstein--Cartan gravity are the
tetrad and spin-connection fields,
$
[e_\mu^{\,\,a}(x),\,\omega_\mu^{ab}(x)],
$
which can be represented in Dirac-matrix form as
$
e_\mu(x)=e_\mu^{\,\,a}(x)\gamma_a,\quad
\omega_\mu(x)=\omega_\mu^{ab}(x)\sigma_{ab},
$
where $\gamma_a$, $\sigma_{ab}$, and $\gamma_5$ denote the Dirac
matrices and their associated generators. We introduce the
diffeomorphism- and local-Lorentz-gauge-invariant holonomy variables
\cite{Xue2009,Xue2010}
\begin{eqnarray}
X_{\mathcal C}(e,\omega)
&=&
{\mathcal P}_{\mathcal C}{\rm tr}
\exp\left\{
ig\oint_{\mathcal C}
v_{\mu\nu}(x)\omega^\mu(x)\,dx^\nu
\right\},
\label{pa0s}
\end{eqnarray}
defined along a closed loop $\mathcal C$, and
\begin{eqnarray}
X_{\mathcal L}(e,\omega,\psi)
&=&
\bar\psi(x_1){\mathcal P}
\exp\left\{
ig\int_{\mathcal L}
v_{\mu\nu}(x)\omega^\mu(x)\,dx^\nu
\right\}
\psi(x_N),
\label{fa0s}
\end{eqnarray}
defined along a path $\mathcal L$ connecting two vertices $x_1$ and
$x_N$ of the Euclidean spacetime.

The open-line variable (\ref{fa0s}) describes the parallel transport of
the fermionic field between its two endpoints in the presence of the
gravitational gauge field. Upon identifying the endpoints,
$x_1=x_N$, and integrating over the fermion fields, the open-line
structure can be related to the corresponding closed-loop variable
(\ref{pa0s}). The vertex field is defined at each point $x$ of the
Euclidean spacetime by
\begin{equation}
v_{\mu\nu}(x)
\equiv
\frac{1}{2}\gamma_5\sigma_{ab}
\left(e^a_\mu e^b_\nu-e^a_\nu e^b_\mu\right).
\label{v}
\end{equation}
Here, $g$ is a dimensionless gravitational gauge coupling. The invariant variables (\ref{pa0s}) and (\ref{fa0s}) are analogous,
respectively, to Wilson-loop and Wilson-line variables in lattice gauge theories.

To formulate a well-defined quantisation of Euclidean gravity with a
background-independent ultraviolet (UV) regularisation, we discretise
spacetime on a simplicial complex with a fundamental spacetime length
$\tilde\ell\gtrsim\ell_{\rm pl}$, as illustrated in
Fig.~\ref{simplex}. To each 1-simplex (edge or link), we associate a
tetrad-valued link
$
\tilde\ell_\mu(x)=\tilde\ell\,e_\mu(x),
$
and an $SO(4)$ spin-connection variable
$
U_\mu(x)=\exp\!\left[ig\tilde\ell\,\omega_\mu(x)\right],
$
whose geometric direction and length are determined by the tetrad
$e_\mu(x)$. The smallest closed loop in the simplicial complex is the
boundary of a triangular 2-simplex, or hinge, $h(x)$. Along this loop,
Eq.~(\ref{pa0s}) defines the corresponding elementary holonomy
$X_h(v,U)$. 

\subsection{\red{Simplicial-regularised action and path-integral quantisation}}

We then regularise four-dimensional Euclidean gravity on the
simplicial complex with lattice spacing $\tilde\ell$
(see Eqs.~(120), (124), and (134) of Ref.~\cite{Xue2010}),
\begin{eqnarray}
{\mathcal A}_{\rm EC}(v,U)
&=&
\frac{1}{8g^2}
\sum_{h\in{\mathcal M}}
\left[
X_h(v,U)+X_h^\dagger(v,U)
\right],
\label{pact}
\end{eqnarray}
where the sum extends over all 2-simplices $h(x)$ of the simplicial
complex. The 2-simplex field 
$X_h^\dagger(v,U)$ is associated with the creation operator
$\hat{\alpha}_i^\dagger$, while $X_h(v,U)$ is associated with the
absorption operator $\hat{\alpha}_i$ in Eq.~(\ref{alphav}), providing
a microscopic description of spacetime foam at short distances \footnote{Different types $i$ of operators $\alpha_i$ and
$\alpha_i^\dagger$ correspond to different types of 2-simplex fields
$X_h(v,U)$. In the present work, we restrict our analysis to the
2-simplex fields $X_h(v,U)$ associated with the Ricci scalar $R$ and
do not consider other types of physical quantities.}.
The coupling
$g$ characterises the strength of the interaction between neighbouring
2-simplices $h(x)$. In the naive continuum limit $\tilde\ell\rightarrow0$, the
action ${\mathcal A}_{\rm EC}(v,U)$ reduces to the Euclidean
gravitational action ${\mathcal A}_{\rm E}$ in Eq.~(\ref{insc}),
without a cosmological-constant term.

The quantisation is defined by the Euclidean path-integral partition
function, cf.~Eq.~(\ref{insc}),
\begin{eqnarray}
{\mathcal Z}_{\rm EC}[\tilde\ell,g(\tilde\ell)]
&=&
\int {\mathcal D}e\,{\mathcal D}U\,
e^{-{\mathcal A}_{\rm EC}},
\label{part}
\end{eqnarray}
which describes quantum fluctuations of the gravitational fields on
the four-dimensional Euclidean simplicial complex, interpreted here
as a foamy Planck lattice. For simplicity, we generally suppress the
fermion fields $\bar\psi$ and $\psi$ in the notation unless otherwise
specified. The partition function (\ref{part}) contains, in the
semiclassical limit, contributions associated with the nucleation and
annihilation of gravitational instantons, whose tunnelling amplitude
was discussed heuristically in Sect.~\ref{instanton}.

The quantum gravitational field theory defined by Eq.~(\ref{part})
admits two distinct regimes, characterised by the gravitational
coupling $g$:
\begin{enumerate}[(1)]
    \item the weak-coupling regime, $g^2\simeq 0$, corresponding to the
    quantum ground state $|0\rangle$, in which quantum excitations are
    absent and
    $
    \langle X_h(v,U)\rangle \sim \alpha_i \simeq 0;
    $
    \item the strong-coupling regime, $g^2>0$, corresponding to
    excited states $|\alpha_i\rangle$, for which
    $
    \langle X_h(v,U)\rangle \sim \alpha_i \neq 0,
    $
    as in Eq.~(\ref{alphav}).
\end{enumerate}
Thus, the theory exhibits two qualitatively distinct phases in the
weak- and strong-coupling regimes, respectively, with a phase
transition separating them.

The two-point correlation function between two 2-simplices
$h(x_1)$ and $h(x_2)$ located at different spacetime points is
defined by
\begin{eqnarray}
\left\langle X_{h}(x_1)X^\dagger_{h}(x_2)\right\rangle
&=& {\mathcal Z}^{-1}_{\rm EC}
\int {\mathcal D}e\,{\mathcal D}U\,
X_{h}(v_1,U_1)
X^\dagger_{h}(v_2,U_2)
e^{-{\mathcal A}_{\rm EC}}
\label{cpart}\\
&\sim&
e^{-|x_2-x_1|/\xi},
\nonumber
\end{eqnarray}
where $\xi$ denotes the correlation length of the 2-simplex fields.
The exponential behaviour in Eq.~(\ref{cpart}) characterises the
decay of correlations between local 2-simplex fields separated by a
distance $|x_2-x_1|$, with $\xi$ determining the characteristic range
over which the quantum-gravitational fluctuations remain correlated.
These correlations are encoded in and provide the microscopic basis
for the connection kernels $C_{ij}(x_1,x_2)$ 
that appear in 
nonlocal Euclidean gravitational action (\ref{nonlocal}).

The gravitational coupling \(g(\tilde\ell)\) depends on the running lattice spacing \(\tilde\ell\), which approaches the Planck length in the ultraviolet limit, \(\tilde\ell\rightarrow\ell_{\rm pl}\). In the naive continuum limit, \(\tilde\ell\rightarrow0\), the simplicial complex approaches a continuous Euclidean manifold, and Eqs.~(\ref{pact}) and (\ref{part}) reduce to the classical Euclidean gravitational action in Eq.~(\ref{insc}), with the gravitational coupling given by
\begin{eqnarray}
g(\tilde\ell)=\frac{4}{3}\frac{\tilde G}{G},
\label{g}
\end{eqnarray}
where \(\tilde G\) denotes the running gravitational coupling, while
$G=\ell_{\rm pl}^{\,2}=M_{\rm pl}^{-2} $
is the Newton constant.

Apart from the background-independent regularisation of the Euclidean gravitational action in Eq.~(\ref{pact}), with a natural ultraviolet (UV) cutoff set by the fundamental length \(\tilde\ell\sim\ell_{\rm pl}\), it is necessary to study its path-integral quantisation in Eq.~(\ref{part}) to determine:
\begin{enumerate}[{(i)}]
\item the different phases of the theory and the phase transitions between them as functions of the running gravitational coupling \(g(\tilde\ell)\) and the characteristic spacetime length \(\tilde\ell\);

\item the critical points associated with these phase transitions and their relation to stable infrared (IR) and ultraviolet (UV) fixed points of the gravitational coupling, denoted by \(g_{\rm ir}\) and \(g_{\rm uv}\), respectively, as the characteristic length scale \(\tilde\ell\) is varied;

\item the scaling-invariant domains governed by stable fixed points, in which the correlation length satisfies \(\xi\gg\tilde\ell\). In these domains, the long-distance behaviour of the theory is governed by the corresponding renormalisation-group (RG) fixed point. Operators that are relevant or marginal at the fixed point determine the long-distance effective theory, whereas irrelevant operators are suppressed by powers of the ratio \(\tilde\ell/\xi\), schematically as \({\cal O}[(\tilde\ell/\xi)^n]\), \(n\geq1\). The resulting effective actions exhibit Wilsonian scaling behaviour under changes of length or momentum scale. Consequently, quantum or classical gravitational field theories at characteristic scales \(\xi\) can be consistently described within the corresponding scaling domain;

\item for systems with a very large characteristic volume, \(\xi\gg\tilde\ell\), the theory must possess a physically sensible infrared (IR) regulator to control possible infrared divergences of the partition function in Eq.~(\ref{part}) and to ensure a well-defined quantum field theory.
\end{enumerate}
If these conditions are satisfied, the resulting framework can provide a consistent effective description of quantum-gravitational dynamics and spacetime foam in four-dimensional Euclidean spacetime \footnote{The qualitative considerations presented here may also apply to other approaches to quantum gravity.}.
In the following sections, we investigate these issues using weak- and strong-coupling expansions in \(g\).

\subsection
{\red{IR-stable fixed point at small coupling}}
\label{irpoint}

For small gravitational coupling, $g\gtrsim 0$, the quantum fields
$\{e,U\}$ are dominated by configurations exhibiting strong
short-range fluctuations on the cutoff scale $\tilde\ell$, with
negligible correlations between neighbouring regions beyond this scale.
The fundamental operators $X_h$ therefore have characteristic
activation energies of order $\tilde\ell^{-1}$. The partition function
in Eq.~(\ref{part}) is dominated by the disordered, symmetry-preserving
phase, for which
$
\langle X_h\rangle=0.
$
Configurations with nonvanishing
$\langle X_h\rangle\sim {\mathcal O}(1)$, are suppressed by the Euclidean action,
schematically as $\exp[-{\cal A}_E]\sim\exp[-{\cal O}(1)/g^2]$.
Consequently, in this disordered phase, the correlation function
$\langle X_{h_1}X^\dagger_{h_2}\rangle$ in Eq.~(\ref{cpart}) decreases
rapidly with the separation between the two fundamental operators.
The corresponding correlation length is therefore of order the cutoff
scale,
$
\tilde\xi \gtrsim \tilde\ell.
$

As the coupling $g\gtrsim 0$ increases, configurations with increasingly
long-range correlations among the $(e,U)$ fields become more important.
In this regime, correlations between the fundamental operators $X_h$
extend over progressively larger distances, and the ratio
$\tilde\xi/\tilde\ell$ increases. This behavior indicates that the
gravitational coupling $g=g(\tilde\ell)$ increases as the running
length scale $\tilde\ell$ decreases toward the Planck length,
$\tilde\ell\rightarrow\ell_{\rm pl}$, corresponding to a flow toward
higher energies. With the $\beta$-function convention
\begin{equation}
\beta(g)\equiv-\tilde\ell\frac{\partial g(\tilde\ell)}
{\partial\tilde\ell},
\label{beta0}    
\end{equation}
this ultraviolet running corresponds to a positive beta function,
$\beta(g)>0$, in the regime
\begin{equation}
g\gtrsim 0,\qquad
\tilde\ell\gtrsim\ell_{\rm pl}.
\label{gl0}    
\end{equation}

Perturbative calculations based on expansions in powers of the small
coupling $g$ are not yet available for the simplicial-complex
formulation of Euclidean quantum gravity. Nevertheless, motivated by
the physical arguments and estimates discussed above, we speculate
that a crossover region may occur near the zero-coupling regime, and associate this region with an infrared fixed point,
\begin{eqnarray}
g_{\rm ir}=\frac{4}{3}\frac{\tilde G_{\rm ir}}{G}\gtrsim0,
\quad
\beta(g_{\rm ir})=0,\quad \frac{\tilde\xi}{\tilde\ell}
\sim |g-g_{\rm ir}|^{-\nu},
\quad \nu>0,
\label{gir}
\end{eqnarray}
in the completely symmetric (disordered) phase, associated
with the early Universe \cite{Xue2015}.

The stability of this fixed point depends on the direction of the RG
flow. With the convention in Eq.~(\ref{beta0}), $g_{\rm ir}$ is
infrared-stable if the RG flow approaches $g_{\rm ir}$ as
$\tilde\ell$ increases toward the infrared, i.e., toward lower
energies. Equivalently, the same fixed point is ultraviolet-unstable
when the flow is followed toward decreasing $\tilde\ell$, corresponding
to higher energies.

\subsection{\red{UV-stable fixed point at large coupling}}
\label{uvpoint}

For large coupling, $g\gg1$, non-perturbative strong-coupling
expansions \cite{Xue2012} show that large-loop holonomy fields
$\langle X_{\mathcal C}\rangle\neq0$, Eq.~(\ref{loop-par1}), are not
exponentially suppressed. In the strong-coupling regime, the
contributions of the smallest-loop holonomy fields $X_h$ can be
combined side by side to form extended surfaces, whose boundaries are
represented by large loops $X_{\mathcal C}$. This behavior is analogous
to the strong-coupling expansion of Wilson loops in lattice gauge
theories. The resulting sum over surfaces leads to an area-law
behavior for the large-loop holonomy,
\begin{eqnarray}
\langle X_{\mathcal C}\rangle
&=&
\int {\mathcal D}e\,{\mathcal D}U\,
X_{\mathcal C}(e,U)
\exp[-{\mathcal A}_{\rm EC}(e,U)]
\label{loop-par0}\\
&\sim&
\int {\mathcal D}A({\mathcal C})\,
\exp[-A({\mathcal C})/\xi^2],
\label{loop-par1}
\end{eqnarray}
where $\int {\mathcal D}A({\mathcal C})$ denotes the functional
measure over surfaces with area $A({\mathcal C})$ bounded by the
large loop ${\mathcal C}$, and $\xi$ is the corresponding correlation
length.

The area law in Eq.~(\ref{loop-par1}) represents the leading
contribution to the partition function in Eq.~(\ref{part}) obtained
after integrating over quantum-gravitational fluctuations associated
with small loops ${\mathcal C}_{\tilde\ell}$ at the short-distance
scale $\tilde\ell$. The correlation length $\xi$ characterises the
spatial extent of the collective dynamics of the simplicial-complex
system in Eq.~(\ref{part}) in the strong-coupling regime. The resulting
area-law behaviour signals an ordered phase, in which the
Planck-lattice configurations develop long-range correlations through
extended surfaces. This phase is contrasted with the disordered,
symmetry-preserving phase at small coupling discussed in the previous
section.

The system undergoes a second-order phase transition from the
symmetric (disordered) phase to a symmetry-breaking (ordered) phase
as the coupling $g$ increases. In the vicinity of the critical
coupling, the running coupling $g(\tilde\ell)$ approaches a
UV-stable fixed point $g_{\rm uv}$ as the cutoff length decreases
toward the Planck scale, $\tilde\ell\rightarrow\ell_{\rm pl}$.
At the transition, extended surfaces bounded by large loops
$X_{\mathcal C}$ proliferate, and their characteristic size becomes
macroscopic. Accordingly, the correlation length satisfies
$\xi\gg\tilde\ell$, with $\xi^2\gg\tilde\ell^2$. Near the UV fixed
point, $g(\tilde\ell)\rightarrow g_{\rm uv}$, the correlation length
and beta function obey the scaling relations \cite{Xue2012}
\begin{eqnarray}
\frac{\xi}{\tilde\ell}
&\propto&
(g-g_{\rm uv})^{-\nu},
\qquad
\beta(g)=-\frac{1}{\nu}(g-g_{\rm uv})<0,
\label{loop-par2}
\end{eqnarray}
for $g\gtrsim g_{\rm uv}$, where the critical exponent is
$\nu=1/2$. The UV fixed point is given by
\begin{eqnarray}
g_{\rm uv}
=\frac{4}{3}\frac{\tilde G_{\rm uv}}{G}
\approx\frac{4}{3},
\qquad
\tilde G_{\rm uv}\approx G.
\label{guv}
\end{eqnarray}
Within the corresponding scaling-invariant domain
${\mathcal D}_{\rm uv}$, the quantum field theory of Euclidean
Einstein--Cartan gravity describing spacetime-foam dynamics coupled
to fermionic fields can be consistently defined according to the
conditions (i)--(iv) listed below Eq.~(\ref{g}). This UV scaling
domain is associated with the later evolution of the Universe
\cite{Xue2015}.

\begin{figure*}[t]
\centering
\begin{center}
%\vspace{-1.5em}
\includegraphics[height=6.5cm,width=10.8cm]{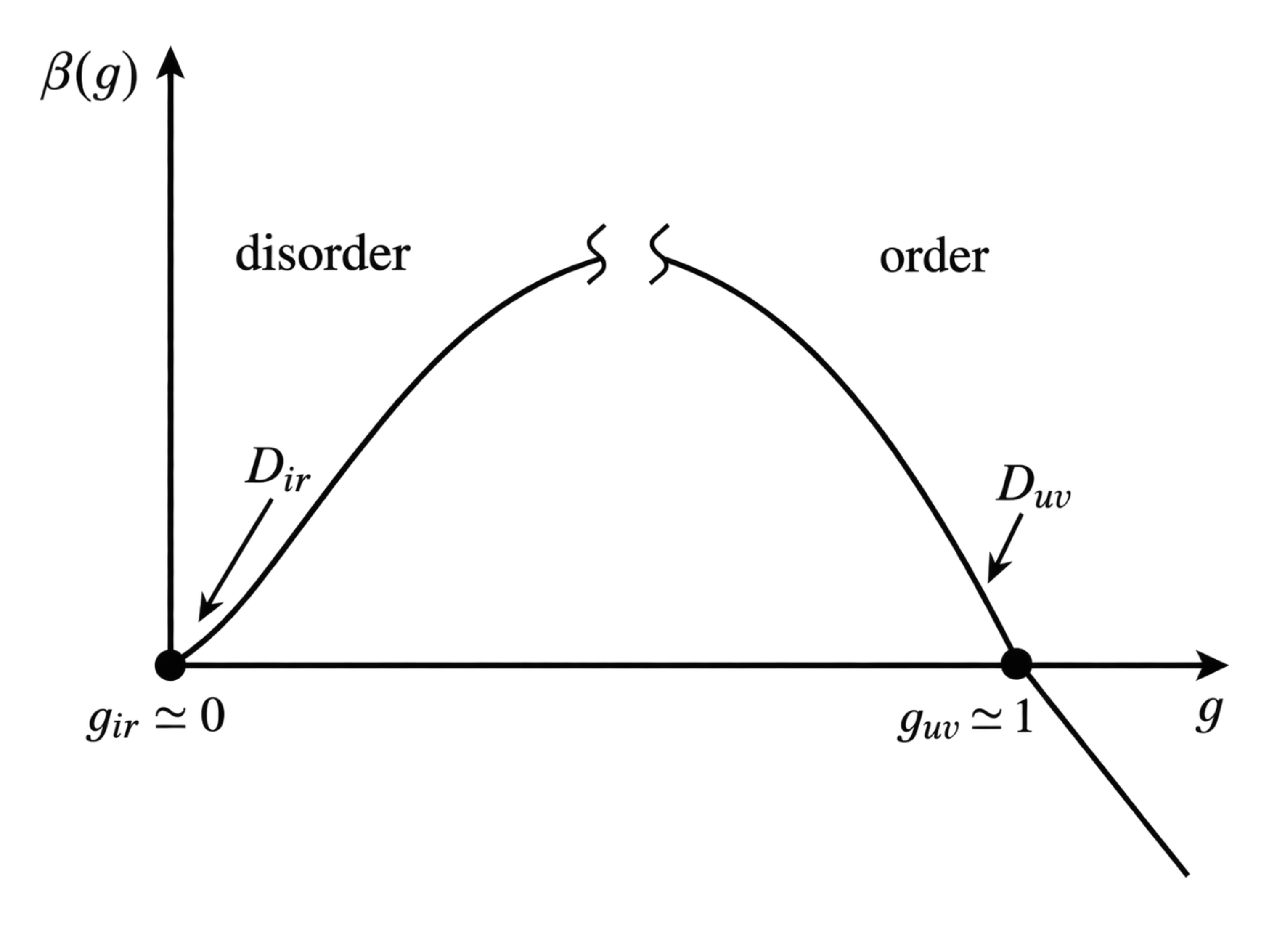}
\caption{Sketch of the $\beta(g)$ function and its fixed points,
$\beta(g_{\rm fix})=0$, for the running gravitational coupling
$g(\tilde\ell)$ as the cutoff length $\tilde\ell$ decreases toward the
Planck length $\ell_{\rm pl}$. Two regimes are illustrated:
(i) the disordered (symmetric) phase, characterised by the
IR-stable (UV-unstable) fixed point $g_{\rm ir}\approx0$ and its
scaling-invariant domain ${\mathcal D}_{\rm ir}$, as discussed in
Eq.~(\ref{gir}); and (ii) the ordered (symmetry-breaking) phase,
characterised by the UV-stable (IR-unstable) fixed point
$g_{\rm uv}\approx4/3$ and its scaling-invariant domain
${\mathcal D}_{\rm uv}$, as discussed in
Eqs.~(\ref{loop-par2}) and (\ref{guv}).}
\label{betaf}
\end{center}
\vspace{-2em}
\end{figure*}

A quantitative determination of the full beta function
$\beta(g)$ as a function of the dimensionless running coupling
$g(\tilde\ell)$ requires numerical simulations. Such simulations
would determine the RG flow connecting the UV-unstable fixed point
$g_{\rm ir}$ in Eq.~(\ref{gir}) to the UV-stable fixed point
$g_{\rm uv}$ in Eq.~(\ref{guv}) as the running length scale
$\tilde\ell$ decreases toward the Planck length
$\ell_{\rm pl}$. They would also allow a quantitative determination
of the scaling-invariant domains, fixed-point values, and critical
exponents. These constitute nontrivial calculations beyond the
qualitative and analytical considerations presented here.
Figure~\ref{betaf} therefore sketches the beta function $\beta(g)$
as a function of $g$, based on the qualitative arguments and
calculations discussed in Sects.~\ref{irpoint} and~\ref{uvpoint}.

\subsection{\red{Cosmological constant emergence from spacetime-foam correlations}}

The area-law result in Eq.~(\ref{loop-par1}) suggests that a
cosmological-constant term can emerge dynamically from the
long-distance correlations of the spacetime foam. In particular, the
dimensionless combination $A(\mathcal C)/\xi^2$ has the same scaling
structure as a contribution to the Euclidean action proportional to
the four-volume, and hence to an effective cosmological term. Such a
term is absent from the regularised microscopic action in
Eq.~(\ref{pact}) at the cutoff scale $\tilde\ell$. Its emergence can
be understood from a Wilsonian coarse-graining procedure as follows.

Following the Wilsonian renormalisation-group approach, rather than
taking the naive continuum limit $\tilde\ell\rightarrow0$, we
integrate out the short-distance quantum degrees of freedom $(e,U)$
associated with small loops
$\mathcal C_{\tilde\ell}$ of characteristic length
$\mathcal C_{\tilde\ell}\sim\tilde\ell$ and area
$A(\mathcal C_{\tilde\ell})\sim\tilde\ell^2$. This procedure produces
an effective partition function
${\mathcal Z}_{\rm EC}[\tilde\xi,g(\tilde\xi)]$ at a larger
coarse-graining scale $\tilde\xi>\tilde\ell$, in which the relevant
degrees of freedom are collective excitations associated with loops
$\mathcal C_{\tilde\xi}$ of characteristic length
$\mathcal C_{\tilde\xi}\sim\tilde\xi$ and area
$A(\mathcal C_{\tilde\xi})\sim\tilde\xi^2$.
The corresponding effective action
${\mathcal A}_{\rm EC}[\tilde\xi,g(\tilde\xi)]$ contains an additional
area-law contribution. Schematically, one may write \cite{Xue2015}
\begin{eqnarray}
\frac{A(\mathcal C_{\tilde\xi})}{\tilde\xi^2}
&\propto&
\sum_{\mathcal C_{\tilde\ell}}
\frac{A(\mathcal C_{\tilde\ell})}{\tilde\xi^2}
\propto
\sum_x
\frac{\tilde\ell^4}
{\tilde\ell^2\tilde\xi^2}
\propto
\sum_x\tilde\ell^4\,
\tilde\rho_{\tilde\Lambda},
\label{vol}
\end{eqnarray}
where the last expression identifies the coarse-grained contribution
with an effective vacuum-energy density,
\begin{eqnarray}
\tilde\rho_{\tilde\Lambda}
\propto
\frac{1}{\tilde\ell^2\tilde\xi^2},
\qquad
\tilde\Lambda
\propto
\frac{1}{\tilde\xi^2}.
\label{cmd}
\end{eqnarray}
After coarse-graining, the area-law contribution to the effective action can be reorganised as a four-volume contribution.

Thus, the cosmological term is not introduced as an independent
microscopic operator at the cutoff scale. Rather, it can emerge as an
effective long-distance contribution generated by the collective
correlations of Planck-scale spacetime-foam degrees of freedom. Eq.~(\ref{cmd}) 
implies
the scaling relation
$
\Lambda\sim\xi^{-2}
$
relates the effective cosmological term to the correlation length of
the Euclidean quantum-gravitational system.
This mechanism differs conceptually from the conventional effective
field-theory treatment, in which a cosmological term
$
\int d^4x\,\sqrt{g}\,\rho_{\Lambda}
$
is included explicitly in the microscopic or bare action. Naive dimensional analysis of a cutoff EFT permits a vacuum-energy contribution of order \(\tilde\ell^{-4}\); the physical cosmological constant is then a renormalised parameter whose observed smallness requires additional explanation. 
In the present
framework, the corresponding long-distance contribution is instead
generated by coarse-graining the correlated quantum fluctuations of
the spacetime foam.

A quantitative demonstration of this mechanism requires numerical
determination of the renormalisation-group flow of the dimensionless
gravitational coupling $g(\tilde\ell)$ and of the correlation length
$\xi(\tilde\ell)$. Such simulations would establish whether the
dimension-two effective cosmological term
$\tilde\Lambda\sim\tilde\xi^{-2}$ is indeed generated dynamically in
the long-distance scaling regime of Euclidean quantum gravity,
starting from the microscopic theory regularised by the fundamental
operators $X_h$ in Eq.~(\ref{pact}) on the Planck-scale simplicial
complex.

These problems are technically challenging. Nevertheless, in order to
gain further insight into the underlying physics, we proceed in an
approximately tractable way by introducing self-interacting scalar
$\Phi_i^4(x)$ theories to describe the long-wavelength collective
modes of spacetime-foam dynamics. Our approach is based on the
following considerations:
\begin{enumerate}[{(i)}]

\item
We describe spacetime-foam dynamics using the Coleman path-integral
framework (Sec.~\ref{nature}) and its simplicial-complex
regularisation (Sec.~\ref{complex}), providing a nonlocal formulation
of quantum Euclidean gravity. Within the simplicial formulation, we
calculate the correlation length $\xi_i$ associated with the
two-point correlation function of the 2-simplex variables,
$
\langle X_h(x_1)X_h^\dagger(x_2)\rangle
$
in Eq.~(\ref{cpart}). 
This correlation function provides a
microscopic realisation of the functional 
two-point kernel 
$C_{ij}(x_1,x_2)$ introduced in Eq.~(\ref{nonlocal}).

\item
We focus on long-range collective excitations satisfying
$\xi_i\gg\tilde\ell$ in the scaling domain associated with the
UV fixed point at gravitational coupling $g\rightarrow g_{\rm uv}=4/3$, as
$\tilde\ell\rightarrow\ell_{\rm pl}$. At these scales, the relevant
degrees of freedom are no longer individual Planck-scale
2-simplex fluctuations but their long-wavelength collective modes.
We describe these collective excitations phenomenologically by
effective scalar fields $\Phi_i(x)$, which may be regarded as
quasiparticle-like modes of the Planck lattice, analogous to phonons
in an ordinary crystal lattice.

\item
Through the Coleman--Hawking interactions in Eq.~(\ref{scalarc2}),
the physical field operators (observables) $O_i$ couple to the collective foamon
fields at distances $\xi_i$ much larger than the Planck scale. We therefore
use the characteristic macroscopic scales $\xi_i$ associated with
these physical $O_i$-processes as infrared scales for the effective
$\Phi_i(x)$ field theories. Within the Wilsonian renormalisation-group
framework, we integrate over the foamon modes $\Phi_i(x)$ in the
Euclidean path integral. The resulting coarse-grained theory
generates an effective action for the physical field operators $O_i$, of
the form given schematically in Eq.~(\ref{nonlocal}), at distances
$\xi_i\gg\ell_{\rm pl}$.

\end{enumerate}

The correlation lengths $\xi_i$ characterise the macroscopic spatial
extent of causally correlated regions of spacetime foams, as probed
through the physical processes described by the field operators $O_i$.
They therefore provide the characteristic infrared scales that
connect the microscopic Planck-scale dynamics of the spacetime foam to the
macroscopic effective field theory of physical operators.

%\footnote{This is reminiscent of the ``vacuum-energy'' density  $\rho^{\rm vac}_{_\Lambda}\approx 1/(\ell^2_{\rm pl} \xi^2)$ rather than $\rho^{\rm vac}_{_\Lambda}\approx 1/(\ell^4_{\rm pl})$ as a candidate for the cosmological term, proposed by V.~G.~Gurzadyan and S.-S.~Xue more than 14 years ago, IJMPA 18 (2003) 561-568, see astro-ph/0105245 and astro-ph/0510495}. 
%follows the RG equation in the theory space of operators. 
%\cite{Reuter1998,ReuterSaueressig2002a,ReuterSaueressig2002b,ReuterSaueressig2012,DouPercacci1998,Souma1999,LauscherReuter2002a,LauscherReuter2002b,PercacciPerini2003a,PercacciPerini2003b,Litim2004,CodelloPercacci2006,CodelloPercacci2009,MachadoSaueressig2008,CodelloPercacciRahmede2008,CodelloPercacciRahmede2009,BenedettiMachadoSaueressig2009,BenedettiMachadoSaueressig2010,Bonanno2012,Niedermaier2009,Niedermaier2010,NiedermaierReuter2006}. 

\section{Foamons describe collective excitations of spacetime foams}\label{eucli}

\red{Using the ultraviolet regulator provided by the simplicial complex,
with fundamental running spacing
$\tilde\ell\rightarrow\ell_{\rm pl}$ and gravitational coupling
$g(\tilde\ell)$, we have briefly discussed the formulation of
Euclidean quantum gravity for spacetime-foam dynamics. In the
scaling domains associated with the fixed points, the correlation
length satisfies $\xi\gg\tilde\ell$, so that short-range modes with
characteristic scale $\tilde\ell$ become irrelevant to the
long-distance dynamics. The corresponding long-wavelength modes,
with characteristic scale $\xi$, are instead important and can be
interpreted as collective excitations of the state
$|\alpha_i\rangle$ in Eq.~(\ref{alphav}) describing the foamy Planck
lattice. We therefore introduce effective ``foamon'' fields
$\Phi_i(x)$ to describe these collective modes, with characteristic
mass scale
$
m_i\sim\xi_i^{-1}
$
and self-interaction terms of the form
$\lambda_i\Phi_i^4(x)$. Here, the index $i$ labels different types of
spacetime-foam excitations, which may be associated with different
microscopic structures or topological sectors. Their two-point
correlation functions,
\[
\langle\Phi_i(x_1)\Phi_j^\dagger(x_2)\rangle,
\]
are introduced as effective long-wavelength descriptions of the
microscopic correlations
$\langle X_h(x_1)X_h^\dagger(x_2)\rangle$ in
Eq.~(\ref{cpart}), and hence of the corresponding two-point
connection kernels $C_{ij}(x_1,x_2)$ in Eq.~(\ref{nonlocal}).
}

\red{Foamon fields are thus self-interacting quasiparticle-like
excitations of the foamy Planck lattice, analogous to phonons,
which describe collective vibrational modes of a crystal lattice.
They are also conceptually analogous to magnons, which describe
collective spin excitations in magnetic systems. In this analogy,
the individual Planck-scale degrees of freedom represented by
$X_h$ constitute the microscopic constituents, while the fields
$\Phi_i$ describe their coherent long-wavelength collective modes.}

\red{The phenomenological $\lambda_i\Phi_i^4$ theories introduced below should therefore be
understood as effective descriptions of the long-wavelength
collective modes of spacetime foam, rather than as direct
microscopic derivations from the underlying simplicial
Euclidean quantum-gravity theory.}

\subsection{Foamons interact with field operators at large distances}

\red{Spacetime foam undergoes quantum fluctuations around the ground
state $|\alpha_i\rangle$, which we heuristically interpret as
gravitational instanton--anti-instanton creation and annihilation
processes, as discussed in Sect.~\ref{instanton}. These quantum
fluctuations are governed by the Euclidean action in
Eq.~(\ref{pact}) on the simplicial complex and, through their
collective dynamics, can give rise to excitations over a broad range
of energy scales, from the Planck scale $M_{\rm pl}$ to much lower
energies. Following the quantisation relations in
Eqs.~(\ref{alphaq}) and (\ref{alphav}), we therefore introduce an
effective self-conjugate scalar field to describe the long-wavelength
collective excitations of spacetime foam,}
\begin{equation}
\Phi_i(x)
=
\hat\alpha_i(x)+\hat\alpha_i^\dagger(x),
\label{scalar}
\end{equation}
\red{in four-dimensional Euclidean spacetime. Here, $i$ labels the
type of spacetime-foam excitation.}

The operators $\hat\alpha_i^\dagger(x)$ and $\hat\alpha_i(x)$ are
interpreted as effective creation and annihilation operators for
type-$i$ foam excitations localized at
$x=(\tau,\mathbf{x})$. At equal Euclidean time, they satisfy
\begin{equation}
[\hat\alpha_i(\tau,\mathbf{x}'),
\hat\alpha_j^\dagger(\tau,\mathbf{x})]
=
\delta_{ij}\,\delta^3(\mathbf{x}'-\mathbf{x}),
\label{comm}
\end{equation}
with all other equal-time commutators vanishing, subject to the
normalisation convention adopted for the Fourier modes.

The field $\Phi_i(x)$ is thus the Hermitian (self-conjugate)
combination of the effective creation and annihilation operators.
It provides a field-theoretic generalisation of the single-mode
quantisation in Eqs.~(\ref{alphaq}) and~(\ref{alphav}) to a
continuum of collective foam excitations propagating over the
simplicial complex. In this effective description, $\Phi_i(x)$
represents the long-wavelength collective degree of freedom
associated with the microscopic spacetime-foam variables $X_h$.

\red{This construction is conceptually related to the third
quantisation framework, in which field operators are introduced for
degrees of freedom associated with quantum spacetime or entire
universes \cite{GIDDINGS1989481}. In the present work, however,
$\Phi_i(x)$ is introduced as an effective collective field for
long-wavelength spacetime-foam excitations, rather than as a
fundamental third-quantised field.}

Foamon field modes at high energies probe the complex topological
structures and strong quantum-gravitational fluctuations of spacetime
foam at the Planck scale. At low energies, the corresponding
long-wavelength foamon modes couple to physical field operators
$O_i$ at macroscopic distances.

\red{We generalise the Coleman--Hawking interaction in
Eq.~(\ref{scalarc2}) to an effective interaction between the quantum
foamon fields $\Phi_i(x)$ in Eq.~(\ref{scalar}) and their associated
physical field operators $O_i(x)$,
\begin{equation}
\int \sqrt{g}\,d^4x\,\Phi_i(x)O_i(x),
\label{scalarc3}
\end{equation}
where the macroscopic scales associated with $O_i$ provide the
relevant infrared (IR) scales for the foamon field theories.
Depending on the quantum numbers and transformation properties of
the microscopic operators $\hat\alpha_i$ or the 2-simplex variables
$X_h$, the effective foamon field $\Phi_i(x)$ may be a scalar,
pseudoscalar, or another appropriate field, according to the type of
operator $O_i$ to which it couples. The microscopic origin and precise
form of the interaction in Eq.~(\ref{scalarc3}) require further
investigation. Here, we use an analogy with condensed-matter systems,
in which collective excitations such as magnons can couple to an
external magnetic field.}

We therefore study an effective self-interacting field theory for the
collective foamon fields $\Phi_i(x)$, coupled through
Eq.~(\ref{scalarc3}) to physical field operators $O_i$. The latter
introduce the relevant infrared scales that regulate the long-distance
behavior of the foamon effective theories. Applying the Wilsonian
renormalisation-group approach, we investigate how integrating over
the foamon modes generates an effective theory for the physical
operators $O_i$ at distances much larger than the Planck scale. In
this description, the correlation length $\xi_i$ of the foamon field
sets the characteristic macroscopic scale of the resulting effective
theory.

The introduction of these physical IR scales provides a possible
resolution of the infrared catastrophe encountered in the original
Coleman--Hawking formulation. In particular, the long-distance
correlations of the foamon fields can generate effective operators
for the physical fields, including the Ricci scalar $R$ and a
cosmological term whose characteristic scale $
\Lambda_R\sim\xi_R^{-2}$.

\red{In this framework, the long-wavelength correlation length of
spacetime foam provides a physical origin for the emergent
cosmological term $\Lambda_R$. Its characteristic scale is determined
by the macroscopic correlation scale associated with the Ricci scalar
$R$ of the spacetime geometry. Thus, the cosmological term is
interpreted as an infrared quantity generated by the collective
long-distance dynamics of Planck-scale spacetime foam, rather than as
an independent microscopic parameter inserted into the bare action.}

\subsection{Foamon field theory for spacetime foam collective excitations}

We consider the following effective Euclidean Lagrangian for the
foamon field \footnote{Gauge symmetries and general covariance are
preserved; the dependence on the background metric $g_{\mu\nu}$ is
not written explicitly.}:
\begin{equation}
{\mathcal L}(\Phi_i,\partial_\mu\Phi_i)
=
\frac{1}{2}(\partial_\mu\Phi_i)^2
+V(\Phi_i),
\label{lw}
\end{equation}
where the effective potential $V(\Phi_i)$ describes the
self-interactions of the foamon field $\Phi_i$ arising from the
collective dynamics of spacetime foam at short distances. Interactions
between different types of foamons, labelled by $i$, are possible but
are not considered here.

We require the potential $V(\Phi_i)$ to be bounded from below. Its
absolute minimum, denoted by $V^{\rm min}$, defines the lowest-energy
and hence stable ground state of the effective theory. For the class
of potentials considered here, the minimum may occur at either (i)
$\Phi_i^{\rm min}=0,$
corresponding to the symmetric (disordered) phase, or (ii)
$
\Phi_i^{\rm min}\neq0,
$
corresponding to the ordered (symmetry-breaking) phase. In our
effective description, these two phases are associated respectively
with $\alpha_i=0$ and $\alpha_i\neq0$ in the ground-state
description $|\alpha_i\rangle$ of Eq.~(\ref{alphav}). We restrict
attention to these two phases and do not consider potentials with
more than two degenerate minima.

Rather than considering the broad class of possible effective
potentials \cite{Palti_2019}, we focus on the relevant and irrelevant
operators that govern the scaling-invariant domains associated with
the fixed points discussed above. We therefore consider the simplest
potential invariant under the discrete transformation
$\Phi_i\rightarrow-\Phi_i$,
\begin{equation}
V(\Phi_i)
=
\frac{1}{2}m_i^2\Phi_i^2
+\frac{\lambda_i}{4}\Phi_i^4
+\cdots ,
\label{vw}
\end{equation}
where the ellipsis denotes higher-order operators. \red{The quartic coupling $\lambda_i$
characterises the strength of self-interactions among the foamon
modes. Its relation to the microscopic gravitational coupling
$g(\tilde\ell)$ in Eq.~(\ref{pact}) is determined only after matching
the effective foamon theory to the underlying simplicial
Euclidean-gravity theory; schematically, one may write
$\lambda_i=\lambda_i[g(\tilde\ell)]$.}

\comment{In four dimensions, the operators $\Phi_i^2$ and $\Phi_i^4$ have
canonical mass dimensions two and four, respectively. Thus,
$\Phi_i^2$ is relevant and $\Phi_i^4$ is marginal by canonical power
counting at the Gaussian fixed point. Higher-dimensional operators
represented by the ellipsis are irrelevant by power counting near
that fixed point. Their precise relevance or irrelevance at an
interacting fixed point is determined by the corresponding RG
scaling dimensions.
The foamon mass parameter $m_i\sim\xi_i^{-1}$ sets the inverse correlation
length of the collective excitation,
where $\xi_i$ characterises the long-wavelength correlations of the
underlying 2-simplex variables $X_h$.
}

\subsection{Foamon field effective action and energy density}

The foamon field ``vacuum-to-vacuum'' transition amplitude is described by the Euclidean path integral
\begin{eqnarray}
e^{-S^{\rm eff}_E(0)}\equiv W_E(0)&=&\prod_i\int {\mathcal D}  \Phi_i ~e^{-\int_x{\mathcal L}(\Phi_i,\partial_\mu \Phi_i)},
\label{genera0}
\end{eqnarray}
where $\int_x\equiv \int_\Omega d^4x =\int_\Omega d\tau d^3x \sqrt{g}$. We define functional integration as 
\begin{equation}
\int {\mathcal D}  \Phi = \prod_i\int^{M_{\rm pl}}_0 {\mathcal D}  \Phi_i= \prod_i\prod_{M_{\rm pl}\ge k >0}\int d  \Phi_i(k).
\label{int}
\end{equation}
The functional integration (\ref{genera0}) is finite and well-defined as a quantum field theory, and
foamon fields $\Phi_i(k)$ fluctuate at all scales $k$ 
(Euclidean momenta) up to the ultraviolet cutoff $M_{\rm pl}$.

In momentum space, the effective Euclidean action is given by \footnote{In the coordinate space, $S_{_E}^{\rm eff}(0) = \sum_i\int_{\rm states}\ln \Delta_i(\alpha_i,x,y)$ and $\int_{\rm states}\equiv \int_x\int_y$.} 
\begin{eqnarray}
S_{_E}^{\rm eff}(0) &=&  \frac{1}{2}\sum_i\int_{\rm states}\ln \Delta_i(\alpha_i,k),
\label{effe}
\end{eqnarray}
where the diagonalised $\Delta_i(\alpha_i,k)$ represents the eigenvalues of the foamon field theory, the sum $\sum_i$ is 
over the topological sector $\alpha_i$, and integration 
$\int_{\rm states}$ is
\begin{eqnarray}
\int_{\rm states}\equiv \int_x \int^{M_{\rm pl}}_{_0} , \quad \int^{M_{\rm pl}}_{_0}\equiv \int^{M_{\rm pl}}_{_0}\frac{d^4k}{(2\pi)^4},
\label{state}
\end{eqnarray}
indicates the sum over all eigenstates of
foamon fields in the momentum phase space $(M_{\rm pl}\ge k>0)$ of a finite spacetime volume $\Omega$. 
Assuming homogeneity on macroscopic scales, we define 
energy ${\mathcal E}=S_{_E}^{\rm eff}(0)
=\Omega \rho_{_W}$ with the energy density
\begin{eqnarray}
\rho_{_W} &=&\frac{1}{2}\int^{M_{\rm pl}}_{_0}\sum_i\ln \Delta_i(\alpha_i,k),
\label{effed_0}
\end{eqnarray}
contributed by foamon fields' fluctuations at all length scales.

The quantity $\rho_{_W}$ is an overall physically irrelevant 
constant; in isolation, it is not directly observable unless foamon-field fluctuations 
couple to observable field operators $O_i$ on given physical scales, through the
Coleman--Hawking interaction in Eq.~(\ref{scalarc3}).

\subsection{Foamon field correlation function and length}

The two-state mixing and two-point connected Green function
(correlation function) of the foamon fields are defined by
\begin{eqnarray}
G^{(2)}_{ij}(x_2-x_1)
&=&
\langle\Phi_i(x_2)\Phi^\dagger_j(x_1)\rangle
-\langle\Phi_i(x_2)\rangle
 \langle\Phi^\dagger_j(x_1)\rangle ,
\label{corr}\\
\langle\cdots\rangle
&=&
W_E^{-1}(0)
\int^{M_{\rm pl}}_{0}{\cal D}\Phi\,
(\cdots)\,
\exp\left[
-\int_x {\cal L}
(\Phi_i,\partial_\mu\Phi_i)
\right].
\nonumber
\end{eqnarray}
Here, we assume translation invariance on macroscopic length scales,
which are much larger than the characteristic microscopic scale of
spacetime foam, taken to be of order the Planck-lattice spacing
$\ell_{\rm pl}$.

In the basis in which the foamon two-point function is diagonal in
the field index, the propagator in the Gaussian approximation takes
the form
\begin{eqnarray}
G^{(2)}_{ij}(x_2-x_1)
&\simeq&
\delta_{ij}
\int\frac{d^4k_E}{(2\pi)^4}
\frac{
e^{ik_E\cdot(x_2-x_1)}
}{
k_E^2+m_i^2
}
\propto
\delta_{ij}\,
\frac{e^{-|x_2-x_1|/\xi_i}}
{|x_2-x_1|^{3/2}},
\label{corr1}
\end{eqnarray}
where 
\begin{equation}
\xi_i\simeq \tilde m_i^{-1}
\label{corr0}    
\end{equation}
is the correlation length of the $i$-th foamon field in the scaling-invariant domain of fixed points, which will be discussed in due course. 
The power-law
prefactor in Eq.~(\ref{corr1}) is characteristic of a massive scalar
propagator in four Euclidean dimensions; for sufficiently large
separations, its exponential dependence is governed by the
correlation length $\xi_i$.

Thus, the two-point correlation function is exponentially suppressed
when
$|x_2-x_1|\gg\xi_i$. The correlation length $\xi_i$ characterizes the
range over which fluctuations of the corresponding foamon field remain
correlated. In the scaling-invariant domains associated with the fixed
points of the foamon field theory, $\xi_i$ provides the characteristic
macroscopic scale of the collective excitation. The effective mass
$\tilde m_i$ may depend on the parameters characterizing the
corresponding $\alpha_i$ sector; the precise relation between
$\tilde m_i$ and $\alpha_i$ depends on the effective foamon-field
dynamics.

More generally, the mixing of multiple foamon states is described by
connected $n$-point Green functions,
\begin{eqnarray}
G^{(n)}_{i_1\cdots i_n}
(x_1,\ldots,x_n)
&=&
\left\langle
\Phi_{i_1}(x_1)\cdots
\Phi_{i_n}(x_n)
\right\rangle_{\rm conn},
\label{ncorr}
\end{eqnarray}
with $n=2,3,4,\ldots$. For theories with a symmetry under which the
effective foamon field is self-conjugate, the odd-point functions may
vanish, in which case the nontrivial connected functions are restricted
to even $n$.

The fluctuating foamon fields $\Phi_i$ are associated with the
ground-state sectors labelled by the parameters $\alpha_i$.
Their correlation functions, Eqs.~(\ref{corr}) and~(\ref{ncorr}),
characterize the propagation and mixing of collective foamon
excitations between different spacetime points and, in the present
interpretation, encode information about transitions between
different $\alpha_i$ sectors.

At macroscopic distances, these foamon correlation functions are
related to the functional kernel
$C_{ij}(x_1,x_2)$ and \(n\)-point kernels \((\cdot\cdot\cdot)\) in Eq.~(\ref{nonlocal}), which describes the
effective coupling induced by spacetime-foam configurations connecting
spacetime points. Thus, the correlation length $\xi_i$ of the
foamon field provides a physical long-distance scale for the
corresponding nonlocal spacetime-foam interaction.

\section{Foamon field regularisation and renormalisation}
\label{coleman}

\subsection{Foamon and field-operator mixing}

At short distances, the quantum dynamics of the foamon fields are
described by a generally interacting effective field theory, with
Lagrangian density ${\mathcal L}$ in Eq.~(\ref{lw}), containing
quadratic and quartic self-interactions as well as higher-dimensional
operators. At length scales much larger than the characteristic
spacetime-foam scale, the long-wavelength foamon fields
$\Phi_i(x)$ couple to physical field operators $O_i(x)$ through a
generalised Coleman--Hawking interaction,
\begin{equation}
\Phi_i(x)O_i(x)
=
\left[
\hat{\alpha}_i(x)+
\hat{\alpha}^{\dagger}_i(x)
\right]O_i(x),
\label{coleman2}
\end{equation}
where, for the operators considered here,
$O_i(x)=O_i^\dagger(x)$. In momentum space, the corresponding
Hermiticity condition is
$O_i(-k)=O_i^\dagger(k)$. The operators $O_i$ are assumed to be
diffeomorphism- and gauge-invariant operators of the effective theory
at length scales much larger than the microscopic spacetime-foam scale.

More generally, interactions can mix different foamon sectors with
different physical operators. We parameterize such mixing by
\begin{equation}
\sum_{i,j} y^{ij}\Phi_i O_j
=
\sum_i \Phi_i O_i
+
\sum_{i\ne j} y^{ij}\Phi_i O_j ,
\label{coleman30}
\end{equation}
where the first term represents the coupling between a foamon field
and the corresponding operator in the same sector, while the second
term describes mixing between different sectors. We normalise the
same-sector coupling to unity, $y^{ii}=1$, and denote a representative
inter-sector mixing coupling by
$y^{ij}\equiv y<1$ for $i\ne j$. The precise values and scale
dependence of the mixing coefficients are determined by the effective
foamon dynamics and the symmetries of the underlying theory.

In the present work, we focus on a foamon field $\Phi\equiv\Phi_i$
that couples to a physical operator
\begin{equation}
O\equiv O(R),
\end{equation}
constructed from the Riemann curvature scalar $R$. We also consider
mixing with fermionic operators, schematically represented by the
left--right bilinear
\begin{equation}
O_j=\bar{\Psi}_L\Psi_R ,
\end{equation}
where $\bar{\Psi}_L$ and $\Psi_R$ denote left- and right-handed
fermionic fields. The corresponding inter-sector coupling is
parameterized by a Yukawa-type coefficient $y<1$. Thus, the relevant
part of the effective interaction can be written schematically as
\begin{equation}
{\cal L}_{\rm int}
=
\Phi O(R)
+
y\,\Phi\,\bar{\Psi}_L\Psi_R
+\cdots .
\label{coleman3}
\end{equation}

\subsection{Field operators at large distances as infrared scales}
\label{irregulator}

The foamon Lagrangian in Eq.~(\ref{lw}) together with the
Coleman--Hawking interaction in Eq.~(\ref{coleman2}) can be incorporated
into the Euclidean generating functional
\begin{equation}
W_E(O_i)
=
\int^{bM_{\rm pl}}_{0}{\cal D}\phi_i
\int^{M_{\rm pl}}_{bM_{\rm pl}}
{\cal D}\hat{\Phi}_i\,
\exp\left\{
-\int_x
\left[
{\cal L}(\Phi_i,\partial_\mu\Phi_i)
+\Phi_iO_i
\right]
\right\},
\label{vacuo0}
\end{equation}
where, schematically, we have separated the foamon field modes into
low- and high-momentum components,
\begin{equation}
\int^{M_{\rm pl}}_{0}
=
\int^{bM_{\rm pl}}_{0}
\int^{M_{\rm pl}}_{bM_{\rm pl}} .
\end{equation}
We introduce the running scale
\begin{equation}
\mu=bM_{\rm pl}<M_{\rm pl},
\qquad 0<b<1 ,
\label{spara}
\end{equation}
which characterises the typical momentum scale associated with the
long-distance field operator $O_i$. The parameter $b$ depends, in
general, on both the foamon sector $\Phi_i$ and the corresponding
operator $O_i$; its sector index is suppressed for simplicity.

Since $\mu=bM_{\rm pl}$ is much smaller than the Planck scale,
the operator $O_i(x)$ varies slowly on the microscopic Planck-length
scale $M_{\rm pl}^{-1}$. Its Fourier components therefore have
dominant support at momenta below the scale $\mu$,
\begin{equation}
O_i(k)\simeq 0,
\qquad
M_{\rm pl}>k\gg\mu ,
\label{rego}
\end{equation}
while
\begin{equation}
O_i(k)\ne0,
\qquad
0\leq k\lesssim\mu .
\label{rego2}
\end{equation}
Thus, $\mu=bM_{\rm pl}$ characterizes the largest momentum scale
relevant to the long-distance operator $O_i$.

Correspondingly, we decompose the foamon field into low- and
high-momentum modes,
\begin{equation}
\Phi_i(k)=
\begin{cases}
\phi_i(k), & 0\leq k<\mu,\\[2mm]
\hat{\Phi}_i(k), & \mu<k\leq M_{\rm pl},
\end{cases}
\label{regp}
\end{equation}
or, equivalently,
\begin{equation}
\Phi_i(x)=\phi_i(x)+\eta_i(x),
\end{equation}
where $\phi_i$ denotes the long-wavelength modes and $\eta_i$
the short-wavelength modes. In a Wilsonian description, the latter
are subsequently integrated out to obtain an effective theory for
the long-wavelength field $\phi_i$.

Because $O_i(k)$ has support predominantly below the scale $\mu$,
its leading interaction with the foamon field involves the
long-wavelength component,
\begin{equation}
\int_x\Phi_i(x)O_i(x)
\simeq
\int_{k<\mu}
\frac{d^4k}{(2\pi)^4}\,
\phi_i(k)O_i(-k),
\label{reg}
\end{equation}
where the momentum-conserving form $O_i(-k)$ has been used.
Consequently, the coupling of $O_i$ to the short-wavelength foamon
modes $\hat{\Phi}_i$ is suppressed when the characteristic momentum
scale of $O_i$ satisfies $\mu\ll M_{\rm pl}$.

Neglecting residual interactions between the low- and high-momentum
sectors at leading order, the generating functional can therefore be
written schematically as
\begin{equation}
W_E(O_i)
\simeq
\int^{\mu}_{0}{\cal D}\phi_i\,
\exp\left[
-\int_x\phi_iO_i
\right]
\int^{M_{\rm pl}}_{\mu}
{\cal D}\hat{\Phi}_i\,
\exp\left[
-\int_x
{\cal L}(\Phi_i,\partial_\mu\Phi_i)
\right].
\label{vacuo}
\end{equation}
The second factor describes the short-distance foamon modes that are
integrated out in the Wilsonian construction, whereas the first factor
contains the long-wavelength foamon modes that remain coupled to the
physical operator $O_i$.

\red{The Planck scale $M_{\rm pl}$ provides the ultraviolet cutoff of the
effective foamon field theory, corresponding to the microscopic
spacetime-foam scale. At large distances, the characteristic momentum
scale
$
\mu=bM_{\rm pl}
$
(\ref{spara}) associated with the physical operator $O_i$ provides a
physical infrared scale for the foamon effective theory. In this
sense, the large-distance field operators act as infrared regulators
of the foamon dynamics. The scale $\mu$ is not an infrared cutoff
imposed externally; rather, it is determined by the characteristic
long-distance variation of the physical operator $O_i$ to which the
foamon field $\Phi_i$ couples. Consequently, the scale $\mu$ selects
the range of foamon correlations that contributes to the effective
interaction. The foamon correlation length $\xi_i$ (\ref{corr0}) is
therefore related to the physical scale $\mu$ through the effective
dynamics of the coupled $\Phi_i$--$O_i$ system. That is, 
$\mu$ sets the physical IR scale relevant to $\xi_i$, 
\begin{equation}
\xi_i(\mu)\sim \frac{c_i}{\mu}\label{muxi}    
\end{equation}
where \(c_i\) is a dimensionless quantity that could be constant in scaling-invariant domains of fixed points of foamon field theories.}

\subsection{General renormalization group equation}\label{app1}

In this section, following the effective-potential method
\cite{Jackiw1974} and the functional renormalization-group approach
\cite{Wetterich1993}, we formulate how the physical field operators
$O_i$ can provide infrared scales for defining an effective action of
renormalized foamon fields and couplings.

Using the generating functional $W_E(O_i)$ (\ref{vacuo}) for the
low-energy foamon modes $\phi_i(k)$ coupled to $O_i(k)$, we introduce
the scale-running parameter
$t=\ln b$. We define the averaged (classical) foamon field by
\begin{eqnarray}
\phi_i^{\rm cl}(k)
\equiv \langle\phi_i(k)\rangle
&=&-\frac{\delta W_E(O_i)}{\delta O_i(k)}.
\label{af1}
\end{eqnarray}
The corresponding effective action is defined through the Legendre
transformation
\begin{eqnarray}
\Gamma(\phi_i^{\rm cl})
&=&-W_E(O_i)-\int_k O_l(k)\phi_l^{\rm cl}(k),
\label{effg1}
\end{eqnarray}
where the repeated index $l$ is summed. The classical field
$\phi_i^{\rm cl}(x)$ and the effective action
$\Gamma(\phi_i^{\rm cl})$ are functionals of the physical operator
$O_i(x)$.

Equation (\ref{af1}) shows that $\phi_i^{\rm cl}$ represents the
average of the low-energy foamon modes $\phi_i(k)$ coupled to the
long-distance operator $O_i$. The inverse relation is
\begin{eqnarray}
O_i(k)
&=&-\frac{\delta\Gamma(\phi_i^{\rm cl})}
{\delta\phi_i^{\rm cl}(k)}.
\label{af2}
\end{eqnarray}

The two-point connected correlation function and its inverse are
\begin{equation}
G^{(2)}_{ij}(k_1,k_2)
=
\frac{\delta^2W_E(O)}
{\delta O_i(k_1)\delta O_j(k_2)},
\qquad
\left[G^{(2)}_{ij}(k_1,k_2)\right]^{-1}
=
\frac{\delta^2\Gamma(\phi^{\rm cl})}
{\delta\phi_i^{\rm cl}(k_1)
 \delta\phi_j^{\rm cl}(k_2)} ,
\label{green}
\end{equation}
with
\begin{equation}
\int_{k_3}
G^{(2)}_{il}(k_1,k_3)
\left[G^{(2)}_{lj}(k_3,k_2)\right]^{-1}
=
\delta_{ij}\delta_{k_1k_2},
\label{greeninv}
\end{equation}
where
$\delta_{k_1k_2}=(2\pi)^4\delta^4(k_1-k_2)$.
The corresponding expressions in coordinate space are obtained by
replacing $k\rightarrow x$, $\int_k\rightarrow\int_x$, and
$\delta_{k_1k_2}\rightarrow\delta_{x_1x_2}
=\delta^4(x_1-x_2)$.

The classical field $\phi_i^{\rm cl}(x)$ and the effective action
generally depend on spacetime. We consider the long-wavelength limit
in which they vary sufficiently slowly over distances of order
$\ell_{\rm pl}$ that they can locally be approximated by constant
background fields. In this approximation, derivative contributions
can be neglected, and the
effective potential determines the effective action,
$
\Gamma(\phi_i^{\rm cl})\simeq
\int_\Omega d^4x\,V_{\rm eff}(\phi_i^{\rm cl}).
$

A variation of the low-energy scale
$\mu=bM_{\rm pl}$ changes the momentum range of the modes coupled to
$O_i$ and hence changes both $\phi_i^{\rm cl}$ and
$\Gamma(\phi_i^{\rm cl})$. Differentiating the Legendre transformation
with respect to $t=\ln b$, we obtain
\begin{eqnarray}
\frac{d\Gamma(\phi_i^{\rm cl})}{dt}
&=&-\frac{dW_E(O)}{dt}
-\frac{d}{dt}\int_k\phi_l^{\rm cl}O_l\nonumber\\
&=&-\int_k
\frac{\delta W_E(O)}{\delta O_l(k)}
\frac{dO_l(k)}{dt}
-\frac{d}{dt}\int_k\phi_l^{\rm cl}O_l\nonumber\\
&=&\int_k\phi_l^{\rm cl}
\frac{dO_l}{dt}
-\frac{d}{dt}\int_k\phi_l^{\rm cl}O_l
=-\int_kO_l(k)
\frac{d\phi_l^{\rm cl}(k)}{dt}.
\label{effv1}
\end{eqnarray}

Using Eqs.~(\ref{af1})--(\ref{green}), the scale variations of the
classical field and the operator satisfy
\begin{eqnarray}
\frac{d}{dt}\phi_l^{\rm cl}(k_1)
&=&-\int_{k_2}
G^{(2)}_{lj}(k_1,k_2)
\frac{dO_j(k_2)}{dt},
\label{rel1}\\
\frac{d}{dt}O_j(k_1)
&=&-\int_{k_2}
\left[G^{(2)}_{jl}(k_1,k_2)\right]^{-1}
\frac{d\phi_l^{\rm cl}(k_2)}{dt}.
\label{rel2}
\end{eqnarray}
Consequently, the scale evolution of the effective action can be
written as
\begin{eqnarray}
\frac{d\Gamma(\phi_i^{\rm cl})}{dt}
&=&
\int_{k_1k_2}
O_l(k_1)G^{(2)}_{lj}(k_1,k_2)
\frac{dO_j(k_2)}{dt}.
\label{effv2a}
\end{eqnarray}
For a translation-invariant foamon theory, the two-point function can
be diagonalised in momentum and sector space. Under this
approximation, Eq.~(\ref{effv2a}) becomes schematically
\begin{eqnarray}
\frac{d\Gamma(\phi_i^{\rm cl})}{dt}
&=&
\frac{\Omega}{2}\sum_j\int_k
G^{(2)}_j(k)\frac{d}{dt}O_j^2(k).
\label{effv2}
\end{eqnarray}
Here, the factor $1/2$ follows after symmetrising the quadratic
variation in $O_j$.

The evolution equation (\ref{effv2}) is structurally analogous to a
functional RG equation of Wetterich type
\cite{Wetterich1993}. In the latter, a scale-dependent quadratic
regulator term of the form
$\phi^\dagger(k){\mathcal R}_k\phi(k)$ is introduced explicitly.
In the present formulation, the physical operator $O_i$, through its
characteristic momentum scale $\mu=bM_{\rm pl}$, plays the analogous
role of introducing an infrared scale into the foamon effective
theory. This analogy should be understood at the level of the
scale-dependent effective description; Eq.~(\ref{effv2}) is not
identical to the standard Wetterich equation.

\red{The general evolution equation (\ref{effv2}), with scaling parameter
$t=\ln b$, describes how the foamon effective theory evolves as the
characteristic scale $\mu=bM_{\rm pl}$ is varied. In the
scaling-invariant domains associated with fixed points, where the
correlation length $\xi_i$ (\ref{corr1}) satisfies
$\xi_i\gg\ell_{\rm pl}$, its solutions are expected to approach
Wilson scaling behavior,
\begin{equation}
\Gamma(\phi_i^{\rm cl})
\simeq
\Gamma^\prime[(\phi_i^{\rm cl})^\prime],
\qquad
\xi_i=\xi_i(O_i)\gg\ell_{\rm pl},
\label{wi}
\end{equation}
where the prime denotes the appropriately rescaled quantities after a
change of scale. In such scaling domains, effective long-distance
field theories of the foamon modes can be defined, and their
correlation lengths
$\xi_i\sim\tilde m_i^{-1}$ (\ref{corr1}) are related to the
characteristic scales selected by the infrared operators $O_i$.}

\red{In general, solving the evolution equation (\ref{effv2}) is difficult,
particularly in the non-perturbative regime of strong foamon
self-interactions. It is therefore useful to identify possible
scaling-invariant domains and their associated fixed points. Motivated
by the qualitative analysis of Euclidean quantum gravity in the
small- and large-coupling regimes in
Sects.~\ref{irpoint} and \ref{uvpoint}, we shall separately study the
corresponding weak- and strong-coupling regimes of the effective
foamon field theory (\ref{lw}), using the Wilson renormalization-group
approach \cite{Wilson1974} as an approximate framework for practical
calculations.}

\subsection{Wilson approach to foamon field renormalisation}
\label{wilsonfoamon}

Since the low-energy (long-wavelength) foamon field $\phi_i(x)$ varies
smoothly on microscopic length scales, we approximate the averaged
classical field by the slowly varying field itself,
$
\phi_i^{\rm cl}(x)\simeq\phi_i(x),
$
and henceforth suppress the superscript ``cl''. We consider one
representative foamon sector and omit the sector subscript ``$i$'' for
simplicity.

Using the Wilson renormalisation-group approach, we integrate out the
high-energy foamon modes $\hat{\Phi}$ in Eq.~(\ref{regp}). The
resulting effective action for the low-energy field $\phi$ can be
written formally as
\begin{eqnarray}
e^{-S_{\rm eff}[\phi]}
&\equiv&
e^{-\int_x{\cal L}_{\rm eff}(\phi,\partial_\mu\phi)}
\nonumber\\
&=&
\int_{bM_{\rm pl}}^{M_{\rm pl}}
{\cal D}\hat{\Phi}\,
\exp\left[
-\int_x
{\cal L}(\Phi,\partial_\mu\Phi)
\right],
\label{wilson}
\end{eqnarray}
where ${\cal L}_{\rm eff}$ is the Wilsonian effective Lagrangian for
the remaining low-energy modes. The Wilson transformation consists of integrating out a high-momentum shell followed by a rescaling of momenta and coordinates.

The integration of the high-energy modes can be implemented
iteratively by integrating out successive thin momentum shells. For
a fixed rescaling factor $b<1$, the $n$-th step integrates over the
momentum shell
\begin{equation}
b^{n+1}M_{\rm pl}<|k|<b^nM_{\rm pl}.
\end{equation}
After integrating out a shell, the remaining theory is rescaled so
that the cutoff is restored to its original value. The corresponding
coordinate and momentum rescalings can be written as
\begin{equation}
x'=bx,
\qquad
k'=k/b,
\qquad b<1.
\label{scaling}
\end{equation}
\comment{or equivalently, depending on the convention for primed and unprimed coordinates,
\begin{equation}
x'=b^{-1}x,\qquad k'=bk.
\end{equation}
The two forms are equivalent provided the convention is used
consistently. Here we adopt the latter convention,
$x'=bx$, $k'=k/b$.}

In coordinate space, the Wilson transformation can be viewed
schematically as a coarse-graining operation: short-distance field
configurations are averaged over cells whose characteristic linear
size is of order
\begin{equation}
\ell_n\sim (b^nM_{\rm pl})^{-1}.
\end{equation}
The coarse-grained field at step $n+1$ is therefore obtained by
averaging the fine-grained field over a cell of characteristic volume
$V_n\sim\ell_n^4$,
\begin{equation}
\phi_{\rm coarse}^{\,n+1}(x)
\simeq
\frac{1}{V_n}
\int_{V_n}d^4y\,
\phi_{\rm fine}^{\,n}(x+y),
\end{equation}
with the corresponding interactions coarse-grained in the same
procedure. This provides the coordinate-space interpretation of the
momentum-shell integration discussed above.

After $n$ successive transformations, the characteristic running
energy scale is
\begin{equation}
\mu_n=b^nM_{\rm pl}.
\end{equation}
The sequence of effective Lagrangians generated in this procedure
defines the RG flow of the foamon effective theory. Near a fixed
point, the correlation length becomes much larger than the
microscopic cutoff scale,
\begin{equation}
\xi\gg\ell_{\rm pl},
\end{equation}
and the theory approaches a scaling-invariant domain. Under
coarse-graining, relevant operators remain important, marginal
operators determine the critical scaling, and irrelevant operators
become progressively suppressed. The effective actions at two
successive scales are then related by Wilsonian scaling,
\begin{equation}
\Gamma'[\phi']
\simeq
\Gamma[\phi],
\label{wiwilson}
\end{equation}
up to the appropriate field and coordinate rescalings.

For a fixed $b<1$, the limit $n\gg1$ gives
$b^n\ll1$ and hence
\begin{equation}
\mu_n=b^nM_{\rm pl}\ll M_{\rm pl}.
\end{equation}
Instead of explicitly keeping the integer number of RG steps $n$, we
may equivalently use a continuously varying parameter
$b_{\rm eff}=b^n$ and write
\begin{equation}
\mu=b_{\rm eff}M_{\rm pl},
\qquad 0<b_{\rm eff}<1.
\label{beff}
\end{equation}
For simplicity, we subsequently denote this effective running
parameter again by $b$. Thus, the notation $\mu=bM_{\rm pl}$ used in
Eq.~(\ref{spara}) represents the continuously varying RG scale rather
than one particular infinitesimal Wilson step.

We emphasise that the Wilson construction adopted here is an
approximate effective-field-theory description. The simple scaling
relations obtained from it are expected to hold only within the
scaling-invariant domains associated with RG fixed points. In general,
a single approximate scaling parameter cannot describe quantitatively
the entire RG trajectory connecting two distinct fixed points, as
schematically illustrated in Fig.~\ref{betaf}. In particular, the
functional form of the beta function and the crossover between the
infrared and ultraviolet fixed points require a complete
non-perturbative treatment of the underlying foamon theory.

For this reason, in studying the scale dependence of the effective
self-interacting $\lambda\Phi^4$ theory in Eq.~(\ref{vacuo0}), coupling to fermions, we apply the
Wilson renormalisation-group construction separately within the
scaling-invariant domains associated with the two fixed points
identified in Sects.~\ref{irpoint} and~\ref{uvpoint}. These domains
are denoted by
\begin{equation}
({\mathcal D}_{\rm ir},g_{\rm ir})
\qquad\text{and}\qquad
({\mathcal D}_{\rm uv},g_{\rm uv}),
\end{equation}
respectively in Fig.~\ref{betaf}.
\red{Note that in the infrared-regulated foamon effective theory, the parameters of
the effective Lagrangian, such as the quartic coupling
$\lambda_i(\mu)$ and the mass parameter $m_i(\mu)$, run with the
renormalisation scale $\mu$. Through the effective description, these
running parameters are related to the gravitational coupling
$g(\tilde\ell)$ and to the correlation length
$\xi(\tilde\ell)$ of the microscopic 2-simplex fields $X_h$, both of
which depend on the gravitational coarse-graining scale $\tilde\ell$,
as introduced in Eqs.~(\ref{pact}) and~(\ref{part}). The relation
between the two scales is defined within the effective theory, with
all dimensionful quantities expressed in units of the fundamental
Planck scale, $M_{\rm pl}=\ell_{\rm pl}^{-1}$. Thus, increasing the
foamon renormalisation scale $\mu$ corresponds to probing higher
energies and shorter distances, whereas decreasing $\mu$ corresponds
to probing lower energies and longer distances. Accordingly,
$
\mu\uparrow~\leftrightarrow{\rm UV},
\quad
\mu\downarrow~\leftrightarrow{\rm IR},
$
while for the underlying simplicial gravitational theory,
$
\tilde\ell\downarrow~\leftrightarrow {\rm UV},
\quad
\tilde\ell\uparrow~\leftrightarrow{\rm IR}.
$
The mapping between $\mu$ and $\tilde\ell$ is therefore an
effective-theory relation rather than an identity.}

\section{Fixed points and scaling-invariant domains}
\label{infrared}

\subsection{Infrared fixed point and scaling-invariant domain at weak coupling}

The field theory in Eq.~(\ref{vw}), evaluated about the symmetric
ground state $\phi=0$, possesses an infrared (IR)-stable Gaussian fixed
point $\lambda_{\rm ir}^{\rm fix}=0$
in the weak-coupling regime. In terms of the running scale
$\mu=bM_{\rm pl}$ introduced in Eq.~(\ref{spara}), the corresponding
renormalisation-group (RG) flow approaches this fixed point as
$\mu$ decreases toward the infrared. 

Within the scaling-invariant domain ${\mathcal D}_{\rm ir}$ of the fixed point $\lambda_{\rm ir}^{\rm fix}=0$, after integrating out short-distance
modes, the long-wavelength foamon field is described by an effective
action of the form
\begin{eqnarray}
\Gamma(\phi_{\rm ir})
&=&
\int d^d x\,{\mathcal L}_{\rm ir}(\phi_{\rm ir})
=
\int_x
\left[
\frac{1}{2}(\partial_\mu\phi_{\rm ir})^2
+\frac{1}{2}m_{\rm ir}^2\phi_{\rm ir}^2
+\frac{\lambda_{\rm ir}}{4}\phi_{\rm ir}^4
+\cdots
\right],
\label{efffree}
\end{eqnarray}
where the mass $m_{\rm ir} \gtrsim 0$ and the coupling $\lambda_{\rm ir} \gtrsim \lambda_{\rm ir}^{\rm fix}$ are RG scaling-invariant.
At the running renormalisation scale
$\mu=bM_{\rm pl}$, the Wilsonian effective action can be expressed in
terms of rescaled coordinates and momenta $(x',k')$ \cite{Peskin1995},
\begin{equation}
\Gamma(\phi)
=
\int d^d x'\,{\mathcal L}_{\rm eff}
=
\int_{x'}
\left[
\frac{1}{2}(\partial'_\mu\phi)^2
+\frac{1}{2}m^2\phi^2
+\frac{\lambda}{4}\phi^4
+\cdots
\right],
\label{domain}
\end{equation}
where $\partial'_\mu=\partial/\partial x'_\mu$. Under the Wilson
rescaling, the field, mass parameter, and quartic coupling transform
according to
\begin{eqnarray}
\phi
&=&
[b^{2-d}(1+\delta Z)]^{1/2}\phi_{\rm ir}
\simeq
b^{(2-d)/2}\phi_{\rm ir},
\nonumber\\
m^2
&=&
(1+\delta Z)^{-1}b^{-2}
(m_{\rm ir}^2+\delta m^2)
\simeq
b^{-2}m_{\rm ir}^2,
\nonumber\\
\lambda
&=&
(1+\delta Z)^{-2}b^{d-4}
(\lambda_{\rm ir}+\delta\lambda)
\simeq
b^{d-4}\lambda_{\rm ir}.
\label{renr}
\end{eqnarray}
Here, $\delta Z$, $\delta m^2$, and $\delta\lambda$ denote radiative
corrections, while the ellipsis in Eq.~(\ref{domain}) represents
counterterms and irrelevant high-dimensional operators. In four dimensions, $d=4$, the mass operator $m^2\phi^2$ is relevant,
whereas the quartic operator $\lambda\phi^4$ is classically marginal. In the scaling-invariant domain, these
corrections may be neglected to leading order, yielding the
Wilson-scaling relation
\begin{equation}
\Gamma(\phi)\simeq\Gamma(\phi_{\rm ir}),
\label{wilsonir}
\end{equation}
independent of $b$, up to the field and parameter rescalings in Eq.~(\ref{renr}), the relevant and marginal
operators determine the long-distance behaviour, whereas irrelevant
operators are progressively suppressed under coarse graining.

For a weakly coupled scalar theory, the perturbative RG equation for
the quartic coupling is
\begin{eqnarray}
\lambda(b)
&=&
\lambda_{\rm ir}
+\beta_\lambda\ln b,
\quad
\beta_\lambda
\equiv
b\frac{\partial\lambda}{\partial b},
\quad
\beta_\lambda^{(1)}
=
\frac{3\lambda_{\rm ir}^2}{16\pi^2}>0.
\label{renr1}
\end{eqnarray}
Since $b$ decreases as the renormalisation scale
$\mu=bM_{\rm pl}$ is lowered, a positive beta function implies that
the quartic coupling decreases toward the infrared. Thus, within the
perturbative weak-coupling regime, the RG flow approaches the Gaussian
IR fixed point,
$
\lambda\longrightarrow
\lambda_{\rm ir}^{\rm fix}=0 
$
and 
$
\beta_\lambda\longrightarrow0.
$
Up to the seven-loop level of small $\lambda$ perturbative calculations, the $\beta$-function 
$\beta^{(7)}_\lambda>0$, 
Ref.~\cite{Shrock:2023xcu} confirms this infrared fixed point and finds no evidence of ultraviolet fixed points at strong couplings $\lambda$. 

We next include a Yukawa interaction (\ref{coleman3}) between the foamon field and
fermions. In the regime of weak $\lambda\phi^4$ interacting and $y\bar\psi_L\psi_R$ Yukawa coupling, using the $\beta$-functions $\beta_\lambda^{(n)}(\lambda, y)$ and $\beta^{(n)}_y(\lambda, y)$ up to two-loop ($n=2$) expansions of small couplings $\lambda$ and $y$, one studies the RG flows of dimensionless couplings, 
\begin{eqnarray}
\lambda(b)=\lambda_{\rm ir}+\beta_\lambda\ln b,\quad y(b)=y_{\rm ir}+\beta_y\ln b ,
\label{rgw}
\end{eqnarray}
where
\begin{eqnarray}
\beta_y
&\equiv&
b\frac{\partial y}{\partial b},
\qquad
\beta_y^{(1)}
=
\frac{5N_f y_{\rm ir}^3}{32\pi^2}>0.
\label{rgwy}
\end{eqnarray}
Here, $N_f$ denotes the total number of fermion species included in
the effective theory.
Starting from sufficiently small initial
values
$\lambda_{\rm ir}\ll1,
$ and $y_{\rm ir}\ll1,
$
the couplings approach the Gaussian fixed point
\begin{eqnarray}
\lambda_{\rm ir}^{\rm fix}=0,
\qquad
y_{\rm ir}^{\rm fix}=0,
\qquad
\beta_\lambda(\lambda_{\rm ir}^{\rm fix},y_{\rm ir}^{\rm fix})
=
\beta_y(\lambda_{\rm ir}^{\rm fix},y_{\rm ir}^{\rm fix})
=0.
\label{fixir}
\end{eqnarray}
The weak-coupling RG flows define the scaling-invariant domain
${\mathcal D}_{\rm ir}$. 

The weak-coupling analysis does not, by itself, determine the behaviour
of the theory at large $\lambda$ and $y$. In particular, as the energy
scale is increased toward the ultraviolet, the RG trajectory may leave
the weak-coupling region. We therefore consider the possibility that
the flow is connected to a nontrivial ultraviolet fixed point,
\begin{eqnarray}
\lambda_{\rm uv}^{\rm fix}\neq0,
\qquad
y_{\rm uv}^{\rm fix}\neq0,
\qquad
\beta_\lambda(\lambda_{\rm uv}^{\rm fix},y_{\rm uv}^{\rm fix})
=
\beta_y(\lambda_{\rm uv}^{\rm fix},y_{\rm uv}^{\rm fix})
=0,
\label{fixuv}
\end{eqnarray}
together with its scaling-invariant domain ${\mathcal D}_{\rm uv}$.
However, the existence and properties of such a nontrivial fixed point
cannot be established reliably within the weak-coupling perturbative
expansion, because $\lambda_{\rm uv}^{\rm fix}$ and
$y_{\rm uv}^{\rm fix}$ would lie outside the perturbative regime.
Nonperturbative calculations are therefore required to determine
whether such a fixed point exists and to establish its critical
properties. We refer the reader to Refs.~\cite{Molgaard2014,
Vladimirov1979} and the literature discussed therein.
In the next section, we investigate, at strong couplings $\lambda$ and $y$, 
the possibility of realising the
ultraviolet fixed point in Eq.~(\ref{fixuv}) by including many
fermionic species, $N_f\gg1$, whose dynamics can generate spontaneous
symmetry breaking (SSB). 

\red{We assume the following associations:
\begin{enumerate}
    \item In the symmetric phase, $\langle\phi\rangle=0$, the IR fixed
    point in Eq.~(\ref{fixir}) and its scaling-invariant domain
    ${\mathcal D}_{\rm ir}$ are associated with the gravitational fixed
    point $g_{\rm ir}\simeq0$ in Eq.~(\ref{gir}). This corresponds to
    the disordered, symmetry-preserving phase of the underlying
    spacetime foam, characterised by
    $\langle X_h\rangle=0$, as discussed in Sec.~\ref{irpoint};
    \item In the symmetry-breaking phase, $\langle\phi\rangle\neq0$,
    the UV fixed point in Eq.~(\ref{fixuv}) and its scaling-invariant
    domain ${\mathcal D}_{\rm uv}$ are associated with the gravitational
    fixed point $g_{\rm uv}\sim\mathcal{O}(1)$ in Eq.~(\ref{guv}).
    This corresponds to the ordered, symmetry-breaking phase of the
    underlying spacetime foam, characterised by
    $\langle X_h\rangle\neq0$, as discussed in Sec.~\ref{uvpoint}.
\end{enumerate}
The detailed relation between the fixed points and scaling-invariant
domains of the foamon field theory and those of the underlying
simplicial Euclidean quantum-gravity theory remains to be established
by a nonperturbative analysis.
}

\subsection{Fermion coupling and non-perturbative effective action}

Due to the mixing between different spacetime-foam sectors in the
interaction (\ref{coleman3}), the foamon field $\Phi$ couples to a
fermion bilinear involving $N_f$ fermionic species,
\begin{eqnarray}
{\cal L}_{\rm int}
=
y\Phi\bar\Psi_L\Psi_R+{\rm h.c.},
\label{lf}
\end{eqnarray}
where $y$ is a Yukawa-type mixing coupling. In the weak-coupling
regime considered here, we take $y<1$.

Neglecting possible bare fermion masses, the fermion fields can be
integrated out in the presence of the slowly varying foamon mean
field $\Phi$. The resulting fermion determinant is schematically
given by
\begin{eqnarray}
\int_{bM_{\rm pl}}^{M_{\rm pl}}
{\cal D}\bar\Psi{\cal D}\Psi\,
\exp\left[-\bar\Psi\Delta\Psi\right]
&\propto&
\exp\left\{
N_f b^4\int_{x'}
\int_{bM_{\rm pl}}^{M_{\rm pl}}
\frac{d^4k}{(2\pi)^4}
\ln\Delta(k^2)
\right\},
\label{fcon}
\end{eqnarray}
where, for the present approximation,
\begin{equation}
\Delta(k^2)=k^2+(y\Phi)^2 .
\end{equation}
Here, the bare fermion masses are neglected, and $y\Phi$ is treated
as a slowly varying external mean field.

Following the Wilsonian procedure discussed in Sec.~\ref{scaling}, the
short-distance modes of the foamon field with characteristic momenta
above $bM_{\rm pl}$ are integrated out. Equivalently, the slowly
varying mean field is coarse-grained over a four-volume
\begin{equation}
V_b\sim(bM_{\rm pl})^{-4}.
\end{equation}
With the corresponding rescaling of the coordinates, this produces
the factor $b^4\int_{x'}$ in Eq.~(\ref{fcon}). Performing the Euclidean
momentum integration gives
\begin{eqnarray}
\frac{N_f b^4}{32\pi^2}\int_{x'}
\Big\{
[k^4-(y\Phi)^4]\ln[k^2+(y\Phi)^2]
-\frac{k^4}{2}
+(y\Phi)^2k^2
\Big\}_{bM_{\rm pl}}^{M_{\rm pl}} .
\label{fcon1}
\end{eqnarray}
Terms that are independent of $\Phi$, together with local quadratic
and quartic terms that can be absorbed into the corresponding
renormalised parameters, are omitted below. The remaining
field-dependent logarithmic contribution can be written, to leading
order in the Wilsonian expansion, as
\begin{eqnarray}
\frac{N_f b^4}{32\pi^2}\int_{x'}
\Big[
-(y\Phi)^4\ln\big(M_{\rm pl}^2+(y\Phi)^2\big)
+(y\Phi)^4\ln\big((bM_{\rm pl})^2+(y\Phi)^2\big)
\Big]
+\cdots .
\label{fcon2}
\end{eqnarray}

For $|y\Phi|\ll M_{\rm pl}$, we may expand the ultraviolet part in
powers of $(y\Phi)/M_{\rm pl}$. Using the Wilson field rescaling (\ref{renr})
\begin{equation}
\Phi\simeq\frac{\phi}{b},
\label{phirescale}
\end{equation}
the leading logarithmic contribution takes the approximate form
\begin{eqnarray}
\Delta\Gamma_f(\phi)
\simeq
\frac{N_f}{32\pi^2}\int_{x'}
(y\phi)^4
\ln\frac{(y\phi)^2}{(bM_{\rm pl})^2},
\label{fcon3}
\end{eqnarray}
up to local polynomial terms and corrections suppressed by powers of
$b$ and/or $(y\Phi/M_{\rm pl})^2$.

Equation~(\ref{fcon3}) is non-perturbative in the background-field
dependence of the fermion determinant in the sense that the logarithmic
dependence has been retained rather than expanded as a finite
polynomial in $y\Phi$. At the same time, it should be understood as a
leading-order mean-field result in the large-$N_f$ approximation.
Corrections beyond leading order in $1/N_f$, as well as fluctuations of
the foamon field around the mean field, are not included here.

For comparison, the perturbative fermion-loop contribution contains
a logarithmic term of the schematic form
\begin{equation}
\Delta\Gamma_f^{\rm pert}
\sim
\frac{N_f}{16\pi^2}
\int_{x'}
(y\phi)^4\ln(1/b),
\end{equation}
with the precise coefficient depending on the field content and
normalisation conventions. The fermionic contribution has the
opposite sign to the corresponding bosonic loop contribution to the
quartic coupling, as indicated by the positive one-loop bosonic
beta function in Eq.~(\ref{renr1}).

Including the leading logarithmic contribution (\ref{fcon3}) in the
effective action (\ref{domain}) gives
\begin{equation}
\Gamma(\phi)
=
\int_{x'}
\left[
\frac{1}{2}(\partial'_\mu\phi)^2
+\frac{1}{2}m^2\phi^2
+\frac{\lambda}{4}\phi^4
+
\frac{N_f(y\phi)^4}{32\pi^2}
\ln\frac{(y\phi)^2}{(bM_{\rm pl})^2}
+\cdots
\right].
\label{domain2}
\end{equation}
The ellipsis denotes counterterms, higher-dimensional operators, and
subleading contributions in the large-$N_f$ and Wilsonian expansions.

As the renormalisation scale
$
\mu=bM_{\rm pl}
$
decreases toward the infrared, $b\rightarrow0$, the dimensionless
couplings $\lambda$ and $y$ approach the IR fixed point in
Eq.~(\ref{fixir}). The effective theory therefore approaches a
quasi-free foamon theory. In this regime, the positive quadratic term
with $m^2>0$ stabilises the symmetric ground state at
$
\phi_c=0.
$

The logarithmic contribution in Eq.~(\ref{domain2}) contains an
explicit dependence on the renormalisation scale $bM_{\rm pl}$ and
therefore represents a scaling correction away from the idealised
IR fixed point. As the renormalisation scale increases,
$b\rightarrow1$, the system can depart from the IR scaling domain and
enter a regime in which the logarithmic fermion contribution becomes
dynamically important. We investigate in the following the resulting
effective potential and the possibility of spontaneous symmetry
breaking, $\phi_c\neq0$, as the system evolves toward the proposed
UV fixed point and scaling-invariant domain.

\begin{figure*}[t]
\centering
\begin{center}
%\vspace{-1.5em}
\includegraphics[height=5.5cm,width=9.8cm]{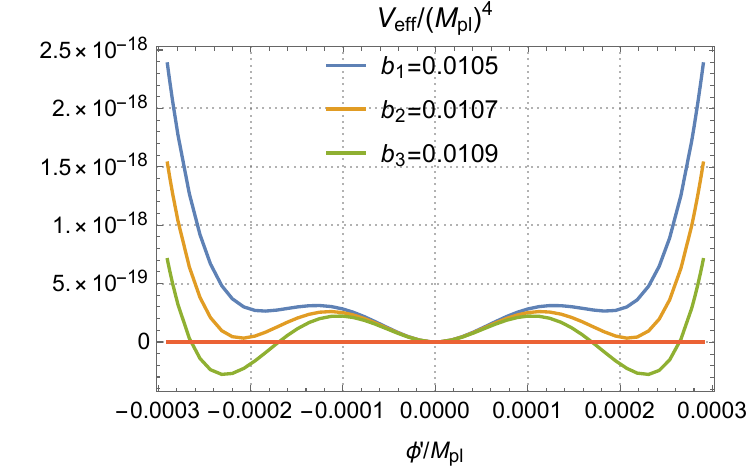}
\caption{We use the one-loop RG flows $\lambda(b)$ and $y(b)$
(\ref{rgw}), with $\lambda_{\rm ir}=10^{-3}$ and $y_{\rm ir}=0.3$,
to illustrate the effective potential (\ref{fv}), in units of the
Planck mass $M_{\rm pl}$, for three values of $b$. Here,
$N_f=10^2$ and $m/M_{\rm pl}=1.135\times10^{-7}$. In the symmetric
phase, the ground state is at the absolute minimum $\phi_c=0$ (blue
curve) for small values of the couplings $\lambda$ and $y$. As the
couplings increase with $b$, the orange curve, determined by Eq.~(\ref{degen}), 
shows the coexistence of
two degenerate minima, at $\phi=0$ and $\phi=\phi_c\neq0$, signaling
the transition to the symmetry-breaking phase. Beyond the transition,
the nontrivial minimum becomes the absolute minimum,
$\phi_c\neq0$ (green curve). The fermion-induced negative logarithmic
contribution competes with the positive quadratic and quartic terms
and generates the nontrivial minimum away from the origin. The
resulting radiatively induced symmetry breaking is qualitatively
analogous to the mechanism of Coleman and Weinberg
\cite{Coleman1973}. The coexistence of two degenerate minima with
$\phi_c\neq0$ indicates that the transition is first-order.} 
\label{wpoten}
\end{center}
\vspace{-2em}
\end{figure*}

\comment{
\begin{eqnarray}
\lambda_{\rm ir} +\frac{3\lambda_{\rm ir}^2}{16\pi^2}\ln b-N_f 
\frac{g'^4_y}{4\pi^2} \ln b,
\label{fmod}
\end{eqnarray}
which is the opposite of the boson one (\ref{renr1}). At a quick glance, the fermion contribution increases the effective quartic coupling as the energy scale $bM_{\rm pl}$ decreases, 
driving it away from the fixed point $\lambda^{\rm fix}_{\rm ir}= 0$, which becomes unstable.
The fermion-loop contribution is opposite to the 
boson-loop contributions (\ref{renr1}). The former increases the effective quartic coupling $\bar\lambda'$ as the energy scale $bM_{\rm pl}$ decreases, 
driving it away from the fixed point $\lambda^{\rm ir}= 0$, which becomes unstable. The fermion-loop modified $\beta$-function is  
\begin{eqnarray}
\beta=b(\partial \bar\lambda'/\partial b) 
&=&\frac{3\lambda_{\rm ir}^2}{16\pi^2}-N_f 
\frac{y'^4}{4\pi^2},
\label{fmod}
\end{eqnarray}
which changes from 
positive sign (\ref{renr1}) to negative one, if $3\lambda_{\rm ir}^2/4-N_fy'^4<1 0$, and its zero point $\beta(\lambda_{\rm uv})=0$ implies another ultraviolet fixed point 
\begin{eqnarray}
\lambda^{\rm uv} 
&=& 2(N_f/3)^{1/2}(y)^2\not=0,
\label{fmod}
\end{eqnarray}
which is obtained by substituting the $\beta=0$ solution $\lambda_{\rm ir}=2(N_f/3)^{1/2}(y)^2$ to Eq.~(\ref{fmod}). As $b<1$ increases and energy scale $bM_{\rm pl}$ increases, $\lambda'$ approaches to $\lambda^{\rm uv}_{\rm fix}$. What is it? If $N_f$ is not very
large and $y\ll 1$, then $\lambda^{\rm uv}_{\rm fix}$ is not very different from $\lambda^{\rm ir}_{\rm fix}$.
}

\subsection{Spontaneous symmetry breaking at the renormalisation scale}
\label{ssb}

The one-loop beta function $\beta_y^{(1)}$ in Eq.~(\ref{rgwy}) and the
fermion-induced logarithmic contribution in Eq.~(\ref{domain2}) both
contain the fermion-number factor $N_f$. Consequently, for
$N_f\gg1$, the Yukawa coupling can evolve more rapidly with the
renormalisation scale than the quartic coupling $\lambda(b)$. As the
scale $\mu=bM_{\rm pl}$ increases, the fermion-induced logarithmic
contribution can become sufficiently important to compete with the
positive quadratic and quartic terms in the effective potential.

The resulting effective potential and its first derivative are
\begin{eqnarray}
V(\phi)
&=&
\frac{1}{2}m^2\phi^2
+\frac{\lambda}{4}\phi^4
+
N_f\frac{(y\phi)^4}{32\pi^2}
\ln\frac{(y\phi)^2}{(bM_{\rm pl})^2},
\label{fv}\\
V'(\phi)
&=&
m^2\phi
+
\left[
\lambda+\frac{N_fy^4}{16\pi^2}
\right]\phi^3
+
\frac{N_fy^4}{8\pi^2}\phi^3
\ln\frac{(y\phi)^2}{(bM_{\rm pl})^2}.
\label{fvd1}
\end{eqnarray}
Here and below, the parameters $m$, $\lambda$, and $y$ are understood
as running parameters evaluated at the scale $\mu=bM_{\rm pl}$.

In Fig.~\ref{wpoten}, we use the one-loop RG flows
$\lambda(b)$ and $y(b)$ in Eq.~(\ref{rgw}) to illustrate the effective
potential $V(\phi)$ in Eq.~(\ref{fv}), in units of $M_{\rm pl}$, for
three representative values $b_1>b_2>b_3$.
In the symmetric phase, corresponding to sufficiently small $b$, the
absolute minimum is located at
$
\phi=0,
$
as shown by the blue curve in Fig.~\ref{wpoten}. In this regime,
$\lambda(b)$ and $y(b)$ remain in the weak-coupling region and are
well described by the perturbative RG equations. Starting from the
small initial couplings
$\lambda_{\rm ir}\ll1$ and $y_{\rm ir}\ll1$ in Eq.~(\ref{fixir}), the
couplings increase as the renormalisation scale
$\mu=bM_{\rm pl}$ is raised. At a critical scale, they reach values
$\lambda_c$ and $y_c$ for which the effective potential develops two
degenerate minima,
\begin{equation}
V(0)=V(\phi_c)=0,
\qquad
\phi_c\neq0,
\end{equation}
as illustrated by the orange curve. This gives a first-order coexistence point.

The coexistence of two degenerate minima at
$\phi=0$ and $\phi=\phi_c\neq0$ indicates a first-order spontaneous
symmetry-breaking transition. At the transition point, the nontrivial
minimum satisfies
\begin{equation}
V(\phi_c)=0,
\qquad
V'(\phi_c)=0.
\label{degen}
\end{equation}
Solving these two equations gives
\begin{eqnarray}
(y_c\phi_c)^2
&=&
(b_cM_{\rm pl})^2
e^{-\zeta_c-1},
\qquad
\zeta_c\equiv
\frac{8\pi^2\lambda_c}{N_fy_c^4},
\label{fc}
\end{eqnarray}
and
\begin{eqnarray}
m_c^2
&=&
\frac{N_fy_c^4}{16\pi^2}\phi_c^2,
\label{fcm}
\end{eqnarray}
at the scale $\mu_c=b_cM_{\rm pl}$.
Equation~(\ref{fc}) can be interpreted as a gap equation for the
nontrivial vacuum expectation value $\phi_c$ and the associated
fermion mass scale
\begin{equation}
M_f=y_c\phi_c.
\end{equation}
The curvature of the effective potential at the nontrivial minimum
then gives the physical mass $m_{\rm eff}$ 
of the scalar excitation. In the present approximation,
\begin{equation}
m_{\rm eff}^2=V''(\phi_c)=2m_c^2,
\qquad
m_{\rm eff}=\sqrt{2}\,m_c.
\end{equation}

The fermion-induced symmetry-breaking scale is therefore
\begin{equation}
M_f=y_c\phi_c
=
b_cM_{\rm pl}\,e^{-(\zeta_c+1)/2}.
\label{mfcritical}
\end{equation}
Thus, when the characteristic physical scale $M_f$ is associated with
the renormalisation scale $\mu_c=b_cM_{\rm pl}$, one expects
\begin{equation}
M_f\sim\mu_c,
\qquad
\phi_c\sim b_cM_{\rm pl},
\end{equation}
up to the dimensionless factor determined by $\zeta_c$. For
$N_f\gg1$, it is possible to have
\begin{equation}
\zeta_c={\cal O}(1),
\qquad
\lambda_c<1,
\qquad
y_c<1,
\end{equation}
so that the transition remains parametrically within the weak-to-
intermediate coupling regime. For a specified physical scale $M_f$
and fixed $\zeta_c$, Eq.~(\ref{mfcritical}) determines a relation
between the critical couplings $\lambda_c$ and $y_c$.

Across the symmetry-breaking transition, the critical couplings can
be regarded as scale-dependent quantities. At the transition scale
they coincide with the running couplings of the symmetric phase,
\begin{equation}
\lambda_c=\lambda,
\qquad
y_c=y,
\end{equation}
while their subsequent evolution in the broken phase is constrained
by the requirement that the physical mass scale $M_f$ be
renormalisation-scale independent. We define
\begin{eqnarray}
\beta_y^{(c)}
&\equiv&
b\frac{\partial y_c}{\partial b},
\qquad
\beta_\lambda^{(c)}
\equiv
b\frac{\partial\lambda_c}{\partial b},
\label{betacd}
\end{eqnarray}
for $b>b_c$.
Although the couplings themselves are continuous at the transition,
their beta functions need not be described by the same perturbative
expressions on the two sides of the transition.

The physical fermion mass $M_f=y_c\phi_c$ must be independent of the
arbitrary renormalisation scale used to parameterise the effective
theory. Thus, the Callan--Symanzik condition is
\begin{eqnarray}
b\frac{\partial M_f}{\partial b}=0,
\label{cs0}
\end{eqnarray}
for $b>b_c$. Using Eq.~(\ref{fc}), this condition gives
\begin{eqnarray}
\beta_y^{(c)}
=
-\frac{N_fy_c^5}{16\pi^2\lambda_c}
+
\frac{y_c}{4\lambda_c}
\beta_\lambda^{(c)}.
\label{cs}
\end{eqnarray}
For $N_f\gg1$, the first term tends to drive $y_c$ toward smaller
values as the scale increases, while maintaining a fixed $M_f$
requires a compensating evolution of $\lambda_c$. In the regime
where the first term dominates, one obtains
$
\beta_y^{(c)}<0,
$
while the corresponding evolution of $\lambda_c$ can be positive, i.e., $\beta_\lambda^{(c)}>0$.
This behaviour is qualitatively analogous to the negative beta
function encountered in large-$N$ analyses of nonlinear sigma
models, although the present scalar--Yukawa system is dynamically
different.

\red{The fermion-induced logarithmic contribution can therefore generate a
nonzero minimum even when $m^2>0$. At the transition point,
Eq.~(\ref{fcm}) gives the curvature mass parameter at the symmetric minimum
$
V''(0)=m_c^2>0,
$
so that the symmetric minimum at $\phi=0$ remains locally stable when
the nontrivial minimum $\phi_c\neq0$ becomes degenerate with it.
Consequently, the order parameter changes discontinuously from
$\phi=0$ to $\phi=\phi_c\neq0$. The transition illustrated in
Fig.~\ref{wpoten} is therefore a first-order, rather than a
second-order, symmetry-breaking transition, as discussed in
Sect.~\ref{uvpoint}. It may be weakly first-order if the discontinuity
$\phi_c\ll M_{\rm pl}$ and the associated mass scale $m_c\ll M_{\rm pl}$ in Eq.~(\ref{fcm}) are
numerically small for the parameter values considered.}

\red{The occurrence of a first-order transition in this simplified
$\lambda_i\Phi_i^4$ foamon effective theory may indicate that the
assumed effective potential does not capture all of the long-distance
collective dynamics of the underlying simplicial Euclidean quantum-
gravity theory. In particular, additional interactions, higher-order
operators, multiple foamon fields, or nonperturbative effects could
modify the effective potential and, in principle, alter the order and
universality properties of the transition. Determining whether the
underlying spacetime-foam dynamics exhibit a genuine second-order
critical regime therefore requires a more complete nonperturbative
treatment.}

\comment{However, one can use this argument qualitatively only. 
This situation is similar to 
the one studied in radiative corrections to the Higgs scalar field 
potential \cite{Coleman1973,Weinberg1973}, the Gaussian effective potential \cite{Barnes1980,Stevenson1985}, the pure $\lambda\phi^4$ theory \cite{Consoli2000,Consoli2020}
and the phase transition of the system with finite matter density \cite{Kleinert2015}. 
%Note that the argument can only be used qualitatively, as been emphasized by Colemen and Weinberg \cite{Coleman1973,Weinberg1973}. The reason is that the new minimum $H^2_m$ lies in a field regime where the perturbation theory is no longer reliable \cite{Fiolhais2013,Kleinert2015}.
}

\subsection{Ultraviolet fixed point and scaling-invariant domain in the symmetry-breaking regime}
\label{uvf}

In the effective theory with the potential $V(\phi)$
(\ref{fv}), the curvature determines the effective scalar mass
of the potential,
\begin{eqnarray}
V''(\phi)
&=&
m^2+
\left[
3\lambda+
N_f\frac{7y^4}{16\pi^2}
\right]\phi^2
+
N_f\frac{3y^4\phi^2}{8\pi^2}
\ln\frac{(y\phi)^2}{(bM_{\rm pl})^2}.
\label{fvd2}
\end{eqnarray}
The corresponding fourth derivative is
\begin{eqnarray}
V''''(\phi)
&=&
\left[
6\lambda+
N_f\frac{17y^4}{8\pi^2}
\right]
+
N_f\frac{3y^4}{4\pi^2}
\ln\frac{(y\phi)^2}{(bM_{\rm pl})^2}.
\label{fvd4}
\end{eqnarray}
%V'''(\phi')&=&\left[6\lambda'+N_f\frac{13y'^4}{ 8\pi^2}\right]\phi^{'}+N_f\frac{3y'^4\phi'}{4\pi^2}\ln \frac{(y'\phi')^2}{(bM_{\rm pl})^2},\label{fvd3}\\
In the symmetry-breaking phase, using the gap equation (\ref{fc}) at
the first-order coexistence point, where
$y_c\phi_c\neq0$, we obtain
\begin{eqnarray}
m^2_{\rm uv}
&\equiv&
V''(\phi_c)
=
m_c^2+
\frac{N_fy_c^4}{16\pi^2}\phi_c^2
=
2m_c^2,
\label{effcm}\\
\lambda_{\rm uv}
&\equiv&
V''''(\phi_c)
=
\frac{11N_f}{8\pi^2}y_c^4 .
\label{effcl}
\end{eqnarray}
These quantities depend on the renormalisation scale
$\mu=bM_{\rm pl}$ and on the critical parameter
$\zeta_c(\lambda_c,y_c)$.

After spontaneous symmetry breaking, we expand the scalar field about
the nontrivial minimum,
\begin{equation}
\phi=\phi_c+\phi_{\rm uv}.
\end{equation}
The low-energy effective theory for the massive scalar fluctuation
$\phi_{\rm uv}$ and the fermionic fields $\psi_{\rm uv}$ can then be
written schematically as
\begin{eqnarray}
{\cal L}_{\rm uv}
&=&
\frac{1}{2}(\partial_\mu\phi_{\rm uv})^2
+\frac{1}{2}m_{\rm uv}^2\phi_{\rm uv}^2
+\frac{\lambda_{\rm uv}}{4!}\phi_{\rm uv}^4
\nonumber\\
&&
+\bar\psi_{\rm uv}
(i\slashed{\partial}+M_f)\psi_{\rm uv}
+y_c\phi_{\rm uv}
\bar\psi_{\rm uv}\psi_{\rm uv}
+\cdots ,
\label{hlag}
\end{eqnarray}
where the normalizations of $\lambda_{\rm uv}$ and $m_{\rm uv}$ follow their definitions
in Eqs.~(\ref{effcm}) and (\ref{effcl}). The ellipsis denotes cubic and higher-order
interactions generated by the expansion about $\phi_c$, counterterms,
and higher-dimensional operators. If the underlying scalar sector possesses a spontaneously broken
continuous symmetry, additional Goldstone modes may also arise. Their
possible description by a nonlinear sigma model, or as longitudinal
modes of massive gauge bosons when the symmetry is gauged, is beyond
the scope of the present work.

Using Eq.~(\ref{effcl}) and the beta function of the Yukawa coupling
in the symmetry-breaking regime, we obtain
\begin{eqnarray}
\beta^{\rm uv}_\lambda
&\equiv&
b\frac{\partial\lambda_{\rm uv}}{\partial b}
=
\frac{11N_f}{2\pi^2}y_c^3\beta_y^{(c)}
<0,
\nonumber\\
\beta^{\rm uv}_y
&\equiv&
\beta_y^{(c)}
<0.
\label{efflb}
\end{eqnarray}
Thus, within the present approximation, both effective beta functions
are negative in the symmetry-breaking regime.

This behaviour is opposite to that obtained in the perturbative
symmetric regime, where the one-loop beta functions
$\beta_\lambda^{(1)}$ and $\beta_y^{(1)}$ are positive for the
couplings considered here. The change in sign suggests that the full
beta functions may possess a zero between the perturbative
small-coupling regime and the nonperturbative symmetry-breaking
regime. However, because the two regimes have been analysed using
different approximations, this observation does not establish that
the zero coincides with the first-order coexistence point
$\zeta_c(\lambda_c,y_c)$. In particular, the first-order phase
transition and an RG fixed point are distinct concepts.

\red{For a large number of fermionic degrees of freedom,
$N_f\gg1$, we therefore consider the possibility that the RG flow
approaches a nontrivial ultraviolet fixed point in the
symmetry-breaking regime. We parameterise this possibility by
\begin{eqnarray}
\lambda_{\rm uv}^{\rm fix}
&\approx&
\lambda_*,
\qquad
y_{\rm uv}^{\rm fix}
\approx
y_*,
\label{fixuv2}
\end{eqnarray}
where
\begin{equation}
\beta^{\rm uv}_\lambda(\lambda_*,y_*)=0,
\qquad
\beta^{\rm uv}_y(\lambda_*,y_*)=0,
\end{equation}
at the scale $\mu_*=b_*M_{\rm pl}$. 
We expect, within the present weak-coupling parametrisation,
$\lambda_*,y_*<1$. The values $(\lambda_*,y_*,b_*)$ should not be
identified with $(\lambda_c,y_c,b_c)$ unless the full nonperturbative RG
flow demonstrates such a relation. We conjecture that the resulting
scaling-invariant domain ${\cal D}_{\rm uv}$ is associated with the
UV scaling-invariant domain around the gravitational fixed point
$g_{\rm uv}\sim{\cal O}(1)$, with
$\beta(g_{\rm uv})=0$, as illustrated in Fig.~\ref{betaf} and
discussed in Sect.~\ref{uvpoint}.
}

Similarly to the effective action and Lagrangian
(\ref{efffree}) in the IR scaling-invariant domain ${\cal D}_{\rm ir}$,
the effective Lagrangian (\ref{hlag}) is expected to describe the
low-energy degrees of freedom within ${\cal D}_{\rm uv}$. At the
renormalisation scale $\mu=bM_{\rm pl}$, we write the corresponding
effective action schematically as
\begin{eqnarray}
\Gamma(\phi,\psi)
&=&
\int d^dx'\,{\cal L}_{\rm eff},
\nonumber\\
{\cal L}_{\rm eff}
&=&
\frac{1}{2}(\partial'_\mu\phi)^2
+\frac{1}{2}m^2\phi^2
+\frac{\lambda}{4}\phi^4
+\bar\psi(i\slashed{\partial}'+M_f)\psi
+y\phi\bar\psi\psi
+\cdots ,
\label{haction}
\end{eqnarray}
where the parameters are understood as scale-dependent effective
couplings and masses.

A quantitative determination of the RG flows of the effective
couplings and field operators in the scaling-invariant domain
${\cal D}_{\rm uv}$ is not currently available, because the proposed
nontrivial UV fixed point lies outside the regime of validity of the
perturbative expansion used in the symmetric phase. In particular,
we have not obtained scaling laws analogous to Eq.~(\ref{renr}) from
a nonperturbative calculation in ${\cal D}_{\rm uv}$. Nevertheless, as
a phenomenological approximation, we assume that the leading
canonical scaling of the fields and dimensionful parameters can be
used as a zeroth-order description,
\begin{eqnarray}
\phi
&\approx&
\phi_{\rm uv}b^{(2-d)/2},
\qquad
\psi
\approx
\psi_{\rm uv}b^{(1-d)/2},
\nonumber\\
m^2
&\approx&
m_{\rm uv}^2b^{-2},
\qquad
\lambda
\approx
\lambda_{\rm uv}b^{d-4},
\qquad
y
\approx
y_{\rm uv}b^{d-4}.
\label{renuv}
\end{eqnarray}
These relations represent canonical dimensional scaling only; they
do not constitute a derivation of the nonperturbative RG flow.
Corrections associated with wave-function renormalisation,
anomalous dimensions, and the running of the couplings,
$\delta Z$, $\delta m$, $\delta\lambda$, and $\delta y$, are neglected
at this level of approximation. Irrelevant operators represented by
the ellipsis are correspondingly suppressed by positive powers of the
coarse-graining parameter.

Within this approximation, the effective action in the
scaling-invariant domain satisfies the Wilsonian scaling relation
\begin{equation}
\Gamma(\phi,\psi)
\simeq
\Gamma(\phi_{\rm uv},\psi_{\rm uv}),
\label{wilsonuv}
\end{equation}
independent of $b$, up to the field and parameter rescalings in Eq.~(\ref{renuv}). The
characteristic correlation length of the massive scalar fluctuation
is
\begin{equation}
\xi_{\rm uv}\simeq m_{\rm uv}^{-1}.
\end{equation}

\red{
Combining the negative beta functions obtained in the
symmetry-breaking regime with the positive perturbative beta functions
in the symmetric regime suggests that the full beta functions may
interpolate between positive and negative values. Consequently, they
may possess nontrivial zeros at intermediate values of the couplings.
For example, one may schematically represent the quartic-coupling flow
$
\beta_\lambda(\mu_*)=0,~
\lambda_* \neq 0,
$
in analogy with the nontrivial gravitational fixed point $
\beta[g(\tilde\ell_*)]=0,~
g_{\rm uv}\neq0,
$ discussed in Sect.~\ref{uvpoint}.
However, the present analysis is insufficient to determine the
complete RG trajectory from the IR fixed point
$(\lambda_{\rm ir},y_{\rm ir})$ to a possible UV fixed point
$(\lambda_*,y_*)$. We therefore regard the existence and
location of the proposed UV fixed point and its scaling-invariant domain ${\cal D}_{\rm uv}$, as well as the complete
interpolation of the beta functions between the IR and UV domains, as
open problems.}

\section{Spacetime foam effects on effective field theories}
\label{effec}

In the IR and UV scaling-invariant domains
${\mathcal D}_{\rm ir}$ and ${\mathcal D}_{\rm uv}$, respectively, the
foamon correlation length $\xi$ provides a physically relevant
long-distance scale. We write
\begin{eqnarray}
\xi\simeq \tilde m^{-1},
\qquad
\tilde m=
\begin{cases}
m_{\rm ir}, & {\mathcal D}_{\rm ir},\\
m_{\rm uv}, & {\mathcal D}_{\rm uv},
\end{cases}
\label{corre}
\end{eqnarray}
where the scaling-invariant $\tilde m$ denotes the corresponding effective mass scale of
the foamon field.

When the microscopic coarse-graining scale satisfies
$\tilde\ell\ll\xi$, the foamon dynamics admit a long-distance scaling
description. Physical observables are then invariant under changes of
the coarse-graining scale when the complete Wilsonian RG
transformation, including the running of fields and couplings, is
taken into account. Following the scaling relations
(\ref{renr}) and (\ref{renuv}), we obtain the approximate canonical
scaling laws for the foamon--operator interaction
$\int_x\phi O$ in Eq.~(\ref{reg}),
\begin{eqnarray}
\phi
\approx
\tilde\phi\,
b^{-(d-2)/2},
\qquad
O
\approx
\tilde O\,
b^{-\left[(d+2)/2\right]} ,
\label{oscaling}
\end{eqnarray}
where
\begin{eqnarray}
\tilde\phi&=&
\begin{cases}
\phi_{\rm ir}, & {\rm in}~~ {\mathcal D}_{\rm ir},\\
\phi_{\rm uv}, & {\rm in}~~{\mathcal D}_{\rm uv},
\end{cases}
\qquad
\tilde O=
\begin{cases}
O_{\rm ir}, & {\rm in}~~{\mathcal D}_{\rm ir},\\
O_{\rm uv}, & {\rm in}~~ {\mathcal D}_{\rm uv}.
\end{cases}
\nonumber
\end{eqnarray}
Here, $(d-2)/2$ and
$(d+2)/2$ are the canonical mass dimensions of the
scalar field $\phi$ and of the operator $O$, respectively, for the
interaction $\phi O$ to have dimension $d$.

The scaling relations in Eq.~(\ref{oscaling}) are based on canonical
dimensions and neglect anomalous dimensions, wave-function
renormalisation corrections, and irrelevant operators. They therefore
provide an approximate description only within the scaling-invariant
domains ${\mathcal D}_{\rm ir}$ and ${\mathcal D}_{\rm uv}$, where
$\xi\gg\tilde\ell$ and the effective foamon theories in
Eqs.~(\ref{domain}) and~(\ref{haction}) provide appropriate
long-distance descriptions. Within these domains, the corresponding
effective actions are approximately invariant under Wilsonian
rescaling, as expressed by Eqs.~(\ref{wilsonir}) and
(\ref{wilsonuv}).

\red{A central feature of this framework is that the foamon fields and
their Coleman--Hawking couplings
$
\int_x\phi_i O_i
$
in Eq.~(\ref{scalarc3}) provide a physical connection between the
long-distance foamon dynamics and the effective theories of the
operators $O_i$. In particular, they provide two potentially important
effects:
\begin{enumerate}[{(i)}]
\item The characteristic long-distance scales associated with the
operators $O_i$ provide physical infrared scales for the foamon
effective theories. These scales regulate the infrared behaviour of
the foamon correlation functions and lead to finite correlation
lengths
$
\xi_i\simeq\tilde m_i^{-1}
$
in the scaling-invariant domains associated with the IR and UV fixed
points. Unlike an externally imposed infrared cutoff, these scales are
dynamically associated with the long-distance physical operators
$O_i$;
\item The resulting long-range foamon correlations generate effective
interactions and, more generally, effective nonlocal contributions for
the physical operators $O_i$. The characteristic scales and effective
couplings of these induced interactions are determined by the
correlation lengths and by the dynamics of the long-wavelength foamon
fields in their respective scaling-invariant domains.
\end{enumerate}
In the following section, we focus on the second effect and investigate
how long-range foamon correlations can generate effective dynamics for
the physical field operators $O_i$, with particular emphasis on the
operator constructed from the Ricci scalar $R$.
}

Thus, the effective framework establishes the following chain of scales:
the long-distance variation of $O_i$ selects an infrared scale $\mu_i$,
which determines the relevant foamon correlation length $\xi_i$.
The latter, in turn, controls the strength and range of the
foamon-induced effective interactions of $O_i$. For the curvature
sector considered below, this provides the basis for relating the
macroscopic correlation scale $\xi_R$ to an emergent cosmological
term, $\Lambda_R\propto\xi_R^{-2}$.

\subsection{Effective action for fundamental field operators at large distances}

Here we restore the subscript ``$i$'' to distinguish the different
sectors of foamon fields $\Phi_i$. In terms of the effective Lagrangian
${\mathcal L}_{\rm eff}$ in Eq.~(\ref{domain}) or Eq.~(\ref{haction}),
valid in the scaling-invariant domains ${\mathcal D}_{\rm ir}$ and
${\mathcal D}_{\rm uv}$, respectively, the Euclidean generating
functional in Eq.~(\ref{vacuo}) can be written as
\begin{equation}
e^{-S_E^{\rm eff}(O_i)}
\equiv
W_E(O_i)
\simeq
\int_0^{bM_{\rm pl}}{\cal D}\phi_i\,
\exp\left\{
-\int_{x'}
\left[
{\cal L}_{\rm eff}(\phi_i,\partial'\phi_i)
+\phi_iO_i
\right]
\right\}.
\label{vacuoe}
\end{equation}
Integrating out the low-energy foamon fields $\phi_i$ therefore
generates an effective action for the physical field operators $O_i$.

To illustrate this procedure, we consider the Gaussian approximation
to the foamon effective action. The quadratic part of Eq.~(\ref{vacuoe})
can be written as
\begin{eqnarray}
W_E(O_i)
&\simeq&
\int_0^{bM_{\rm pl}}{\cal D}\phi_i\,
\exp\left[
-\frac12\phi_i\Delta_{ij}\phi_j
-\phi_iO_i
\right]
\nonumber\\
&\simeq&
[\det\Delta_{ij}]^{-1/2}
\exp\left[
\frac12 O_i
(\Delta^{-1})_{ij}
O_j
\right],
\label{eff}
\end{eqnarray}
where repeated indices imply integration or summation over the
corresponding spacetime, momentum, and field-sector variables.

The operator $\Delta_{ij}$ is the inverse propagator of the foamon
fields. In a basis in which the two-point function is diagonal in the
foamon-sector index, we have
\begin{equation}
\Delta_{ij} = \delta_{ij} G_i^{(2)}(x'_1-x'_2)
\simeq
\delta_{ij}
\left(-\partial'^2+m_i^2\right).\label{delik}
\end{equation}
In momentum space,
$
\Delta_i(k',m_i)
=
k_E^{\prime\,2}+m_i^2
$,
$
G^{(2)}_{ij}(k')
\simeq
\delta_{ij}\,
(k_E^{\prime\,2}+m_i^2)^{-1}
=
(\Delta^{-1})_{ij},
$
consistent with the two-point correlation function in
Eq.~(\ref{corr1}).
The indices $(i,j)$ in coordinate space collectively denote
$(x'_1,i)$ and $(x'_2,j)$, respectively, whereas in momentum space
they denote $(k'_1,i)$ and $(k'_2,j)$. The diagonal quantities
$\Delta_i(k',m_i)$ are therefore the approximate eigenvalues of the
inverse two-point correlation function of the foamon effective theory.

Taking the logarithm of Eq.~(\ref{eff}), the induced effective action
for the physical operators becomes
\begin{eqnarray}
S_E^{\rm eff}(O_i)
&\simeq&
\frac12\sum_i {\rm Tr}\ln\Delta_i
-\frac12
\sum_i
O_i\Delta_i^{-1}O_i
+\cdots ,
\label{effactionOi}
\end{eqnarray}
where the ellipsis denotes field-independent normalisation terms,
interactions beyond the Gaussian approximation, and possible mixing
between different foamon sectors. In coordinate space, the second
term can be written schematically as
\begin{equation}
-\frac12
\sum_i
\int d^4x'_1\,d^4x'_2\,
O_i(x'_1)
G_i^{(2)}(x'_1-x'_2)
O_i(x'_2).
\label{nonlocalOi}
\end{equation}
Consequently, the long-distance foamon correlation function directly
determines the nonlocal kernel of the induced effective action for the
physical operators $O_i$. The proposed foamon field theory naturally gives $G_i^{(2)}(x'_1-x'_2)\sim C_{ij}(x_1,x_2)$ in the Coleman action (\ref{nonlocal}).

\subsection{Foamon field energy densities relevant to field operators}
\label{foamedensity}

In the IR- and UV-scaling-invariant domains, the quadratic foamon
effective action can be diagonalised in terms of foamon sectors $i$ and
momentum modes $k'$. The corresponding eigenvalues of the inverse
two-point correlation function are given by Eq.~(\ref{delik}). The
$O_i$-field independent part of the Gaussian functional integral is
\begin{eqnarray}
[\det\Delta_i]^{-1/2}
&=&
\exp\left[
-\frac{1}{2}{\rm Tr}\ln\Delta_i
\right]
\nonumber\\
&\simeq&
\exp\left[
-\frac{1}{2}\int_{x'}
\int_0^{bM_{\rm pl}}
\frac{d^4k'}{(2\pi)^4}
\ln\left(k'^2+m_i^2\right)
\right],
\label{vacuu}
\end{eqnarray}
and
\begin{eqnarray}
\int^{bM_{\rm pl}}_0 \frac{d^4k'}{(2\pi)^4} \ln \Delta_i(k'^2)&\simeq &\frac{1}{32\pi^2}\Big\{[(bM_{\rm pl})^4-m^4_i]\ln [(bM_{\rm pl})^2+m^2_i]\nonumber\\
&-&\frac{(bM_{\rm pl})^4}{2}+m^2_i (bM_{\rm pl})^2+m^4_i\ln m^2_i\Big\}
%&\approx & \frac{1}{32\pi^2} m^2_i(bM_{\rm pl})^2 = \frac{1}{32\pi^2} \tilde m^2_iM_{\rm pl}^2,
\label{cosmo}
\end{eqnarray}
Here, $bM_{\rm pl}$ is the Wilsonian momentum scale associated with
the long-distance physical operator $O_i$, as discussed in
Sect.~\ref{irregulator}.

Following the low-energy effective Lagrangians in
Eqs.~(\ref{domain}) and (\ref{haction}), together with the Wilsonian
scaling relations in Eqs.~(\ref{wilsonir}) and
(\ref{wilsonuv}), we identify the scale($b$)-independent contribution
associated with the relevant long-wavelength foamon modes.  In a
Wilsonian scaling estimate, the corresponding foamon energy density
has the form
\begin{eqnarray}
\tilde\rho_i
&\sim&
m_i^2(bM_{\rm pl})^2
=
\tilde m_i^2 M_{\rm pl}^2,
\label{vacue}
\end{eqnarray}
where the dimensionful foamon mass parameter is related to the
long-distance mass scale $\tilde m_i$ by the Wilsonian rescaling
\begin{equation}
m_i=\frac{\tilde m_i}{b},
\label{massscaling}
\end{equation}
as in Eqs.~(\ref{renr}) and (\ref{renuv}). Here, the scaling-invariant mass $\tilde m_i$ (\ref{corre}) characterises the mass scale in the corresponding scaling-invariant
domain.

Thus, within the Wilsonian scaling description, the relevant foamon
contribution can be represented schematically as
\begin{eqnarray}
[\det\Delta_i]^{-1/2}
&\simeq&
\exp\left[
-\int_{x'}\tilde\rho_i+\cdots
\right],
\label{vacuf}
\end{eqnarray}
where the ellipsis denotes scale-dependent contributions, including
field-independent counterterms and contributions from irrelevant
operators that are not retained in the long-distance effective
description.

The quantity $\tilde\rho_i$ represents the characteristic energy
density associated with the relevant long-wavelength foamon field
modes $\tilde\phi_i$ coupled to the physical operator $\tilde O_i$
at the corresponding scale $\tilde m_i$. It is therefore a
long-distance, Wilsonian contribution rather than the complete
vacuum-energy density $\rho_{\rm UV}\sim M_{\rm pl}^4$ (\ref{effed_0}) obtained by summing over all microscopic foam
degrees of freedom  associated with all short-distance foamon
modes with characteristic wavelengths of order $M_{\rm pl}^{-1}$ .

\subsection{Nonlocal field actions originating from spacetime-foam correlations}
\label{nonlocalaction}

We now consider the second, the field operator $O_i$-dependent part of the effective
action in Eq.~(\ref{eff}) within the scaling-invariant domains
${\mathcal D}_{\rm ir}$ and ${\mathcal D}_{\rm uv}$. Using the
diagonalised two-point foamon correlation function in
Eq.~(\ref{corr1}), the corresponding nonlocal contribution can be
written, in coordinate space, as
\begin{eqnarray}
\exp\left[
\frac{1}{2}
O_i(\Delta_{ij})^{-1}O_j^\dagger
\right]
&\simeq&
\exp\left\{
\frac{1}{2}
\int_{x'_1,x'_2}
O_i(x'_1)O_i^\dagger(x'_2)
\right.
\nonumber\\
&&\left.
\times
\int_0^{bM_{\rm pl}}
\frac{d^4k'}{(2\pi)^4}
\frac{
e^{ik'\cdot(x'_2-x'_1)}
}{
k'^2+m_i^2
}
\right\}.
\label{racci}
\end{eqnarray}
Here and below, repeated foamon-sector indices are summed. The
inverse operator
$
(\Delta_{ij})^{-1}
$
is the foamon propagator and couples the physical operators
$O_i(x'_1)$ and $O_i^\dagger(x'_2)$ at different spacetime points.
Equation~(\ref{racci}) therefore represents a nonlocal effective
interaction generated by long-distance foamon correlations.

This nonlocal structure is analogous, at the effective level, to the
connection functional $C_{ij}(x_1,x_2)$ in Eq.~(\ref{nonlocal}),
which describes the coupling of field operators at separated
spacetime points induced by wormhole configurations. The
difference is that $C_{ij}$ characterises the underlying
wormhole-induced interaction, whereas the kernel in
Eq.~(\ref{racci}) is generated by the propagation (\ref{corr1}) of the effective
long-wavelength foamon fields.
More precisely, in four Euclidean dimensions, its large-distance
behaviour is
\begin{equation}
G_i^{(2)}(r)
\propto
\frac{e^{-\tilde m_i r}}{r^{3/2}},
\qquad
r=|x'_2-x'_1|,
\label{propasymp}
\end{equation}
so that the dominant exponential suppression occurs for
$r\gg\xi_i=\tilde m_i^{-1}$. Consequently, the nonlocal interaction is effectively
local on length scales much larger than the foamon correlation length,
provided that the physical operators $O_i$ vary sufficiently slowly
over this scale.

To obtain the leading local contribution, we expand the physical
operator about $x'_1$,
\begin{eqnarray}
O_i(x'_2)
&=&
O_i(x'_1)
+
(x'_2-x'_1)^\mu\partial'_\mu O_i(x'_1)
\nonumber\\
&&+
\frac{1}{2}
(x'_2-x'_1)^\mu(x'_2-x'_1)^\nu
\partial'_\mu\partial'_\nu O_i(x'_1)
+\cdots .
\label{dexp}
\end{eqnarray}
Because the kernel depends only on the separation
$x'_2-x'_1$ and is invariant under
$x'_2-x'_1\rightarrow-(x'_2-x'_1)$, the term linear in the separation
vanishes after integration. The leading term is therefore
\begin{eqnarray}
\frac{1}{2}
\int_{x'_1}
O_i(x'_1)O_i^\dagger(x'_1)
\int d^4r\,
G_i^{(2)}(r),
\label{local0}
\end{eqnarray}
where $r=x'_2-x'_1$. Higher-derivative terms are suppressed by powers
of the ratio between the foamon correlation length $\xi_i$ and the
characteristic variation scale 
$L_i$ of the physical operator $O_i$, i.e., 
\begin{equation}
\xi_i/L_i\ll 1, 
\label{xiLi}
\end{equation}
and $O_i$ varies smoothly on length scale $\xi_i$. 
They will be
neglected in the leading-order local approximation.

Equivalently, using the momentum-space representation of the
propagator, the leading local contribution can be expressed
schematically as
\begin{eqnarray}
\frac{1}{2}
(m_i)^{-4}
\int_{x'}O_i^2(x')
\int_0^{bM_{\rm pl}}
\frac{d^4k'}{(2\pi)^4}
\frac{1}{k'^2+m_i^2},
\label{local}
\end{eqnarray}
where
\begin{equation}
O_i^2\equiv O_iO_i^\dagger.
\end{equation}
In the neighbourhood of the point $x'_1$, we approximately have the four-dimensional volume
$\int_{x'_2}\approx %(\pi^2/2)
(m_i)^{-4}$ in Eq.~(\ref{racci}) \footnote{Note that the phase factor $e^{ik'(x'_2-x'_1)}\sim 1$ and $k'(x'_2-x'_1)$ is invariant under the $b$ scaling (\ref{scaling}).}. The factor $(m_i)^{-4}$ represents the characteristic four-volume
over which the massive foamon correlation kernel has appreciable
support. Numerical factors associated with the precise definition of
this correlation volume are not essential for the scaling argument.

Using the Wilsonian scaling relations in
Eqs.~(\ref{renr}) and (\ref{renuv}), together with the scaling of the
physical operators in Eq.~(\ref{oscaling}), we obtain
\begin{eqnarray}
(m_i)^{-4}\int_{x'}O_i^2
&\longrightarrow&
\frac{1}{b^2}
\int_{x'}\frac{\tilde O_i^2}{\tilde m_i^4}.
\label{o2scale}
\end{eqnarray}
Here we have used the mass 
scaling (\ref{massscaling})
and the canonical scaling (\ref{oscaling}) of $O_i$ appropriate to the interaction
$\phi_iO_i$.

The remaining momentum integral is
\begin{eqnarray}
\frac{1}{b^2}
\int_0^{bM_{\rm pl}}
\frac{d^4k'}{(2\pi)^4}
\frac{1}{k'^2+m_i^2}
&=&
\frac{1}{16\pi^2b^2}
\left[
(bM_{\rm pl})^2
-
m_i^2
\ln\left(
1+\frac{(bM_{\rm pl})^2}{m_i^2}
\right)
\right].
\label{newton}
\end{eqnarray}
Substituting
$m_i=\tilde m_i/b$ gives
\begin{eqnarray}
\frac{1}{b^2}
\int_0^{bM_{\rm pl}}
\frac{d^4k'}{(2\pi)^4}
\frac{1}{k'^2+m_i^2}
&=&
\frac{1}{16\pi^2}
\left[
M_{\rm pl}^2
-
\frac{\tilde m_i^2}{b^4}
\ln\left(
1+\frac{b^4M_{\rm pl}^2}{\tilde m_i^2}
\right)
\right].
\label{newton2}
\end{eqnarray}
\comment{In the regime
\[
\frac{b^4M_{\rm pl}^2}{\tilde m_i^2}\ll1,
\]
the logarithm can be expanded, yielding
\begin{eqnarray}
\frac{1}{b^2}
\int_0^{bM_{\rm pl}}
\frac{d^4k'}{(2\pi)^4}
\frac{1}{k'^2+m_i^2}
&\simeq&
\frac{M_{\rm pl}^2}{16\pi^2}
\left[
1+\mathcal O
\left(
\frac{b^4M_{\rm pl}^2}{\tilde m_i^2}
\right)
\right].
\label{newton3}
\end{eqnarray}
}
Thus, to leading order in the Wilsonian scaling-invariant domain, we identify the scale ($b$) independent contribution,
\begin{eqnarray}
\exp\left[
\frac{1}{2}
O_i(\Delta_{ij})^{-1}O_j^\dagger
\right]
&\simeq&
\exp\left[
\frac{M_{\rm pl}^2}{32\pi^2}
\int_{x'}
\frac{\tilde O_i^2}{\tilde m_i^4}
+\cdots
\right].
\label{racci1}
\end{eqnarray} 
The ellipsis denotes higher-derivative operators, subleading
corrections in the ratio of the Wilsonian variation scale $bM_{\rm pl}$ to the foamon correlation-length scale $\tilde m_i$, and contributions from interactions beyond the Gaussian
approximation.

Equation~(\ref{racci1}) shows that the nonlocal interaction generated
by foamon propagation reduces, at leading order in the
long-wavelength expansion, to a local quadratic operator
$\tilde O_i^2$. The Planck scale and the foamon correlation length control the coefficient of this induced operator. The precise numerical coefficient can receive corrections from the full
interacting foamon theory.

\subsection{Effective field theories emerging from spacetime-foam dynamics}
\label{effeca}

Within the IR- and UV-scaling-invariant domains
${\mathcal D}_{\rm ir}$ and ${\mathcal D}_{\rm uv}$, respectively, the
long-wavelength foamon dynamics generate an approximately
scale-independent effective action for the physical field operators
$\tilde O_i$ at characteristic scales
$\tilde m_i\ll M_{\rm pl}$. Combining Eqs.~(\ref{effactionOi}),
(\ref{vacuf}), and (\ref{racci1}), we write this effective action
schematically as
\begin{eqnarray}
S^{\rm eff}_E(\tilde O_i)
&\simeq&
-\int_{x'}\tilde\rho_i
+
\frac{M_{\rm pl}^2}{16\pi^2}
\int_{x'}
\frac{\tilde O_i^2}{\tilde m_i^4}
+
c_n\int_{x'}
\frac{\tilde O_i^{2n}}{\tilde m_i^{4n}}
+\cdots ,
\label{efff}
\end{eqnarray}
where numerical coefficients and normalisation factors depend on the
precise definition of the foamon--operator coupling and on the
Wilsonian coarse-graining scheme. The quadratic term
$\tilde O_i^2$ originates from the foamon two-point correlation
function in Eq.~(\ref{corr1}), in close analogy with the two-point
connection functional $C_{ij}(x_1,x_2)$ appearing in
Eq.~(\ref{nonlocal}).

More generally, higher-order operators in Eq.~(\ref{efff}) are
generated by connected higher-point foamon correlation functions,
Eq.~(\ref{ncorr}). Their coefficients are suppressed by appropriate
powers of the ratio between the microscopic foam scale and the
macroscopic correlation scale. Schematically,
$
c_n\sim
{\cal O}\left[
\left(\ell_{\rm pl}/\xi_i\right)^{p_n}
\right],
$
where the exponent $p_n>0$ depends on the operator dimension and on
the detailed foamon dynamics. Such higher-dimensional operators are
irrelevant in the Wilsonian sense near the corresponding fixed point
when their scaling dimensions make their couplings decrease toward
long distances. We therefore retain only the leading operators in the
scaling-invariant domains considered here.

The characteristic foamon mass scale $
\tilde m_i\simeq \xi_i^{-1}
$
is determined by the long-distance dynamics of the
foamon field coupled to the physical operator $O_i$. In this sense,
the characteristic scale of $O_i$ provides a physical infrared scale
for the foamon effective theory, rather than an externally imposed
infrared cutoff. The resulting effective description therefore avoids
the uncontrolled infrared limit of the original Coleman
construction within the regime in which the foamon--operator
effective theory is applicable. We do not explicitly consider the
matter-sector energy-momentum tensor $T_{\mu\nu}$ in the present
article.

The first term in Eq.~(\ref{efff}) represents the field-independent
energy-density contribution associated with the relevant
long-wavelength foamon modes coupled to $\tilde O_i$. From
Eq.~(\ref{vacue}), its characteristic scaling is
\begin{equation}
\tilde\rho_i
\sim
M_{\rm pl}^2\tilde m_i^2
\sim
\frac{M_{\rm pl}^2}{\xi_i^2}.
\label{mM}
\end{equation}
Thus, this contribution combines the microscopic ultraviolet scale
$M_{\rm pl}$ of the spacetime foam with the macroscopic infrared
scale $\xi_i$ associated with the corresponding physical operator
$O_i$. In four dimensions, the scaling
$\tilde\rho_i\propto M_{\rm pl}^2/\xi_i^2$ has the same dimensional
form as the energy density associated with a cosmological term whose
curvature scale is $\xi_i^{-1}$. It may therefore be regarded as
holographic-type scaling, although an explicit holographic
interpretation requires an additional entropy or area-law argument.

Since
$\tilde m_i\ll M_{\rm pl},$
one has parametrically
\begin{equation}
\tilde\rho_i
\sim
\left(\frac{\tilde m_i}{M_{\rm pl}}\right)^2
M_{\rm pl}^4
\ll
M_{\rm pl}^4.
\end{equation}
In this Wilsonian sense, integrating out the short-distance foamon
degrees of freedom leaves a long-distance contribution whose
characteristic energy scale is parametrically smaller than the
Planck-scale vacuum-energy density.

By Wick rotation of the effective Euclidean theory in
Eq.~(\ref{efff}), we obtain the corresponding effective action
$S^{\rm eff}_M(\tilde O_i)$ in Minkowski spacetime. The Wick rotation
is performed at the level of the effective foamon-field theory and
its coupling to the long-distance operators $O_i$, rather than being
applied directly to the original Euclidean gravitational path
integral
in Eq.~(\ref{insc}). In other words, the effective Minkowski theory
is obtained only after the microscopic Euclidean quantum-gravity
degrees of freedom have been coarse-grained into the effective
foamon--operator description.

\red{The resulting effective actions
$S^{\rm eff}_E(\tilde O_i)$ and
$S^{\rm eff}_M(\tilde O_i)$ describe the long-distance dynamics of
the physical field operators $\tilde O_i$, including induced
operators such as $\tilde O_i^2$ and higher-order terms. Their
characteristic length scale is $\xi_i=\tilde m_i^{-1}$, which is
assumed to satisfy
$
\xi_i\gg\tilde\ell\gtrsim\ell_{\rm pl}.
$
At such scales, the microscopic discreteness of the Planck-scale
foam is not resolved, and the effective description can be
formulated in terms of continuous spacetime fields. Provided that
the underlying microscopic theory and coarse-graining preserve the
relevant spacetime symmetries, Lorentz symmetry and translation
invariance can emerge as approximate macroscopic symmetries of the
effective theory.}

\red{The effective action in Eq.~(\ref{efff}) shares an important
structural feature with the nonlocal action in Eq.~(\ref{nonlocal}).
In the Coleman--Hawking formulation, the background-wormhole
connection functional $C_{ij}(x_1,x_2)$ determines the nonlocal
coupling between physical operators $O_i$ at separated spacetime
points. In the present framework, we instead interpret the
long-distance kernel as arising from collective foamon correlations
associated with the microscopic 2-simplex fields $X_h$ and their
effective collective fields $\Phi_i$. The physical operators $O_i$
couple to these collective modes through Eq.~(\ref{reg}) and
introduce characteristic long-distance scales that regulate the
foamon correlations. We therefore model the collective modes and
their interactions by an effective foamon quantum field theory and
study their fixed points and scaling behaviour approximately.}

\red{The preceding analysis provides a qualitative mechanism by which
effective classical or quantum field theories for the physical
operators $O_i$ can emerge from long-wavelength collective
excitations of spacetime foam. The characteristic foamon scales
$\tilde m_i$ determine the coefficients of the induced operators
through the generalised Coleman--Hawking interaction
Eq.~(\ref{scalarc3}), thereby generating effective couplings and
parameters of the $O_i$ field theories. In this sense, the effective
action in Eq.~(\ref{efff}) represents the leading-order long-distance
contribution when the RG flow lies within either the IR- or
UV-scaling-invariant domain.
}

\red{The approximation is expected to be less reliable in the crossover
region connecting the two scaling-invariant domains. As the running
parameter $b$ changes between the IR and UV regimes, higher-dimensional
and higher-derivative operators, including those represented by the
ellipsis in Eq.~(\ref{efff}) and generated by the derivative
expansion in Eq.~(\ref{dexp}), can become increasingly important.
Determining their coefficients and the detailed interpolation between
the two fixed-point regimes requires a nonperturbative analysis,
which we leave for future numerical investigations.
}

We now apply the resulting effective action to the Einstein equation
and investigate its implications for cosmological evolution and
astrophysical systems.

\section{Einstein constant in cosmology}\label{effeic}

\subsection{Ricci scalar renders cosmological constant}

\red{In Einstein's theory, the Ricci scalar curvature $R=g^{\mu\nu}G_{\mu\nu}$ is a fundamental geometric measure derived from spacetime curvature. It is a single scalar value that describes how the volume (size) of a small region of curved spacetime deviates from that of flat spacetime due to matter and gravity. 
In this case, $R$ is the field operator at large distances, as an infrared regulator 
for the quantum field theory of the foamon $\phi_R$
that couples (\ref{scalarc3}).} 

The field operator $\tilde O$ is related to $R/(16\pi G)$, and the characteristic 
scale $\tilde m$ of low-energy fields $\phi_R$ is related to the cosmological term $\Lambda_R/(8\pi G)$: 
\begin{equation}
\tilde m^2 \sim 8 \pi \Lambda_R,\quad \quad \tilde O^2 \sim \pi (\tilde m)^4R. 
\label{eins} 
\end{equation}
\red{The effective action  (\ref{efff}) leads to the emergence of the Einstein action (\ref{eaction}) with the cosmological constant $\Lambda_R$. 
Subscript $_R$ specifies the scale $\Lambda_R$ (length  
$\xi_R\simeq \tilde m^{-1}$) that is determined by the characteristic scaling behaviour of the Ricci scalar $R$,
\begin{equation}
\Lambda_R\sim \frac{1}{8\pi}\xi^{-2}_R, \quad \xi_R\gg \ell_{\rm pl}=M^{-1}_{\rm pl}.
\label{lambdar}
\end{equation}
It thus explains the smallness of the cosmological constant $\Lambda_R\ll M^2_{\rm pl}$, getting rid of the problem of cosmological constant  
$\Lambda\propto M^2_{\rm pl}$, which is attributed to the vacuum energy up to the ultraviolet Planck scale.} 

\red{Beside its value, the scale $\Lambda_R$ (\ref{lambdar}) determined by the Ricci scalar $R$ has the following features: 
\begin{enumerate}[{(i)}]
    \item $\Lambda_R$ is attributed to the largest correlation length $\xi_R$ (lowest-lying energy mode $\tilde m$) of quantum foamon fields $\phi_R$ ``boxed'' by the macroscopic field $R$ at an infrared scale which is much smaller than the Planck scale. The $R$ varies smoothly over the spacetime of the length $\xi_R$, 
    see Eq.~(\ref{xiLi}). The $\Lambda_R$ value increases (decreases) as the characteristic scaling behaviour of the Ricci scalar $R$ decreases (increases).
     \item $\Lambda_R$ is not a homogeneous constant in spacetime. Instead, it depends on the sizes of different gravitating systems, whose $R$ scale varies from the universe and galaxies to the solar system;
    \item $\Lambda_R=\Lambda_R(x)$ is a smooth function in the curved spacetime described by $R$. From the effective action (\ref{efff}) and Eq.~(\ref{eins}), we approximately obtain the Einstein equation $G_{\mu\nu}+\Lambda_R(x) g_{\mu\nu}=8\pi G T_{\mu\nu}$ and generalized Bianchi identity \cite{Xue2015}
\begin{equation}
\big[8\pi G T_{\mu\nu}-\Lambda_R(x) g_{\mu\nu}\big]_{;\mu}=0,
\label{gbi}
\end{equation}
for the energy-momentum conservation including $\Lambda_R(x)$. Thus, when $\Lambda_R$ varies in spacetime, the matter sector is
not separately conserved: the variation of the effective
cosmological term corresponds to an exchange of energy-momentum
between the matter sector and the sector responsible for the
emergent cosmological contribution. The resulting coupled dynamics
also generate nonlinear back-reaction on $R$ through the Einstein equation.
This implies the interaction between $\Lambda_R$ (dark energy) and matter.
\end{enumerate}
These features are distinct from the Einstein cosmological constant $\Lambda$.} 

\red{In the $\Lambda$CDM model for the universe evolution, the Einstein equation reduces to the Friedmann equation for a homogeneous universe with energy density $\rho=\rho_{_M}+\rho_{_R}+\rho_{_\Lambda}$ of matter, radiation and $\Lambda_R$. %and pressure $p=p_{_M}+p_{_R}+p_{_\Lambda}$, $p_{_M}\approx 0,\quad p_{_R}=\rho_{_R}/3,\quad p_{_\Lambda}=\omega \rho_{_\Lambda}$. 
The Ricci scalar reads 
\begin{equation}
R_H=6(H^2+\dot H)\approx 12H^2(1-\frac{3}{4}\Omega_M),
\label{riccif}
\end{equation}
where
$\dot H=-H^2(1+q)$ and deceleration parameter 
$q=\frac{1}{2}\Omega_M-
\Omega_\Lambda$ for $p_{_\Lambda}=-\rho_{_\Lambda}$ of the equation of state. 
In the matter-dominated epoch, the Friedmann equation 
$\Omega_M+\Omega_\Lambda\approx 1$, where $\Omega_{M,\Lambda}=\rho_{_{M,\Lambda}}/\rho_c$,the critical density $\rho_c=3H^2/(8\pi G)$ and the Hubble function $H^2=(8\pi G/3)\rho$.}

\red{In this circumstance, the Universe size $H^{-1}$ is the infrared cutoff that the field operator $R$ provides for the quantum foamon field $\phi_R$ theory. The correlation length $\xi_R=\tilde m^{-1}$ of the quantum foamon field $\phi_R$ theory is thus proportional to $H^{-1}$. 
Therefore, Eqs.~(\ref{lambdar}) and (\ref{riccif}) yield the cosmological constant, 
\begin{equation}
\Lambda_R \propto R_H\propto H^2,
\label{lambdaf}
\end{equation}
which is set by the cosmic scale $H$. We cannot quantitatively determine the proportionality at this preliminary stage, since non-linear 
back-reaction via Eq.~(\ref{gbi}) cannot be not 
easily calculated.} 

The Planck length $\ell_{\rm pl}$ and the horizon size $H^{-1}$ determine the foamon field energy density (dark-energy density)
\begin{equation}
\rho_{_\Lambda}=\frac{\Lambda_R}{8\pi G}\sim \frac{\xi_R^{-2} M_{\rm pl
}^2}{(8\pi)^2}
\propto \frac{H^2 M_{\rm pl
}^2}{(8\pi)^2},
\label{lambda} 
\end{equation}
whose value is consistent in order of magnitude with today's observations. It is reminiscent of the one derived from simply counting degrees of freedom \cite{Gurzadyan2003}. The result (\ref{lambda}) is many orders of magnitude smaller (larger) than the naive estimate $\ell^{-4}_{\rm pl}$ of the foamon field energy density.

\subsection{Equation of state and holographic entropy}

As the Universe volume $V\propto H^{-3}$ expands adiabatically $\delta V>0$, conserving the total number of degrees of freedom $\delta N=0$, the positive particle's repulsive energy $E$ (mean kinetic energy) decreases $\delta E<0$, because the energy density $\rho$
decreases.  
It obeys the first law of thermodynamics in the comoving frame,
\begin{equation}
p\delta V+\delta E =0,\quad p=-(\delta E/\delta V) >0,\quad E=\int dV \rho .
\label{vvp} 
\end{equation}
The spectra and distributions of 
particles determine the pressure $p$ and the energy density $\rho$. It leads to
the equation of state $p=(\gamma-1)\rho$ with thermal index $\gamma>1$ and $\gamma=4/3$ for free relativistic particles.  

Unlike particles kinematically moving in the spacetime manifold, spacetime foams (wormholes at the Planck length) and their correlations are geometrically embedded in the stretching manifold and thus have a different equation of state \cite{Xue2009a}. 
The gravitational energy of spacetime foamons is positive  
\begin{equation}
E_{_\Lambda}=\int dV \rho_{_\Lambda}=\frac{M^2_{\rm pl}}{16 \pi}\xi_R>0, \quad dV = 4\pi \xi^2_R d\xi_R,
\label{lvve} 
\end{equation}
increasing with $\xi_R$ and the volume expansion $d V>0$ of the Universe, namely the variation $\delta E_{_\Lambda}>0$. Thus, using Eq.~(\ref{vvp}) $p_{_\Lambda}=-(\delta E_{_\Lambda}/\delta V) <0$, we obtain the negative pressure $p_{_\Lambda}$ and the equation of state
\begin{equation}
p_{_\Lambda}= - (\delta E_{_\Lambda}/\delta V)= \omega_{_\Lambda}\rho_{_\Lambda},\quad \omega_{_\Lambda}=-1,
\label{lvvp} 
\end{equation}
of foamon field modes, which accounts for the cosmological constant $\Lambda_R$ (dark energy).

The basic observations on $\omega_{_\Lambda}=-1$ (\ref{lvvp}) are that, contrary to the case $\delta E<0$ of particles, 
a stretched manifold increases the energy $E_\Lambda\propto\xi_R$ (\ref{lvve}) of spacetime foamons, which are long-range ($\xi_R$) collective excitations  
of the foamy spacetime at the Planck scale, namely the Planck lattice. \red{While 
the fundamental reasons for $\omega_{_\Lambda}=-1$ (\ref{lvvp}) are that
the resulting energy density $\rho_{_\Lambda}$ (\ref{lambda}) not only 
gives a constrain on the UV ($M_{\rm pl}$) 
and IR ($\xi_R^{-1}$) cutoffs of quantum foamon field theories, which are free from catastrophes of UV and IR divergences; but also
is in accordance with the holographic principle, i.e., the 
energy and entropy cannot scale extensively with volume $\xi_R^3$, vastly over-counting degrees of freedom for a very large IR cutoff \cite{Cohen1999}. The interaction of dark energy and matter can effectively modify the equation of state $\omega_{_\Lambda}\approx -1$ \cite{Begue2019}.
}

The resulting energy density (\ref{lambda}) of the quantum foamon field can be formally written as $\rho_{_\Lambda}=M_{\rm pl}n_{_\Lambda}$ 
in terms of the effective foamon number density $n_{_\Lambda}$ \red{(entropy density $s_{_\Lambda}$)}, 
\begin{equation}
n_{_\Lambda}= \frac{\ell^{-1}_{\rm pl}\xi^{-2}_R}{(8\pi)^2}\propto s_{_\Lambda},
\label{nsden} 
\end{equation}
as if the long-range modes of the spacetime foamon field were holographically distributed as a thin layer of width $4\pi\xi^2_R\ell_{\rm pl}$ near the 
universe horizon $\xi_R\sim H^{-1}$. \red{Total number or entropy of spacetime foamon modes  
increases with area $\xi_R^2$, instead of 
volume $\xi_R^3$, following the holographic principle.} 
The holographic configuration minimises the total energy and number (entropy) of spacetime foamon modes, whose wavelength is on the order of the horizon. \red{The possible interpretation could be that, on macroscopic average in strengthening manifold, the short-range (high-energy) foamon field fluctuating modes \footnote{Here, the local foamon field fluctuation exhibits similar dynamics to the “microcyclic universes” on small scales, shown by the local scale factor oscillation in Figure 1 of Refs.~\cite{Wang2020, Wang2020a}.} cancel each other, while the long-range (low-energy) modes on the horizon $H$ scale survive collectively.}

\subsection{Dark energy interaction with matter in cosmological evolution}

Through the foamon-particle interaction $y\phi\bar\psi\psi$
(\ref{coleman3}) or (\ref{haction}), 
high-energy foamon-field fluctuations produce massive particle-antiparticle pairs \cite{Parker1974}. 
\red{We have found that these pairs oscillate coherently with the foamon field fluctuation, forming a holographic layer of massive pairs near the horizon, interacting with the (dark) energy density $\rho_{_\Lambda}$ (\ref{lambda}) of the foamon fields. Following the Friedmann equations for the Hubble function $H$, $\rho_{_\Lambda}$ converts to an energy density of massive pairs. On the other hand, massive pairs' back-reaction on the foamon field 
converts their energy to the $\rho_{_\Lambda}$ of the foamon field. Such
back-and-forth processes lead to the interaction between matter $M$ and dark energy $\Lambda$ in the entire evolution of the Universe.
The $\rho_{_\Lambda}$ on the inflation scale $H\sim 10^{-(4\sim 5)}M_{\rm pl}$ drives inflation, which ends due to the $\rho_{_\Lambda}$ decrease for massive pair productions \cite{Xue2023}. It vanishes during reheating by producing a large number of 
massive pairs, which then decay to relativistic particles \cite{Xue2023a}, resulting in a Big Bang. Afterwards, in standard cosmology, the $\rho_{_\Lambda}$ increases to today's value due to the back-reaction from radiation 
and matter energy densities, and cosmological
fine-tuning and coincidence problems can be resolved \cite{Xue2024}.} 

\red{This scenario was tested with supernovae and Baryon Acoustic Oscillations data \cite{Begue2019} and it can be
a solution to the problem of the $H_0$ and $S_8$ tensions \cite{Gao2021, Gao:2022ahg}. Moreover, it is worthwhile in future to study the measurable signature of gravitating features of such an interacting dark energy $\Lambda_R$ at the $R$ scale, which may imprint on the 
integrated Sachs-Wolfe effect \cite{Sachs1967} from varying gravitational potentials, and background gravitational waves from the production of primordial black holes. Undoubtedly, this dark energy $\Lambda_R$ requires further in-depth 
research in many aspects. Research on this issue requires
theoretical studies and analysis using full
CMB angular power spectrum data. Moreover, the dark energy $\Lambda_R$ 
may also lead to modifications in the
mid-scale clustering patterns of large-scale structures,
which will potentially affect the clustering strength at
various scales. Therefore, such modifications may also
affect the mass function and profile of dark matter halos, the statistical properties of galaxy clusters, the alignments of galaxies, the structure of the cosmic web, and
so on. The study of these effects is a complicated issue,
requires the use of N-body simulations, and also relies
on assumptions about the nature of dark matter (such
as cold or fuzzy dark matter). 
%All these aspects deserve further in-depth discussion. 
Nevertheless, we will use a simple model to study the $\Lambda_R$ effects on self-gravitating systems in Sect.~\ref{effeia}.}

\subsection{Scale factor and renormalisation-group flow}

In Friedmann--Lemaître--Robertson--Walker (FLRW) cosmology, the scale
factor $a(t)$ relates physical and comoving coordinates and momenta,
\begin{equation}
x_{\rm phys}=a(t)x_{\rm com},
\qquad
k_{\rm phys}=\frac{k_{\rm com}}{a(t)} .
\label{uscaling}
\end{equation}
\red{The 
$(x_{\rm phys},k_{\rm phys})$ and $(x_{\rm com},k_{\rm coms})$ are dimensional variables, coordinating the causal patch of the Universe. 
The characteristic physical length scale of the universe is
identified here with the foamon correlation length $\xi_R(t)$. Thus, the scale factor may be related to the
ratio of the macroscopic correlation length to the microscopic
Planck-lattice scale,
\begin{equation}
a(t)\propto\frac{\xi_R(t)}{\tilde\ell(t)},
\label{ascale}
\end{equation}
up to a normalization factor $a_0(t_0)$ \cite{Xue2015}.}

As the universe expands, short-distance foamon fluctuations are
integrated out through the Wilsonian coarse-graining procedure,
yielding effective fields, masses, couplings, and Lagrangians
appropriate to progressively larger physical scales. The comparison between 
the scaling parameter $b$ in Eq.~(\ref{scaling}) and the scale factor $a(t)$ in Eq.~(\ref{uscaling}) motivates
the correspondence
\begin{equation}
(x',k')\leftrightarrow(x_{\rm phys},k_{\rm phys}),
\qquad
(x,k)\leftrightarrow(x_{\rm com},k_{\rm com}),
\label{ccaling}
\end{equation}
\red{and, phenomenologically,
\begin{equation}
b\simeq\frac{a(t)}{a_0(t_0)}
=\frac{1}{1+z},
\label{bascaling}
\end{equation}
relating to the redshift $z$, where we normalize $a_0(t_0)=1$ at the present time $t_0$.}

With this identification, the RG parameter $b$ provides a convenient
measure of the cosmological evolution of the effective foamon theory:
the limit $b\rightarrow0$ corresponds to the early-Universe regime,
while $b\rightarrow1$ corresponds to the present epoch. The proposed
relation between the Wilsonian RG flow and cosmic expansion is
schematic and requires further investigation to establish whether a
dynamical correspondence between $b$ and $a(t)$ can be derived from
the underlying theory.

\section{\red{Emergent cosmological term in self-gravitating systems}}\label{effeia}

\subsection{\red{Virial equilibrium configuration}}\label{virs}

Self-gravitating systems of matter density $\rho_{_M}$ and dark energy density $\rho_{_{\Lambda}}$ are described by the Einstein equation, which gives the Ricci scalar \footnote{Using the observed value of the cosmological constant $\rho_{_\Lambda}$ \cite{Aghanim2020} 
and different equations of states $p_{_\Lambda}=\omega_{_\Lambda}\rho_{_\Lambda}$,
References \cite{Ho:2015nsa,Roupas2014,He2017, Zhang2023} study dark energy effects based on the hydrostatic equilibrium of a self-gravitating gas in the presence of dark energy or the Schwarzschild-de Sitter (SdS) solution, respectively.}
\begin{equation}
R=g^{\mu\nu}R_{\mu\nu}=8\pi G\rho_{_M} - 16\pi G\rho_{_\Lambda},
\label{ricciv}
\end{equation}
for negligible matter pressure $p_{_M}\approx 0$. In spherical symmetry, it yields the Poisson equation for the gravitational potential $\phi(r)$,
\begin{equation}
\nabla^2\varphi(r)=4\pi G \rho_{_M}-8\pi G\rho_{_\Lambda},\quad  R\approx 2 \nabla^2\varphi(r),
\label{poisson}
\end{equation}
in the Newtonian limit.
The Poisson equation admits the solution $\varphi=\varphi_{_M}+\varphi_{_\Lambda}$,
\begin{eqnarray}
    \varphi_{_M}(r)=-G \int_V\frac{\rho_{_M}(\vec r')}{|\vec r-\vec r'|}d^3\vec r',\quad
\varphi_{_\Lambda}(r)\approx \frac{4\pi G}{3}\rho_{_\Lambda}r^2,\label{gpl}
\end{eqnarray}
assuming that $\rho_{_\Lambda}$ varies smoothly in the volume $V=\frac{4\pi}{3}r^3$ of the radial size $r$.
The negative $U_M$ 
(positive $U_\Lambda$) are attractive 
(repulsive) energies, respectively, 
\begin{eqnarray}
    U_M=\frac{1}{2}\int_V\rho_{_M}\varphi_{_M}d^3 r,\quad U_\Lambda=\int_V\rho_{_M}\varphi_{_\Lambda}d^3r.\label{gpe}
\end{eqnarray}
The total gravitational potential energy $U=U_{_M}+U_{_\Lambda}<0$, i.e., $|U_M|\gg U_\Lambda$ and $\rho_{_M}\gg \rho_{_\Lambda}$ for bound systems.
The total kinetic energy $K_M$ of the matter is
\begin{eqnarray}
    K_M=\frac{1}{2}\int_{V}dm ~v^2_{_M}= \frac{1}{2}\int_{V}d^3 r\rho_{_M}v^2_{_M},\label{ke}
\end{eqnarray}
where $v_{_M}$ is the velocity of the mass element $dm=d^3 r\rho_{_M}$.

While the Universe is expanding,  self-gravitating systems dynamically evolve into a virial equilibrium configuration via equilibrating potential and kinetic energies in a finite volume, obeying the virial relation
\begin{eqnarray}
    2K_M + U_M+ U_\Lambda =0, \quad K_M=-(U_M+ U_\Lambda)/2.\label{vir}
\end{eqnarray}
The total energy $E=K_M+U_M+ U_\Lambda=U/2 <0$ and the kinetic energy $K_M=|U|/2$.
These relationships effectively describe 
the complex dynamics 
of self-gravitating systems in astrophysics, 
from galaxies to the solar system. 

We will use the subscript $_{\rm vir}$ 
to indicate the quantities of the virial equilibrium configuration (\ref{vir}) of 
self-gravitating systems. 
The virial mass energies are
\begin{eqnarray}
M_{\rm vir}&=&\int_{V_{\rm vir}}d^3r\rho_{_M}= \rho^{\rm vir}_{_M} V_{\rm vir},\label{virm}\\
M^\Lambda_{\rm vir}&=&\int_{V_{\rm vir}}d^3r\rho_{_\Lambda}= \rho^{\rm vir}_{_\Lambda} V_{\rm vir},
\end{eqnarray}
where $\rho^{\rm vir}_{_M}$ and $\rho^{\rm vir}_{_\Lambda}$ are the mean densities of matter and dark energy within a virial volume $V_{\rm vir}=\frac{4\pi}{3}r^3_{\rm vir}$ of radius $r_{\rm vir}$. Using Eq.~(\ref{ricciv}) and $\rho^{\rm vir}_{_M}\gg \rho^{\rm vir}_{_\Lambda}$,
we define an averaged Ricci scalar in the virial volume
\begin{eqnarray}
R_{\rm vir}\approx 8\pi G \rho^{\rm vir}_{_M}.\label{virr}
\end{eqnarray}
at the redshift $z$. Following (\ref{lambdar}) and (\ref{lambdaf}), the dark energy $\Lambda_{\rm vir}$ should be proportional to the scale given by 
the averaged Ricci scalar (\ref{virr})
\begin{eqnarray}
\Lambda_{\rm vir}\propto R_{\rm vir},\quad \rho^{\rm vir}_{_\Lambda}= \frac{\Lambda_{\rm vir}}{8\pi G}\propto \rho^{\rm vir}_{_M}.\label{virrl}
\end{eqnarray}
Thus, unlike a strictly homogeneous cosmological constant, the
effective cosmological term in a self-gravitating system depends on
the characteristic curvature and hence on the mean matter density of
the system. This relation in an equilibrium approximation will be used below to study the effects of
the emergent $\Lambda_R$ on virialised astrophysical systems such as
galaxies and the Sun.

\subsection{\red{Rotation curve affected by dark energy}}

In a galactic system, the matter density $\rho_{_M}(r)$ in principle contains stellar disk, bulge, gas and dark-matter
halo contributions. 
Since the dark energy contribution 
$U_{\rm \Lambda}$ (\ref{gpe}) is dominated by the dark energy potential $\varphi_{\rm \Lambda}\propto r^2$ (\ref{gpl}) at large radii, we neglect the small-radii matter density distribution of the stellar disk, bulge
and gas. Consider only the NFW dark matter halo contribution \cite{Navarro1996}, 
\begin{eqnarray}
    \rho_{_M}(r)\approx\rho_{_{\rm NFW}}(r)= \frac{\rho_s}{(r/r_s)(1+r/r_s)^2},\label{nfw}
\end{eqnarray}
with halo characteristic scales $\rho_s$ 
and $r_s$. The circular velocity $v(r)$ is given as
\begin{eqnarray}
    v^2(r)=
    \frac{2K_M}{M_{\rm vir}}= \frac{1}{M_{\rm vir}}\int_{V}d^3 r\rho_{_{\rm NFW}}v^2_{_M},\label{vcd}
\end{eqnarray}
relating to the velocity dispersion $\sigma^2= v^2/2$ in the isothermal configuration.  

We calculate the virial mass $M_{\rm vir}$, gravitational potentials $U_M$, and $U_\Lambda$ in 
Eq.~(\ref{gpe}). 
As a result, the virial relation (\ref{vir}) yields the model rotation curve in terms of dimensionless radius $x=r/r_{\rm vir}$
\begin{eqnarray}
    \frac{v^2_{_{\Lambda} }(x)}{v^2_{\rm vir}}&=& \frac{1}{x}\frac{\ln (1+cx)-\frac{cx}{1+cx}}{\ln (1+c)-\frac{c}{1+c}}\nonumber\\
    &+&\left(\frac{\rho^{\rm vir}_{_\Lambda}}{\rho^{\rm vir}_{_M}}\right)\frac{\Big[\frac{cx^3}{1+cx}+\frac{3x}{c}\big(1-\frac{cx}{2}\big)- \frac{3}{c^2}\ln (1+cx)\Big]}{\ln (1+c)-\frac{c}{1+c}},\label{vc}
\end{eqnarray}
where the virial velocity 
$v^2_{\rm vir}=GM_{\rm vir}/r_{\rm vir}$ and the concentration parameter $c=r_{\rm vir}/r_s$. 
The first term is the NFW result, which is Kepler's law for astrophysical systems of an extended mass distribution.
The second term is a small correction from the dark energy contribution, characterised by
the ratio of dark energy and matter densities
\begin{equation}
    \epsilon= \frac{\rho^{\rm vir}_{_\Lambda}}{\rho^{\rm vir}_{_M}}\ll 1.
    \label{xpara}
\end{equation}  

\begin{figure}
\centering
\includegraphics[width=3.2in,height=2.2in]{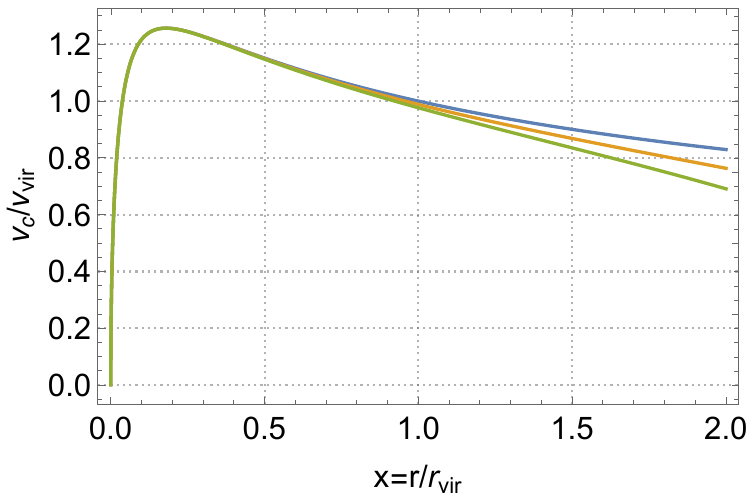}
\caption{To illustrate qualitatively the effect of the emergent cosmological
term, we plot the rotation curve in Eq.~(\ref{vc}) and the deflection
angle in Eq.~(\ref{gl}) using representative Milky Way parameters,
$c=12$, $r_{\rm vir}=220,{\rm kpc}$,
$r_s=r_{\rm vir}/c\simeq18,{\rm kpc}$,
$M_{\rm vir}\simeq1.1\times10^{12}M_\odot$, and
$v_{\rm vir}^2=GM_{\rm vir}/r_{\rm vir}$.
At large radii, the standard NFW contribution ($\epsilon=0$) decreases
approximately as $v\propto r^{-1/2}$, reflecting the declining
gravitational acceleration of the extended halo. The repulsive
contribution of the effective dark-energy density further reduces the
circular velocity, with the deviation increasing as the ratio
$\epsilon$ in Eq.~(\ref{xpara}) becomes larger. Further measurements
of rotation curves at large radii, such as those discussed in
Ref.~\cite{Jiao2023}, could therefore constrain this effect.
If $\epsilon$ increases with redshift, as suggested by the
curvature-dependent dark-energy scenario, the model predicts a
stronger suppression of the outer rotation curves for higher-redshift
galaxies. The examples $\epsilon=0.1$ and $0.2$ illustrate this trend.
It is qualitatively consistent with the observed evolution of galaxy
rotation curves reported in Fig.~2 of Ref.~\cite{Genzel_2017}, although a
quantitative comparison with the data is required to test the model. 
}\label{rotationc}
\end{figure}

Considering the expansion of the Universe in time, the average dark-matter density $\rho^{\rm vir}_{_M}$ (\ref{virm}) is parameterised as in \cite{Navarro1996}
\begin{eqnarray}
    \rho^{\rm vir}_{_M}=\frac{M_{\rm vir}}{V_{\rm vir}} =\Delta_c(z)\rho_c(z),\quad \rho_c(z)=\frac{3H^2(z)}{8\pi G}\label{mvir}
\end{eqnarray}
in terms of the critical density $\rho_c$ in the homogeneous universe with redshift $z$. The over-density factor in the presence of the cosmological constant \cite{Bryan_1998} 
\begin{eqnarray}
\Delta_c(z)=18\pi^2+82(\Omega_M-1)-39(\Omega_M-1)^2 >1,\label{overl}
\end{eqnarray}
which measure the deviation of the local density $\rho^{\rm vir}_{_M}$ from the global density $\rho_c$. This describes dark-matter self-gravitating clusters together, leading to galactic halo formation in the evolution of the Universe. Analogously to Eq.~(\ref{mvir}), we introduce the over-density factor $\Delta^{\Lambda}_c(z)$ 
\begin{eqnarray}
    \rho^{\rm vir}_{_\Lambda}=\Delta^{\Lambda}_c(z)\rho_c(z),\label{lvir}
\end{eqnarray}
describing the deviation of the dark energy density from the globally homogeneous density $\rho_c(z)$, 
as a consequence of local dark energy and matter gravitating together. 
We speculate that
as redshift $z$ increases, $\Delta_c(z)$ (\ref{mvir}) becomes smaller, i.e., halos are less dense, due to the attractive gravitating nature of dark matter $\rho^{\rm vir}_{_M}$; while $\Delta^{\Lambda}_c(z)$ (\ref{lvir}) becomes larger, due to the repulsive gravitating nature of dark energy $\rho^{\rm vir}_{_\Lambda}$. Thus, the ratio of dark energy and matter densities
$\epsilon=\Delta_c^{\Lambda}/\Delta_c$ (\ref{xpara}) 
should increase with redshift. This needs to be studied in a cosmological model. 
Using the results (\ref{vc}), we show in Fig.~\ref{rotationc} the dark-energy contribution to the rotation curve (\ref{vc}) and deflection angle (\ref{gl}), in particular at large radii 
$r>r_{\rm vir}$.

An important consequence of Eq.~(\ref{virrl}) is that the effective
dark-energy density need not be spatially homogeneous. Due to the attractive self-gravitating nature, the 
universe is locally inhomogeneous and made up of local halos  $\rho^{\rm vir}_{_M} $ and voids 
$\rho^{\rm voi}_{_M} \approx 0$. Equations (\ref{virr}) and 
(\ref{virrl}) show that the dark energy is also not 
homogeneous in spacetime, and its distribution approximately follows the clumpy distribution of matter and 
the Ricci scalar. 
The same curvature-dependent mechanism can therefore
operate in other stabilised self-gravitating systems, including the
Solar System. In the following, we use this framework to investigate
the resulting modification of the gravitational potential and the
deflection of light.

\subsection{\red{Deflection angle affected by dark energy}}

Following Eq.~(\ref{vc}),
the deflection angle due to gravitational lensing is given by \cite{Narayan1996}
\begin{eqnarray}
   \hat\alpha_{_\Lambda}(r)=2\pi v_{_\Lambda}^2(r),\quad 
    \frac{\hat\alpha_{_\Lambda}(x)}{\hat\alpha_{\rm vir}}=\frac{v^2_{_\Lambda}(x)}{v^2_{\rm vir}}
    \label{gl}
\end{eqnarray}
where $\hat\alpha_{\rm vir}=2\pi v^2_{\rm vir}$ 
and $v^2_{\rm vir}=v^2(r_{\rm vir})$.

A particular value of the deflection angle (\ref{gl}) on the surface of halo cores $r=r_{\rm vir}=r_s$ ($x=c=1$) is given by
\begin{eqnarray}
    \frac{\hat\alpha_{_\Lambda}}{\hat\alpha_{\rm vir}}&=& 1-0.41\epsilon,\quad\hat\alpha_{_\Lambda}= 2\pi \frac{GM_{\rm vir}}{r_{\rm vir}}(1-0.41\epsilon).\label{gls}
\end{eqnarray}
The deflection angle due to gravitational lensing becomes smaller 
because of the repulsive gravitational nature of dark energy. Similarly, we consider light grazing the Sun's surface. 
In the weak-field Schwarzschild--de Sitter approximation, where the geometry can be approximately described by the de Sitter and Schwarzschild-de Sitter (SdS) solution
\begin{eqnarray}
   g_{00}&=& 1-\frac{2GM_\odot}{r_\odot}-\frac{\Lambda_\odot r^2_\odot}{3}=1-\frac{2GM_\odot}{r_\odot}\left(1+\frac{\rho^\odot_{_\Lambda}}{\rho^\odot_{_M}}\right),\label{sun}
\end{eqnarray}
where, in the present framework, $\Lambda_\odot\propto R_\odot$ (\ref{virrl}) and the Ricci scalar $R_\odot\approx 8\pi G \rho^\odot_{_{M}}$ (\ref{virr}). The average densities of matter and dark energy of 
the Sun are $\rho^\odot_{_{M}}=M_\odot/(4\pi r_\odot^3/3)$ and $\rho^\odot_{_{\Lambda}}=\Lambda_\odot/(8\pi G)$, and the ratio $\epsilon_\odot=\rho^\odot_{_{\Lambda}}/\rho^\odot_{_{M}}\ll 1$. Up to the leading order of $\epsilon_\odot$, our effective-potential approximation gives
\begin{equation}
\hat\alpha_\odot^\Lambda
\simeq
\hat\alpha_\odot(1-\epsilon_\odot),
\label{glsun}
\end{equation}
where $\hat\alpha_\odot\simeq1.75$ arcsec is the standard
Schwarzschild prediction. 
Precise tracking using radio waves and spacecraft such as the Cassini mission has verified these measurements to a $0.03\%$ margin of error \cite{Fomalont_2009}, which constrains $\epsilon_\odot<10^{-5}$.

This solar-system result should be regarded as a model-dependent
estimate rather than a standard Schwarzschild--de Sitter lensing
formula. 
Consequently, a quantitative constraint on $\epsilon_\odot$ requires
a fully relativistic calculation of null geodesics and observable
angles in the spacetime generated by the spatially varying
$\Lambda_R$. 

\comment{https://en.wikipedia.org/wiki/Eddington experiment   The three classical tests of general relativity proposed by Albert Einstein are the precession of Mercury's orbit, the deflection of light by the Sun, and the gravitational redshift of light.
https://en.wikipedia.org/wiki/Tests_of_general_relativity}

\section{Summary and remarks on vacuum energy}

We have developed a framework in which the observed cosmological
energy scale may emerge from the collective dynamics of Planck-scale
spacetime foam. At microscopic scales, Wheeler-type spacetime foam is
assumed to undergo quantum creation and annihilation, while its
long-wavelength collective excitations interact with physical field
operators at macroscopic scales. We describe these collective modes
phenomenologically as quasiparticle-like ``foamons'', whose
Coleman--Hawking couplings provide physical infrared scales for their
correlation lengths.

Using a Wilsonian renormalisation-group description, we associate the
foamon dynamics with scaling-invariant domains around infrared and
ultraviolet fixed points. The corresponding correlation lengths
provide characteristic scales for the effective theories of physical
operators at large distances. In this framework, the cosmological
term is not introduced as an independent Planck-scale vacuum
contribution, but emerges from long-wavelength correlations,
\begin{equation}
\Lambda_R\sim\xi_R^{-2}.
\end{equation}
For cosmology, identifying $\xi_R$ with the characteristic Hubble
scale gives
\begin{equation}
\Lambda_R\sim H^2,
\qquad
\rho_{\Lambda_R}\sim M_{\rm pl}^2H^2,
\end{equation}
which has the observed order of magnitude while being parametrically
smaller than the microscopic Planck-scale estimate
$\rho_{\rm UV}\sim M_{\rm pl}^4$.

The long-wavelength foamon contribution behaves effectively as a
dark-energy component with
$p_\Lambda\simeq-\rho_\Lambda$. Because the Ricci scalar is generally
inhomogeneous, however, the resulting effective cosmological term
$\Lambda_R(x)$ need not be spatially or temporally constant. Its
variation can lead to energy-momentum exchange with matter and to
nonlinear back-reaction during cosmological evolution.

The same curvature dependence implies that the effective dark-energy
density can vary in self-gravitating systems. Regions with larger
matter density and Ricci curvature can therefore be associated with
larger $\Lambda_R$. This provides possible corrections to galactic
rotation curves and gravitational lensing, and may lead to additional
effects in other gravitational, astrophysical and cosmological observables.

The framework presented here is phenomenological and relies on several
assumptions concerning the microscopic spacetime-foam dynamics,
foamon effective field theory, and the relation between Wilsonian
scales and cosmological evolution. A quantitative derivation from a
nonperturbative theory of quantum gravity, together with a
self-consistent treatment of cosmological and astrophysical
observations, remains an important direction for future work.

Consider the foamon field $\Phi_f$ associated with the fermion
operator $O_f=\bar\Psi_L\Psi_R$ introduced in Eq.~(\ref{coleman3}).
From the effective action (\ref{efff}), one obtains an induced
four-fermion interaction of the form
\begin{equation}
\frac{M_{\rm pl}^2}{16\pi^2 M_f^4}
(\bar\Psi_L\Psi_R)(\bar\Psi_L\Psi_R),
\label{ein4}
\end{equation}
together with the characteristic energy-density scale
\begin{equation}
\rho_f\sim\frac{M_f^2}{64\pi^2G},
\end{equation}
where $\xi_f=M_f^{-1}$ is the correlation length associated with the
fermion mass scale $M_f$. Such an interaction may provide an effective
beyond-the-Standard-Model operator \cite{Xue:2016dpl,Xue2017}, and is
consistent with the general conjecture that the parameters of
low-energy field theories may be influenced by spacetime wormhole
fluctuations \cite{Coleman1988,Preskill:1988na,Preskill1989}.
It is also interesting to consider foamon
sectors with nontrivial topology. In particular, an axion-associated
foamon field $\Phi_a$ could generate an effective axion action, mass
scale, and coupling to photons.

The spacetime foam dynamics, due to quantum gravity, naturally provides the ultraviolet cutoff $M_{\rm pl}$ for 
quantum field theories for matter constituents, i.e., particles of masses $m_{\rm phy}$. The ground state of field theories is the vacuum that does not contain
real particles with the relativistic energy-momentum relation 
$p^2 = m^2_{\rm phy}$. However, the vacuum is made up of 
virtual particles, which are fluctuations in quantum fields. 
Real particles
are energy-momentum excitations upon vacuum. Only the energy-momentum difference between the real particle state and the vacuum 
contributes to the energy-momentum tensor $8\pi GT_{\mu\nu}$ of the matter in the RHS of the classical Einstein equation. The 
vacuum energy density $M^4_{\rm pl}$ from quantum fluctuations of matter constituents
is conceptually and quantitatively not identified with the observed cosmological term in the LHS of the classical Einstein equation \cite{Xue2009a}. 
Instead, we show that the cosmological constant $\Lambda_R$ is attributed to the spacetime foamon field (dark) energy density 
$\rho_{_\Lambda}\propto M^2_{\rm pl}H^2$ (\ref{lambda}), which is of purely  
spacetime origin, appears in the LHS of the Einstein equation and interacts with the 
energy-momentum of physical particle excitations upon vacuum.
This picture differs from several recent proposals \cite{Carlip2019,Wang2020,Wang2020a,Wang2020b,Carlip2023,Blitz2025} that study the attractive (negative) 
energy of oscillating modes of the gravitational field at small scales, compensating for the positive vacuum energy of matter (particle) fields at larger scales, as heuristically discussed \cite{Wheeler1955, Misner1973} using Wheeler’s “spacetime foam”. 
In the present approach, however,
the observed dark-energy scale is instead associated with the
long-wavelength collective correlations of the spacetime foam.

%\section{Acknowledgment}
The author appreciates the
advice and suggestions of the reviewers. In this work, 
no AI tool is used except to check English and improve presentation.

%\bibliography{OscillationRef}
%\bibliography{../OscillationRef}
%\bibliography{../../MyBibFiles/MyBibFile.bib}

\providecommand{\href}[2]{#2}\begingroup\raggedright\endgroup

\end{document}